\documentclass[12pt]{article}
\usepackage{amsfonts,amsmath}
\usepackage{amssymb}
\usepackage[T2A]{fontenc}

\usepackage[hypertex] {hyperref}

\newcommand{\CH}{contracting homotopy }
\newcommand{\CHs}{contracting homotopies }

\newcommand{\Sp}{{\mathcal H}}

\newcommand{\dr}{{{\rm d}}}
\renewcommand{\theequation}{\thesection.\arabic{equation}}

\makeatletter \@addtoreset{equation}{section} \makeatother
\newcommand{\hhmt}{h}

\newcommand{\gt}{\tau}
 \newcommand{\eq}{\eqref}
\newcommand{\ie}{{\it i.e.,} }
{\vspace{3mm} }

\def\al{\alpha}

\def\*{\star}

\def\E2{\mathbf{E}}

\newcommand{\rhs}{{\it r.h.s.} }
\newcommand{\rhss}{{\it r.h.s.'s} }
\newcommand{\lhs}{{\it l.h.s.} }
\newcommand{\be}{\begin{equation}}
\newcommand{\ee}{\end{equation}}
\newcommand{\bee}{\begin{eqnarray}}
\newcommand{\beee}{\begin{array}}
\newcommand{\eee}{\end{eqnarray}}
\newcommand{\eeee}{\end{array}}
\newcommand{\gr}{\rho}
\newcommand{\ga}{\alpha}
\newcommand{\gb}{\beta}
\newcommand{\gbzer}{{\beta}}
\newcommand{\gga}{\gamma}

\newcommand{\ls}{\!\!\!\!\!\!}

\newcommand{\gd}{\delta}
\newcommand{\gl}{\lambda}

\newcommand{\gep}{\epsilon}
\newcommand{\gvep}{\varepsilon}
\newcommand{\gs}{\sigma}
\newcommand{\go}{\omega}

\newcommand{\by}{{\bar{y}}}
\newcommand{\dal}{\dot \alpha}
\newcommand{\dgb}{\dot \beta}

\newcommand{\q}{\,,\qquad}
\newcommand{\nn}{\nonumber}
\newcommand{\half}{\frac{1}{2}}

\newcommand{\p}{\partial}

\newcommand{\ff}{\frac}
\newcommand{\hmt}{\vartriangle}
\newcommand{\pa}{{  \dot{\ga}}}

\begin{document}

\begin{flushright}
FIAN/TD/10-2019\\
\end{flushright}

\vspace{0.5cm}
\begin{center}
{\large\bf Limiting Shifted Homotopy in Higher-Spin Theory and Spin-Locality }

\vspace{1 cm}

\textbf{V.E.~Didenko${}^1$, O.A.~Gelfond${}^{1,2}$, A.V.~Korybut${}^1$ and  M.A.~Vasiliev${}^1$}\\

\vspace{1 cm}

\textbf{}\textbf{}\\
 \vspace{0.5cm}
 \textit{${}^1$ I.E. Tamm Department of Theoretical Physics,
Lebedev Physical Institute,}\\
 \textit{ Leninsky prospect 53, 119991, Moscow, Russia }\\

\vspace{0.7 cm}\textit{ ${}^2$ Federal State Institution
"Scientific Research Institute for System Analysis
of the Russian Academy of Science",}\\
\textit{Nakhimovsky prospect 36-1, 117218, Moscow, Russia}

\par\end{center}

\begin{center}
\vspace{0.6cm}

\par\end{center}

\vspace{0.4cm}

\begin{abstract}
\noindent Higher-spin vertices containing up to quintic
interactions at the Lagrangian level are explicitly calculated in
the one-form sector of the non-linear unfolded higher-spin
equations using a $\gb\to-\infty$--shifted contracting homotopy
introduced in the paper.  The problem is solved in a background
independent way and for any value of the complex  parameter $\eta$
in the higher-spin equations. All obtained vertices are shown to
be spin-local containing a finite number of derivatives in the
spinor space for any given set of spins.  The  vertices
proportional to $\eta^2$ and $\bar \eta^2$ are in addition
ultra-local, \ie zero-forms that enter into the vertex in question
are free from the dependence on at least one of the spinor
variables $y$ or $\bar y$. Also the $\eta^2$ and $\bar \eta^2$
vertices are shown to vanish on any purely gravitational
background hence not contributing to the higher-spin current
interactions on $AdS_4$. This implies in particular that the
gravitational constant in front of the stress tensor is positive
being proportional to $\eta\bar \eta$. It is shown that the
$\gb$-shifted homotopy technique developed in this paper can be
reinterpreted as the conventional one but in the $\gb$-dependent
deformed star product.

\end{abstract}
\newpage
\tableofcontents

\newpage

\section{Introduction}

In this paper we continue our analysis of non-linear corrections resulting from
higher-spin (HS) equations of \cite{more} and their
locality properties. In the preceding papers
\cite{Gelfond:2018vmi,Didenko:2018fgx} a proper formalism based on
the Pfaffian
Locality Theorem (PLT) \cite{Gelfond:2018vmi} that allows one extracting
physical vertices in the local frame was proposed and set in
motion. Using it we have checked some   low-order vertices and
found them local. Here we extend this program to higher orders.

One of the most convenient ways of studying HS interaction problem is
by using the unfolded approach \cite{Vasiliev:1988xc,Vasiliev:1988sa} that makes HS symmetry
manifest. The equations of motion schematically read
\begin{align}
&\dr_x\go=-\go*\go+\Upsilon(\go,\go,
C)+\Upsilon(\go,\go,C,C)+\dots\,,\label{ver1}\\
&\dr_x C=-[\go,C]_{*}+\Upsilon(\go,C,C)+\dots\,,\label{ver2}
\end{align}
where master fields, differential one-form $\go(Y|x)$ and
zero-form $C(Y|x)$, depend on auxiliary spinor variables $Y^A
=(y^\ga,\bar y^{\dot\ga}) $ that encode HS physical fields along
with their on-shell descendants. These span naturally
representations of the HS algebra being realized as star-product
operation $*$ acting on $Y$'s. {By $\Upsilon(\go,C,\dots)$ and
$\Upsilon(\go, \go, C,\dots)$ we schematically denote higher order
corrections to the interaction vertices.} First vertices on the
\rhss, $-\go*\go$ and $-[\go,C]_*$, are fully determined by the
global HS symmetry. This allows one extracting all of the
remaining $\Upsilon$'s where the HS symmetry gets deformed from
the mere consistency requirement $\dr_x^2=0$. In practice this
procedure gets increasingly complicated with the order of
perturbation as can be seen for instance from
\cite{Vasiliev:1989yr}. Yet, at each order one faces cohomological
ambiguity in the form of field redefinition that one should
dispose one way or another. The freedom in field redefinitions is
therefore a natural freedom in the form of $\Upsilon$-vertices.
Algebraically, the process is equivalent to the resolution of the
Chevalley-Eilenberg cohomological problem for the HS algebra
\cite{Vasiliev:2007yc}. In \cite{Vasiliev:1988sa} the machinery of
Hochschild cohomology was used based on the observation that the
HS field equations remain consistent with all fields valued in any
associative algebra.

The deformation problem for \eqref{ver1},  \eqref{ver2} was
explicitly solved to all orders in \cite{more} where a  simple
generating system for \eqref{ver1},  \eqref{ver2} was proposed. This
system proves that there is no obstruction for system
\eqref{ver1},  \eqref{ver2} and allows one extracting HS vertices at
any order. The same time it naturally captures field redefinition
ambiguity. To be more specific, the generating equations of
\cite{more} are written in a bigger space large enough to include
representative freedom for fields $\go$ and $C$. As shown in
\cite{Vasiliev:2015wma} and in the forthcoming paper \cite{GV},
the setup of \cite{more} allows one to distinguish between
different star-product functional classes being of great
importance in the context of locality problem for HS equations
including the analysis of this paper.

As was originally
shown in \cite{bbb,bbd} (see also \cite{Metsaev:1991mt,Metsaev:1991nb})
HS vertices contain higher derivatives of the order
increasing with spin.
HS equations \eqref{ver1},  \eqref{ver2} are based on the
infinite-dimensional HS symmetry \cite{hsa4,OP1,KV1} that mixes fields of any spin.
 The transformation law includes higher derivatives. It is
therefore natural to expect some sort of non-locality resided in
equations \eqref{ver1},  \eqref{ver2}. It has long been known
however \cite{bbb,bbd,Metsaev:1991mt,Metsaev:1991nb} that at least
at cubic order (quadratic at the level of equations) interaction
of three massless fields is local, \ie any vertex $s_1-s_2-s_3$
contains at most   finite number of space-time derivatives. Put it
differently, while the number of derivatives in cubic interactions
grows with spin and hence is unbounded for the infinite tower of
massless fields, still for three given spins their interaction is
local. This means that at least at lowest interaction order the
structure of non-locality is very specific admitting local
decomposition in terms of individual spin vertices. An optimistic
point of view is that at any order HS vertex $s_1-s_2-\dots-s_n$
is local. More realistic option suggested in
\cite{Gelfond:2018vmi} and discussed in more detail below is that
some sort of spin-locality can occur in the twistor variable space
that controls space-time locality via the HS unfolded equations.
Whether it is so or not remains to be seen but some indication
that HS locality might break down at quartic level was given in
\cite{Metsaev:1991mt} using light cone formalism in Minkowski
space (see also \cite{Metsaev:2018x ip}).

More recently, the holographic reconstruction of HS bulk
interactions from free boundary theory in accordance with the
original HS holographic conjecture \cite{Klebanov:2002ja},
\cite{Sezgin:2002rt} was argued to point out certain non-locality
of the quartic vertex
\cite{Bekaert:2015tva,Sleight:2017pcz,Ponomarev:2017qab}.
Interpretation of these results however is not straightforward as
the holographic reconstruction lends itself to a specific choice of
field variables and it is not clear if the non-locality so observed is
universal or an artifact of the formalism. A related comment is
that the proper definition of locality in $AdS$ background with
non-commuting derivatives may require  identification of a specific ordering
prescription.

It is therefore
important to study HS   locality problem by  $AdS/CFT$--independent
 means directly in the bulk. The only option of this kind available so far is to
study nonlinear HS equations of \cite{more}. As demonstrated, in
particular, in this paper, our approach works not only for the
parity-even HS theories holographically dual to free boundary
theories but also for  HS theories with the arbitrary complex
parameter $\eta$ conjectured to be holographically
dual to the boundary models with Chern-Simons interactions in
accordance with \cite{Aharony:2011jz,Giombi:2011kc}. Note that
application of the idea of holographic reconstruction to the latter
models is far more involved.

Our approach rests on the analysis of HS generating equations
\cite{more} that reproduce system \eqref{ver1},  \eqref{ver2}
along with its field redefinition freedom. It makes it very
convenient to analyze its effect on HS locality. Technically, one
or another field representation for \rhss of \eqref{ver1},
\eqref{ver2} is accounted for by using one or another contracting
homotopy operator in the generating system \cite{more}. From this
perspective local (minimally non-local) interaction results from
the appropriate choice of \CH in the generating system.
Particularly, in \cite{Gelfond:2018vmi} a class of the so-called
shifted contracting homotopies \footnote{Called resolutions in
that reference.} was introduced and it was shown to reduce the
degree of non-locality in all orders of the theory governed by
PLT. Properties of these contracting homotopies were studied in
\cite{Didenko:2018fgx} where they were shown to reproduce local
lower-order vertex $\Upsilon(\go,C,C)$ found originally in
\cite{Vasiliev:2016xui,Vasiliev:2017cae} for $AdS_4$ provided the
PLT conditions are respected. The analysis of
\cite{Didenko:2018fgx} alongside revealed an important concept of
ultra-locality missing in other approaches. Namely, it was shown
that while $\Upsilon(\go,\go, C)$--vertex is always local it can
be reduced to the form where it contains no dependence on the
generating $y$ or $\bar y$ variables in the zero-forms $C(Y|x)$.
Such vertices were called ultra-local in \cite{Didenko:2018fgx}.
The concept of ultra-locality plays great role in locality in
general. Particularly, it can be shown that if one starts with
some local but not ultra-local vertex $\Upsilon(\go,\go,C)$ then
it entails non-locality of $\Upsilon(\go,C,C)$.

A natural question is whether spin ultra-locality persists at higher
order. To answer this question, in this paper we examine next-to-leading order
vertex $\Upsilon(\go,\go,C,C)$ that captures part of the quintic
interaction. The vertex has the following structure
\be\label{verin}
\Upsilon(\go,\go,C,C)=\Upsilon^{\eta\eta}+
\Upsilon^{\bar\eta\bar\eta}{}+
\Upsilon^{\eta\bar\eta}\,,
\ee
where $\eta$ is an arbitrary complex parameter of the $d=4$ HS
theory \cite{Vasiliev:1988sa,more} and the labels carried by the vertices
refer  to the $\eta$, $\bar \eta$--dependent overall factor in  the vertex
in question.
From the PLT perspective the vertices $\Upsilon^{\eta\eta}$, $\Upsilon^{\bar\eta\bar \eta}$  and
 $\Upsilon^{\eta\bar \eta}$ originate from
different PLT classes and thus may have different locality
properties. Remarkably, the   homotopy  technique developed in
this paper implies that, in accordance with PLT, the most
nontrivial part residing in $\Upsilon^{\eta\eta}$ and
$\Upsilon^{\bar\eta\bar \eta}$ turns out to be ultra-local, the
fact being of great importance for the locality properties at
higher orders.

 We  show that, in agreement with
\cite{Gelfond:2017wrh},  $\Upsilon^{\eta\eta}$ and
$\Upsilon^{\bar\eta\bar \eta}$ vanish if $\go$ is set to its
$AdS_4$ vacuum value. Moreover they vanish if $\go$ only contains spin $s\leq 2$
fields, particularly for any gravitational background.
In this case, the only contribution is from the vertex
$\Upsilon^{\eta\bar\eta}$ which is local at any $\go$. This
implies in turn that the resulting currents
 on the \rhs of the HS equations are local and the coupling constants in front
of them have definite signs independent of the phase of $\eta$. In particular,
the gravitational  constant as a coefficient in front of the stress tensor
is positive.


To arrive at these results we generalize shifted   \CH  of
\cite{Gelfond:2018vmi}  and
adopt them for higher-order analysis adding one more $\gbzer \frac{\partial}{\partial y}$ shift
with an arbitrary parameter $\gbzer$.  We show that the $\gbzer$-dependent
 \CH operators are well defined for $-\infty<\gbzer<1$. The
limit $\gbzer\to -\infty$ is then conjectured to correspond to the
local frame of HS theory. Using the introduced $\gb$--shifted \CH we
calculate vertices
$\Upsilon(\go,\go,C,C|\gbzer)$ and find their local limit at
$\gbzer\to-\infty$. In \cite{GV}, the effect of
parameter $\gbzer$ on HS generating equations is analyzed
in the language of classes of functions introduced in the context of
locality in \cite{Vasiliev:2015wma}.

One of the remarkable properties of the $\gb$--shifted \CH is that
the lower-order  vertices $\Upsilon^{\eta}$ and
$\Upsilon^{\bar\eta}$ found in
\cite{Vasiliev:2016xui,Didenko:2018fgx} as well as  the vertex
$\Upsilon^{\eta\bar\eta}$ found in this paper turn out to be
$\gb$-independent. This explains in particular why the  proposed
modified scheme  does not affect the lower-order results of
\cite{Vasiliev:2016xui,Didenko:2018fgx}. On the other hand, being
spurious at lower orders, $\gbzer$  becomes essential at higher
orders so that the limit $\gb\to-\infty$ becomes crucial for
finding local expressions for $\Upsilon^{\eta\eta}$ and
$\Upsilon^{\bar\eta\bar\eta}$.

In calculation of (anti)holomorphic vertices we faced a remarkable
cancellation. While our approach leads to the ultra-local
$\Upsilon(\go,\go, C,C)$ its consistency condition involves some
lower order vertices, particularly $\Upsilon(\go,C,C)$. The latter
is obtained using shifted homotopies. Though local it does not
however have a form of a minimally derivative vertex. A local
field redefinition can be carried out to make it the minimal one.
This redefinition was explicitly found in \cite{Vasiliev:2016xui}.
Remarkably, it is exactly this field redefinition that leads to a
dramatic cancellation in (anti)holomorphic structures,
particularly making $\Upsilon^{\eta\eta}(\go,\go, C,C)$ to vanish
on $AdS$ background. This makes us argue that the maximally local
vertices require maximal locality at each perturbation order.

Studying the properties of the $\gb$--shifted \CH  we show
that their effect is equivalent to certain $\gb$-induced
reordering of the original HS star product.\footnote{We are
grateful to Carlo Iazeolla, David De Filippi and Per Sundell for
the stimulating discussion of this issue.}
 An interesting remaining question is whether  the local
results for HS vertices can be equivalently obtained by virtue of
the conventional homotopy with no $\gb$-shift but from the
$\gb$-reordered HS equations at $\gb=-\infty$.

The paper is organized as follows: in Section \ref{HS} we recollect  necessary
background on nonlinear HS equations. The new $\gb$--shifted \CH is defined
in Section \ref{reso}  where also its properties
 are derived.  The final results for  vertices are presented in Section \ref{results}
while  technical details of their derivation are presented in
Sections \ref{derivation} and \ref{S2sec}. An alternative interpretation of the
$\gb$--shifted \CH is presented in Section \ref{order} where
it is shown to be equivalent to conventional homotopy  in the appropriate
$\gb$-dependent star product and then is used in Section \ref{S2sec} to
demonstrate cancellation of certain second-order contribution to
HS vertices. Conclusion contains discussion of the obtained results and
future directions.
Further technical detail are presented in
Appendices A  and B. Also Appendices~B  and C collect some useful
formulae.

\section{Higher-spin generating equations}
\label{HS}
\subsection{Nonlinear equations}
Nonlinear  HS equations \eqref{ver1},  \eqref{ver2} originate from
the following generating system \cite{more}
\be
\dr_x W+W*W=0\,,\label{HS1}
\ee
\be
\dr_x S+W*S+S*W=0\,,\label{HS2}
\ee
\be\dr_x B+[W,B]_*=0\,,\label{HS3}
\ee
\be S*S=i(\theta^{A} \theta_{A}+ B*\Gamma) \q \Gamma =\eta \gga +\bar \eta \bar\gga\,,
\label{HS4}\ee
\be
[S,B]_*=0\,.\label{HS5}
\ee
Master fields $W(Z;Y;K|x)$, $S(Z;Y;K|x)$ and $B(Z;Y;K|x)$
 depend on generating spinor variables $Z_A=(z_{\al},
\bar z_{\dal})$, $Y_A=(y_{\al}, \bar y_{\dal})$, where spinorial
indices range over two values, and discrete involutive elements
$K=(k,\bar k)$
\be\label{hcom}
\{k,y_{\al}\}=\{k,z_{\al}\}=0\,,\qquad [k,\bar y_{\dal}]=[k,\bar
z_{\dal}]=0\,,\qquad k^2=1\q [k\,,\bar k]=0\,.
\ee
Similarly for $\bar k$.

Star product is defined as follows
\be\label{star}
(f*g)(Z, Y)=\ff{1}{(2\pi)^4}\int \dr^4 U \dr^4 V f(Z+U; Y+U)g(Z-V;
Y+V)\exp(iU_{A}V^{A})\,.
\ee
Indices are raised and lowered with the aid of antisymmetric form
$\gep_{AB}=-\gep_{BA}$ as follows $X^{A}=\gep^{AB}X_{B}$ and
$X_A=X^{B}\gep_{BA}$. There are two types of anticommuting
differentials in \eqref{HS1},  \eqref{HS5}, space-time
$\mathrm{d}x^{n}$ and  spinor differentials
$\theta_{A}=(\theta_{\al}, \bar{\theta}_{\dal})$. Master fields
belong to different gradings with respect to these differentials,
$W=W_{n}\mathrm{d}x^{n}$,
$S=S_{\al}\theta^{\al}+\bar{S}_{\dal}\theta^{\dal}$, and $B$ is a
zero-form. $\theta$-differentials have the following commutation
rules
\be
\{\theta_{A}\,,\theta_{B}\}=0\q
\{\theta_{\al},k\}=\{\bar\theta_{\dal},\bar k\}=0\,,\qquad
[\theta_{\al},\bar k]=[\bar\theta_{\dal},k]=0\,.
\ee
Lastly, taking into account that $\theta^3=\bar\theta^3=0$,
\be\label{klein}
\gga=\exp({iz_{\al}y^{\al}})k\theta^{\al} \theta_{\al}\,,\qquad
\bar\gga=\exp({i\bar{z}_{\dal}\bar{y}^{\dal}})\bar
k\bar\theta^{\dal}\bar\theta_{\dal}\,
\ee
turn out to be central with respect to the star product, \ie
\be
\gga * f = f*\gga\q\bar \gga * f = f*\bar \gga\q\forall f=f(Z;Y;K;\theta)\,.
\ee

\subsection{Perturbative expansion}
\label{pert} To set up perturbation theory one starts with vacuum
solution
\begin{align}
&B_0=0\,,\label{B0}\\
&S_0=\theta^\al z_{\al}+\bar{\theta}^{\dal}\bar
z_{\dal}\,.\label{S0}
\end{align}
Plugging it into \eqref{HS1}-\eqref{HS5} we find that $W_0$ should
be $Z$-independent, $W_0=\go(Y; K|x)$, and satisfy \eqref{HS1}.
Similarly, at next order one gets $B_1=C(Y; K|x)$ from $[S_0,
B_1]=0$ and $C$ satisfies \eqref{HS3}. This way we find the first
terms on the \rhss
 of \eqref{ver1},  \eqref{ver2}. As in \cite{Vasiliev:1988sa},
our perturbative
expansion is in powers of the zero-forms $C$ with one-forms $\go$ treated as
being of zero order.

Since
\be
[S_0\,,]_* = -2i \mathrm{d}_Z\q \mathrm{d}_Z:= \theta^A \frac{\p}{\p Z^A}\,,
\ee
typical equation that one has to solve to determine the
field $Z$--dependence at any order  is
\be\label{steq}
\mathrm{d}_{Z} f(Z;Y;\theta)=J(Z;Y;\theta)\,,
\ee
where $J$ originates from the lower-order terms. For example, if
$f=S$, then (discarding the $\bar \theta, \bar z$ sector for brevity) most
general form that follows from equations at order $n$ is
\be\label{nord}
J_n=\theta^2\!\int\! \dr\tau \dr\gs C(y_1)\dots C(y_n) \rho \exp({i\tau
z_{\al}y^{\al}\!-\!A^{j}z^{\al}\p_{j\al}\!-\!B^{j} y^{\al}\p_{j\al}\!-\!
\ff i2P^{ij} \p_{i}^{\al}\p_{j\al}})\Big|_{y_i= 0}k\,,
\ee
where $\p_{i\al}=\ff{\p}{\p y^{\al}_{i}}$, $\tau$ and $\gs$ are
integration parameters over some compact domain,  $A^i$, $B^i$,
$P^{ij}$ are some $\tau,\gs$-dependent coefficients (it is
convenient to identify the coefficient in front of $z_\ga y^\ga$
with one of the integration variables $\tau$) and $\rho$ is some
polynomial in $\p_{i\ga}$ with $\tau,\gs$--dependent coefficients.
{Note also that while $C(y,\bar y)$ depends on $\bar y$ as well as
on $y$, here being restricted to purely holomorphic sector we
highlight only the $y$--dependence of $C$.} For instance, it is
easy to find condition on $S_1$ from \eqref{HS4}
\be
-2i\dr_z S_1=i\eta
C*\gga=i\eta\theta^2\int\dr\tau\gd(1-\tau)C(y_1)\exp({i\tau
z_{\al}y^{\al}-\tau z^{\al}\p_{1\al}})\Big|_{y_1=0} k\,.
\ee

Particular solution to \eqref{steq} can be written as
\be
f=\hmt J\,,
\ee
where $\hmt$ is some \CH operator. The  challenge is to find such
contracting homotopies that lead to HS vertices being as local as
possible. Contracting homotopies used in
\cite{Gelfond:2018vmi,Didenko:2018fgx} with respect to $z$ have
the form
\bee\label{oldres}
\hmt_{q}J(z;y;\theta)&=&(z+q)^{\al}\ff{\p}{\p\theta^{\al}}\int_{0}^{1}\dr\tau\ff1\tau
J(\tau z-(1-\tau)q; y; \tau\theta)\,,
\eee
where $q$ in principle can be any $z$-independent spinor
(operator). The form of \eqref{nord} suggests the following
natural choice for $q$
\be\label{dershift}
q_\al=-i\sum_j v^{j}\partial_{j\al}
\ee
with {\it a priori} arbitrary coefficients $v^i$. Conventional
\CH of \cite{more}, that
 leads to local field equations at
the linearized level, corresponds to $v^i=0$ and is known to be
inconsistent with locality at the interaction level
\cite{Giombi:2012ms,Boulanger:2015ova}. The role of field derivative shifts
in \eqref{dershift} is to set some control over non-local
contractions $\p_{i}^{\al}\p_{j\al}$ in vertices.

Let us clarify our notion of spin-locality  attributed to
spinor space rather than space-time, \ie to derivatives in
auxiliary spinor  rather than space-time variables. The two are
related via the unfolding routine that generally assumes
non-linear and infinite derivative one-to-one map. Therefore, what
we call local  should be understood as spin-local. In the
lowest order these two notions coincide.

The precise relation of spin-locality and the corresponding
space-time derivative behavior can be worked out on $AdS_4$
background. In that case in \eqref{ver1},  \eqref{ver2} one has
\be
\go=\Omega+\go'\,,
\ee
where $\Omega$ is the bilinear $AdS_4$ flat connection
\be\label{W0}
\Omega=\ff i4(\go_{\al\gb}y^{\al}y^{\gb}+\bar\go_{\dal\dgb}\bar
y^{\dal}\bar y^{\dgb}+2e_{\al\dal}y^{\al}\bar{y}{}^{\dal})
\ee
with $\go_{\al\gb}$, $\bar\go_{\dal\dgb}$ and $e_{\al\dal}$ being
the $AdS_4$ background fields while $\go'(Y|x)$ stands for
perturbative fluctuations. Plugging $\Omega$ into HS equations
\eqref{ver1},  \eqref{ver2} using \eqref{star} results in
\begin{align}
&\dr_x\go'-\go^{\al\gb}y_{\al}\ff{\p}{\p
y^{\gb}}\go'-\bar{\go}^{\dal\dgb}\bar{y}_{\dal}\ff{\p}{\p \bar
y^{\dgb}}\go'-e^{\al\dgb}\left(y_{\al}\ff{\p}{\p \bar
y^{\dgb}}+\bar{y}_{\dgb}\ff{\p}{\p
y^{\al}}\right)\go'=-\go'*\go'+\Upsilon(\Omega, \Omega,
C)+\dots\,,\label{ver1ads}\\
&\dr_x C-\go^{\al\gb}y_{\al}\ff{\p}{\p
y^{\gb}}C-\bar{\go}^{\dal\dgb}\bar{y}_{\dal}\ff{\p}{\p \bar
y^{\dgb}}C+ie^{\al\dgb}\left(y_{\al}\bar{y}_{\dgb}-\ff{\p}{\p
y^{\al}\p\bar{y}^{\dgb}}\right)C=-[\go', C]_{*}+\Upsilon(\Omega,
C, C)+\dots\label{ver2ads}\,.
\end{align}
Equations \eqref{ver1ads},  \eqref{ver2ads} show that  space-time
derivatives entering via De Rham derivative $\dr_x$ are related to
spinor derivatives. Particularly, at lowest order, space-time
derivative of the field $C(Y|x)$ is expressed via  second
derivative with respect to $\ff{\p}{\p y}$ and $\ff{\p}{\p\bar
y}$. At higher orders the map is more involved but still available
from \eqref{ver1ads},  \eqref{ver2ads}. As HS equations
\eqref{HS1}-\eqref{HS5} are naturally formulated in spinor terms
everywhere in this paper the locality is understood as
spin-locality. {On the other hand, as discussed in more detail in
\cite{GV}, spin-locality implies space-time locality of
higher-order interactions rewritten in terms of the original
constituent fields like $\go$ and $C$ and  local currents of
various ranks (degrees in fields)  built from these fields.}

Let us stress the important difference between the one-forms $\go$ and
zero-forms $C$ manifested by equations  (\ref{ver1ads}), (\ref{ver2ads}).
The \lhs  of (\ref{ver1ads}) is homogeneous in $Y$. As a result, $\go(Y)$
contains at most a finite number of derivatives of the dynamical frame-like
component in $\go$ for any given spin $s$
(the degree of homogeneity in $Y$ is $2(s-1)$; for more detail see e.g.
\cite{Vasiliev:1999ba}). That is why the presence of $\go$ can only add a finite
number of derivatives for a fixed spin,  not affecting general aspects of
the analysis of spin-locality.
On the other hand, equation (\ref{ver2ads}) relates infinitely many components
of $C(y,\bar y;K|x)$ with the space-time derivatives for any given spin $s$ because
\be
\label{sc}
2s =|n_y - n_{\bar y}|\,,
\ee
where $n_y$ and $ n_{\bar y}$ are  the numbers of unbarred and
barred variables $Y$ in $C(y,\bar y;K|x)$ \cite{Vasiliev:1999ba}.
As a result, contractions between zero-forms can produce
non-localities even  for fixed spins. Correspondingly, in the
analysis of this paper we will only control the form of
contractions between zero-forms $C$ neglecting contractions
involving one-forms $\go$. From (\ref{sc}) it also follows that,
if spins are fixed, to prove spin-locality it suffices to show
that the number of contractions of either unbarred or barred
$y,\bar y$-variables is finite.

The important result of \cite{Gelfond:2018vmi} is that the degree
of non-locality of HS vertices which depends on field variable
choice via \CH operators reduces provided a linear condition on
$v^i$ \eqref{dershift} from PLT is imposed. For the even sector
that (anti)-holomorphic parts of $W$ and $S$ belong to the
locality condition is
\be\label{even}
\sum_i(-)^iv^i_C=0\,,
\ee
while for the odd one where (anti)holomorphic part of $B$ resides
it reads
\be\label{odd}
\sum_i(-)^iv^i_C=1\,,
\ee
where by $v^i_C$ we label those coefficients in \eqref{dershift}
that are attributed to zero-forms $C$ derivatives only and index
$i$ runs over the values $i=1,\dots, n$, where $n$ is the
perturbative order (the amount of $C$'s). Shifts $q$
\eqref{dershift} with $v^i_C$ satisfying \eq{even} (\eq{odd}) will
be referred to as {\it PLT-even}({\it odd}).

Shifted \CH  with $q$ \eqref{dershift} contains shifts  that
differentiate fields $C$ and $\go$ but not exponential kernel like
in \eqref{nord}. It works perfectly fine at lowest interaction
order $\Upsilon(\go,\go,C)$ and $\Upsilon(\go,C,C)$ considered in
\cite{Didenko:2018fgx} but when it comes to higher orders it
becomes insufficient.  One reason for this is that at higher
orders homotopy field derivatives act on fields from lower orders
that show up in combinations with $y$-dependent kernels. This
suggests  that the corresponding $\p_{i\al}$ acts on the field
along with explicit $y$-dependence of the kernel as well,
demanding an extension of shifted \CH \eqref{dershift} to include
explicit differentiation over the $y$-variable
\be\label{yshift}
q'=q+i\gbzer\ff{\p}{\p y}\,,
\ee
where $\gbzer$ is a  parameter. As shown below, the respective
 \CHs are well defined for $-\infty<\gbzer <1 $. Since $\ff{\p}{\p y}$
and $y$ do not commute the form of \CH operator
\eqref{oldres} becomes ambiguous demanding an ordering choice. To make it well defined we will
use the integral representation discussed in the next section.

\section{$\gb$-shifted contracting homotopy }
\label{reso}
\subsection{Definition}
Contracting homotopy operator well suited for the higher-order
analysis of HS equations has the  form
\begin{align}\label{res}
\hmt_{q,\gbzer}J= \int \ff{\dr^2 u \dr^2
v}{(2\pi)^2}\exp (iu_{\al}v^{\al})\int_{0}^{1}\ff{\dr\tau}{\tau}(z+q-v)^{\al}\ff{\p}{\p\theta^{\al}}
J(\tau z-(1-\tau)(q-v); y+\gbzer u;
\tau\theta)\,,
\end{align}
where $q$ is a $y$- and $z$-independent spinor. $q$ can be an
operator that acts on fields $C$ and $\go$ as in \eqref{dershift}.
For $\gbzer=0$, operator $\hmt_{q,\gbzer}$ reduces down to
\eqref{oldres}, while for $\gbzer\neq 0$ it accounts for shift
\eqref{yshift}. Note that, by integration by parts,
 $v$ effectively differentiates over $u$, \ie $y$.

Contracting homotopy \eqref{res} gives resolution of identity
allowing one solving \eqref{steq}:
\be
\label{ur} \{\dr_z, \hmt_{q,\gbzer}\}=1-h_{q,\gbzer}\,,
\ee
where
\be\label{h}
h_{q,\, \gbzer}J(z; y; \theta)=\int \ff{\dr^2 u\, \dr^2
v}{(2\pi)^2}\, \exp (iu_{\al}v^{\al})J(-q+v; y+\gbzer u;0)\,
\ee
 is the cohomology projector to the $z,\theta$-independent part.

The proof of \eqref{ur} is similar to the conventional resolution of identity. Consider
\begin{align}
&(\dr_z\hmt_{q,\,\gb}\!+\!\hmt_{q,\,\gb}\dr_z)J(z;y;\theta)=\int
\ff{\dr^2 u \dr^2 v}{(2\pi)^2}\exp (iu_{\al}v^{\al})\int_{0}^{1}\ff{\dr\tau}{\tau}\Big(
\theta^{\al}\ff{\p}{\p\theta^{\al}}\!+\!(z\!+\!q\!-\!v)^{\al}\theta^{\gb}
\ff{\p}{\p\theta^{\al}}\ff{\p}{\p z^{\gb}}\nn\\
&\!+\!(z\!+\!q\!-\!v)^{\al}\ff{\p}{\p\theta^{\al}}\theta^{\gb}\ff{\p}{\p
z^{\gb}}\Big) J(\tau z\!-\!(1\!-\!\tau)(q\!-\!v); y\!+\!\gbzer u;
\tau\theta)=\nn\\
& \int \ff{\dr^2 u \dr^2
v}{(2\pi)^2}\exp (iu_{\al}v^{\al})\int_{0}^{1}\ff{\dr\tau}{\tau}\Big(
\theta^{\al}\ff{\p}{\p\theta^{\al}}\!+\!(z\!+\!q\!-\!v)^{\al}\ff{\p}{\p
(z\!+\!q\!-\!v)^{\al}}\Big)J(\tau z\!-\!(1\!-\!\tau)(q\!-\!v); y\!+\!\gbzer u;
\tau\theta)=\nn\\
&\int \ff{\dr^2 u \dr^2
v}{(2\pi)^2}\exp (iu_{\al}v^{\al})\int_{0}^{1}\dr\tau\,\ff{\dr}{\dr\tau} J(\tau
z\!-\!(1\!-\!\tau)(q\!-\!v); y\!+\!\gbzer u;
\tau\theta) =\nn\\
& J(z;y;\theta)\!-\!\int \ff{\dr^2 u\, \dr^2 v}{(2\pi)^2}\,
\exp (iu_{\al}v^{\al})J( v\!-\!q; y\!+\!\gbzer u;0)\,.
\end{align}

\subsection{Properties}
\label{prop}
\subsubsection{General}
$\gb$--shifted \CHs share standard properties derived in \cite{Didenko:2018fgx} at $\gbzer=0$.
First,  they
anticommute
\be\label{acom}
\hmt_{q_1,\,\gbzer_1}\hmt_{q_2,\,\gbzer_2}=-\hmt_{q_2,\,\gbzer_2}\hmt_{q_1,\,\gbzer_1}\,,\qquad
\hmt_{q,\gb}^2=0
\ee
and obey
\be\label{hhmt}
h_{q,\,\gbzer}\hmt_{q,\,\gbzer}=0\,.
\ee
Also they satisfy the following useful property
\be \label{simJakobi}
\hmt_{q_3,\,\gb_3}  \hmt_{q_2,\,\gb_2} -\hmt_{q_3,\,\gb_3} \hmt_{q_1,\,\gb_1}
+\hmt_{q_2,\,\gb_2}  \hmt_{q_1,\,\gb_1} =
 h_{q_3,\,\gb_3} \hmt_{q_2,\,\gb_2}\hmt_{q_1,\,\gb_1}\,
\ee
for any $q_i$ and $\gb_i$. Applying $h_{q_4,\,\gb_4}$ to both
sides of this relation and using that
\be
h_{q_2,\,\gb_2}h_{q_1,\,\gb_1} =h_{q_1,\,\gb_1}
\ee
one recovers {\it triangle identity} of \cite{Vasiliev:1989xz} in the form
\be \label{tri}
h_{q_4,\,\gb_4}  \hmt_{q_3,\,\gb_3}  \hmt_{q_2,\,\gb_2}-
h_{q_4,\,\gb_4}\hmt_{q_3,\,\gb_3} \hmt_{q_1,\,\gb_1} +h_{q_4,\,\gb_4}
\hmt_{q_2,\,\gb_2}
\hmt_{q_1,\,\gb_1}=h_{q_3,\,\gb_3}\hmt_{q_2,\,\gb_2}\hmt_{q_1,\,\gb_1}.
\ee
Though slightly modified, \CHs \eqref{res} share
star-exchange properties studied in \cite{Didenko:2018fgx}:
\begin{align}
&\hmt_{q,\,\gbzer}(f(y;k)*J(z;y;k;\theta))=f(y;k)*\hmt_{q+(1-\gbzer)p,\,
\gbzer}J(z;y;k;\theta)\,,\label{leftd}\\
&h_{q,\, \gbzer}(f(y;k)*J(z;y;k;\theta))=f(y,k)*h_{q+(1-\gbzer)p,\,
\gbzer}J(z;y;k;\theta)\,\label{lefth},\\
&\hmt_{q,\,\gbzer}(J(z;y;\theta)* k^\nu*
f(y;k))=\hmt_{q+(-1)^\nu (1-\gbzer)p,\,
\gbzer}\Big(J(z;y;\theta)* k^\nu\Big) * f\left(y,k\right)\,,\label{rightd}\\
&h_{q,\,\gbzer}(J(z;y;\theta)* k^\nu* f(y;k))=h_{q+(-1)^\nu
(1-\gbzer)p,\, \gbzer}\Big(J(z;y;\theta)* k^\nu\Big) *
f\left(y,k\right), \,\label{righth}
\end{align}
where  we use the notation \be\label{pi=} p_{\al}f(y;k)\equiv
f(y;k)p_{\al}:=-i\ff{\p}{\p y^{\al}}(f_{1}(y)+f_{2}(y)k)\,. \ee
Note that star-exchange formulae \eqref{leftd}-\eqref{righth}
literally acquire the form of those studied in
\cite{Didenko:2018fgx} for $\gb=0$ upon the  redefinition
\be\label{redef} \hat q=(1-\gb)q\,,\qquad \hat{p}=(1-\gb)p\,, \ee
which gives for example \be\label{lexch} \hmt_{\hat
q,\,\gbzer}(f(y;k)*J(z;y;k;\theta))=f(y;k)*\hmt_{\hat q+\hat p,\,
\gbzer}J(z;y;k;\theta)\,. \ee The class of  $\gb=0$ \CHs was
considered in \cite{Didenko:2018fgx} for which we adopt the
simplified notation \be\label{oldhom}
\hmt_{q}:=\hmt_{q,\,0}\,,\qquad h_{q}:=h_{q,\,0}\,. \ee The
following important combination of these operators typically shows
up at lower orders \be\label{def1.0dok0} \hhmt_c\hmt_b\hmt_a
f(z,y) \theta^\gb\theta_\gb = 2 \int {\dr^3_{\triangle} \tau\,}
(b-c)_\gga
 ( a - c)^\gga f( -\tau_1 c-\tau_3 b-\tau_2 a ,y)\,,
\ee
where the $\tau_{1-3}$ integration is carried over a simplex
\be \label{mera3simplex}
\int\dr_{\Delta}^3\tau:=\int_{[0,1]^3}\dr\tau_1\,\dr\tau_2\,\dr\tau_3\,
\gd(1-\tau_1-\tau_2-\tau_3)\,.
\ee
Eq. \eqref{def1.0dok0} entails the scaling property
\be\label{scale}
h_{\gl a}\Delta_{\gl b}\Delta_{\gl c} f(z,y)= \gl^2
h_{a}\Delta_{b}\Delta_{c} f(\gl z,y)\,\qquad \forall \gl\ne0\,.\ee

\subsubsection{Special property of $\hmt_{0,\gbzer}$ }
An important property of $\hmt_{0,\gbzer}$ is that it
anticommutes with the space-time differential $\mathrm{d}_x$
\begin{equation}\label{0gb}
\lbrace{ \mathrm{d}_x,\hmt_{0,\,\beta}\rbrace}=0\,.
\end{equation}
 Indeed,
consider for example a one-form in $\theta$
\be
\int \mathrm{d} \tau \theta^\alpha
g_\alpha(\partial_1,\dots,\partial_N,y,z)
\exp\Big(i\tau z_\alpha y^\alpha-A^j(\tau)z^\alpha
\partial_{j\alpha}-B^j(\tau)y^\alpha
\partial_{j\alpha}-\frac{i}{2}P^{ij}(\tau)\partial^\alpha_i\partial_{j\alpha}\Big)
\underbrace{\Phi\dots
\Phi}_N\,,
\ee
where $\Phi$ can be either $C$ or $\omega$. As a result of
application of $\mathrm{d}_x$  each $\Phi$ will turn into some
$\Upsilon$ from \eqref{ver1},  \eqref{ver2} according to equations
of motion. Corresponding derivative  with respect to $Y$ of $\Phi$
will be replaced by the derivative of $\Upsilon$. Since
operator $\hmt_{0,\,\beta}$ does not contain  derivatives over
arguments of $\Phi$ (the shift parameter $q=0$) (\ref{0gb})
follows.

As a consequence of resolution of identity (\ref{ur}) and (\ref{0gb}) we also
have
\begin{equation}
\label{dh}
[ \mathrm{d}_x\,,h_{0,\,\beta}]=0\,.
\end{equation}

\subsubsection{Action on $\gamma$}
\label{gammact}

An important property of the $\gb$--shifted \CH  is its action on the central element $\gga$
\eqref{klein}
\be\label{kleinpr}
\hmt_{q,\,\gbzer}\gga=\hmt_{  (1-\gbzer)^{-1}q,\,0}\gga\,.
\ee
That is when applied to $\gga$ it acts as shifted \CH \eqref{oldres} with a rescaled parameter $q$. The proof of
\eqref{kleinpr} is straightforward. Applying $\hmt_{q,\,\gbzer}$
to $\gga$ \eqref{klein} using \eqref{res} one finds that
\be
\hmt_{q,\,\gbzer}\gga=2\left(z+
{q}(1-\gbzer)^{-1}\right)^{\al}\theta_{\al}
\int_{0}^{1}\dr\tau\,\ff{(1-\gbzer)\tau}{(1-\gbzer(1-\tau))^3}
 \exp\left({\ff{i}{1-\gbzer(1-\tau)}(\tau
z-(1-\tau)q)_{\al}y^{\al}}\right)k\,.
\ee
Changing integration variable
\be
\tau'= {\tau}(1-\gbzer(1-\tau))^{-1}\in[0,1]
\ee
we obtain
\be
\hmt_{q,\,\gbzer}\gga=2(z+
\ff{q}{1-\gbzer})^{\al}\theta_{\al}\int_{0}^{1}\dr\tau'\,\tau'
 \exp({i(\tau'
z-(1-\tau') {q}(1-\gbzer)^{-1})_{\al}y^{\al}})k=\hmt_{
\ff{q}{1-\gbzer},\,0}\gga\,.
\ee

Using the anticommutativity property (\ref{acom}) from here it
follows also
\be\label{kleinpr1}
\hmt_{q_1,\,\gbzer_1}\hmt_{q_2,\,\gbzer_2}\gga=\hmt_{
(1-\gbzer_1)^{-1}q_1,\,0} \hmt_{  (1-\gbzer_2)^{-1}q_2,\,0}\gga\,.
\ee
Recall that \cite{Didenko:2018fgx}
\bee \label{dg}
\hmt_{q  }\gga\equiv\hmt_{q,\,0 }\gga&=&2(z^{\gb}+q^{\gb}
)\theta_{\gb}\int_{0}^{1}\dr\gt \gt \exp({i(\gt
z_{\al}\!-\!(1\!-\!\gt)q_{\al})y^{\al}})k\,,
 \\ \nn\label{ddg}
\hmt_{  q_1 }
\hmt_{  q_2}
\gga\equiv\hmt_{  q_1,\,0}
\hmt_{  q_2,\,0}
\gga &=& 2 (z+q_1)_\gga
 ( z+q_2)^\gga \int\dr_{\Delta}^3\tau\exp({i (\tau_1 z \!-\!\tau_2 q_2\!-\!\tau_3
 q_1)_{\al}y^\ga})k\,.
\eee

Since, HS vertices eventually are driven by $\gb$--shifted \CHs
applied to $\gga$ one might think of $\gbzer$ deformation as
unnecessary just leading to some rescaling of the homotopy
parameter. This turns out to be the case at lower orders but
drastically departs at higher orders where structures like
$\hmt(\hmt\gga*\hmt\gga)$ show up. Moreover, this fact explains
why the necessity  of the $\gb$--shifted \CH  was not seen in the
lower-order analysis of \cite{Didenko:2018fgx}. Property
\eqref{kleinpr} plays crucial role in the $\gb$--independence of
the lower-order HS vertices.

\subsection{$\gb$ dependence}
\subsubsection{Contracting homotopy operator  and cohomology projector}
To appreciate the role of the parameter $\gbzer$ it is useful to
apply the $\gb$--shifted \CH  to the function of the form
\be
\label{f} f (z,y, \theta)= \int_{[0,1]^2} \dr^2\tau\, \delta
(1-\tau_1 - \tau_2) \exp [i\tau_1 z_\ga y^\ga] \phi (\tau_1
z,\tau_2 y, \tau_1 \theta,\tau_1)\,,
\ee
which naturally results from the perturbative analysis at higher orders \cite{Vasiliev:2015wma}. In this case, formula (\ref{h}) yields
\be
\label{hgb0} h_{0,\,\gbzer} (f) =\int_0^1 \dr\,\tau
\frac{1}{(1-\gbzer \tau)^2}\int\ff{\dr^2 v \dr^2 u}{(2\pi)^2}\,
\exp i[ v_\ga u^\ga] \phi\Big(
u,\frac{(1-\tau)}{(1-\tau\gbzer)}y+\frac{\tau(1-\tau)\gbzer}{(1-\tau\gbzer)}
v ,0,\tau\Big)\,.
\ee

Derivation of the expression for   \CH is more involved. Upon the change of homotopy integration
variables described in Appendix~A  
it yields
\bee
\label{hmt0} &&\ls\ls\hmt_{0,\,\gbzer}(f) = \ff{1}{(2\pi)^2}\int
\!\dr^2 u\, \dr^2 v\!\int\dr_{\Delta}^3\tau \Big [\frac{(1-\gbzer)
\tau_1}{1-\gbzer(1-\tau_2)} \Big ]^{p-1}
\ls\,\exp i[ v_\ga u^\ga + \tau_1 z_\ga y^\ga]\times \nn\\
&&\ls\frac{(1-\gbzer \tau_1) z^\gbzer -\gbzer \tau_3
u^\gbzer}{1-\gbzer(1-\tau_2)}\frac{\p}{\p \theta^\gbzer} \phi\Big
(\tau_1 z +\frac{\tau_2 \tau_3 \gbzer}{1-\gbzer(1-\tau_2)} u, v +
\tau_3 y,\theta, \frac{1-\tau_3
-\gbzer\tau_1}{1-\gbzer(1-\tau_2)}\Big) \,,\label{hmtgb0}
\eee
where 
 $p$ is the degree of $f$ in $\theta$,
\be
f(w,u,\mu \theta,\tau)= \mu^p f(w,u, \theta,\tau)\,.
\ee
The last argument of $\phi$ in (\ref{hmtgb0}) results from the change of
integration variables.

Star product of the two functions of the form (\ref{f}) has the form
\cite{Vasiliev:2015wma}
\bee
\label{f12} f_1*f_2 && = \ff{1}{(2\pi)^2}\int_0^1 \dr\tau_1
\int_0^1 \dr\tau_2 \int \dr^2 s\dr^2 t \exp i [\tau_1\circ\tau_2
z_\ga y^\ga
 +s_\ga t^\ga]\nn\\
 && \times\phi_1 (\tau_1((1-\tau_2)z -\tau_2 y +s),(1-\tau_1)((1-\tau_2)y-\tau_2 z +s),\tau_1\theta,\tau_1)\nn\\
 &&\times\phi_2 (\tau_2((1-\tau_1)z +\tau_1 y -t),(1-\tau_2)((1-\tau_1)y+\tau_1 z +t),\tau_2\theta,\tau_2)\,,
\eee
where the product law
\be\label{tot}
\tau_1\circ \tau_2 = \tau_1(1-\tau_2) + \tau_2(1-\tau_1)\,
\ee
 is commutative and associative.
Note that $0\leq\tau_1\circ \tau_2\leq 1$ as well as $ 1- \tau_1\circ \tau_2$,
\be\label{1tot}
1- \tau_1\circ \tau_2 = \tau_1 \tau_2 + (1-\tau_1)(1-\tau_2)\,.
\ee
Formula (\ref{f12}) is heavily used in the computation of higher-order
corrections to HS equations.

The following comments are now in order. Formulae (\ref{hgb0}), (\ref{hmtgb0}) contain
 prefactors and rational dependence on the integration homotopy parameters $\tau$
due to the Gaussian integration resulting from the substitution of
the dependence on $u$ and $v$ into the exponential factor in (\ref{f}). The resulting expressions
are well defined for
\be
-\infty < \gbzer < 1\,.
\ee

Beyond this region,  they may contain divergencies due to the degeneracy of the quadratic form in the
 Gaussian integral. At $\gbzer=0$, these formulae reproduce those of the conventional homotopy
 introduced in \cite{more}. The presence of $\gbzer$
 in the denominators of (\ref{hgb0}) makes the limit $\gbzer\to -\infty$ nontrivial having the effect of suppressing
 many of the terms including those responsible for contractions between zero-forms $C$ as well
 as the $Y$-dependence in their arguments.

\subsubsection{Local limit}
 Let us briefly explain the idea of the limiting mechanism.
 Consider the following integral
 \be
\int_0^1\dr\tau \frac{1}{(1-\gb\tau)^{2+n}}\,.
\ee
Setting $\gb =-\gvep^{-1}$ we obtain
\be\label{hlim}
\int_0^1 \frac{\dr\tau}{(1-\gb\tau)^{2+n}} =(n+1)^{-1} \gvep
+O(\gvep^2)\,.
\ee
This implies that the contributions to HS vertices containing such
or further suppressed factors all vanish in the limit $\gb\to
-\infty$. For instance,
\be\label{hlim'}
\lim_{\gb\to-\infty}\int_0^1\dr\tau
\frac{(\gb\tau)^m}{(1-\gb\tau)^{2+n}} =0\q m\leq n\,.
\ee
On the other hand, expressions of the form
\be
\lim_{\gb\to -\infty}\int_0^1\dr\tau \frac{\gb (\gb
\tau)^m}{(1-\gb\tau)^{2+n}}
\ee
with $m\leq n$ remain  finite  in the limit $\gb\to-\infty$ (for more detail
see Appendix~B3). This fact admits an important interpretation: addition of
at least one power of $\tau$ to the integrand of an integral remaining finite
in the limit $\gb\to -\infty$ sends the expression to zero in the limit.
As will be illustrated in the next section (for more detail see \cite{GV}) this mechanism makes the limit
$\gb \to -\infty$ appropriate for locality.

 \subsubsection{Example}
\label{Example}

Let us outline how the limit $\gb\to-\infty$ leads to local results
considering as an example the $W_1*W_1$ part of the
 HS vertex in the one-form sector (\ref{ver1}).
Here it is important that the process starts from functions (\ref{f}) with a $y$-independent
 function $\phi$ since (see also Appendix~B) 
 \be
 C(y)* \gamma = C(-z) \exp(iz_\ga y^\ga) k \theta^{\al}\theta_{\al}\,.
 \ee
Then \cite{Didenko:2018fgx} 
\be\label{S1om0}
S_1=s_1 +\bar s_1=-\ff{\eta}{2}\hmt_{0}(C*\gga)+h.c. \,
\ee
with
\be\label{s11}
s_1= \eta  \int_0^1 \dr\tau \tau z_\ga \theta^\ga \exp (i\tau
z_\ga y^\ga) C(-\tau z) k\,
\ee
and the convention that $h.c.$ (Hermitean conjugation) swaps
barred and unbarred variables along with dotted and undotted
indices\footnote{We avoid using symbol $c.c.$ since HS equations
are invariant under the Hermitean conjugation  that not only
conjugates complex numbers but also reverses the order of product
factors including the  differentials $\theta$ that may give
additional sign compensating the effect of complex conjugation of
$i$ (for more detail see e.g. \cite{Vasiliev:1999ba}).}. Then
\be\label{W1conv0}
W_{1}= \ff{1}{2i}\hmt_{0}(\dr_x S_{1}+\go *S_1 +S_1 *\go)+h.c.\,.
\ee
For instance, the term $\hmt_{0} ( \go *s_1)$ has the structure
(discarding $\go$ along with all its derivatives)
\be
\hmt_{0} ( \go *s_1 )=\eta\int_0^1 \dr\tau_1
(1-\tau_1) \exp i(\tau_1 z_\ga y^\ga  ) 
   C(- \tau_1 z +\ldots)k\,.
\ee
The computation of $\hmt_{0} ( \go *s_1 )*\hmt_{0} ( \go *s_1 )$
is straightforward. Referring for the explicit final result to the
next sections  here we focus on the form of the  arguments of
zero-forms $C$.

By virtue of (\ref{f12}) this yields upon evaluation of the integration over $s$ and $t$
\bee
\hmt_{0} ( \go *s_1 ) *\hmt_{0} ( \go *s_1 )\sim\eta^2\int_0^1
\dr\tau_1 \int_0^1 \dr\gs_1 (1-\tau_1)(1-\gs_1)
\exp i[\tau_1\circ\gs_1 z_\ga y^\ga +\tau_1\gs_1\p_{1\ga}
\p_2^{\ga} ]\nn\\   C(y_1+\tau_1\gs_1 y - \tau_1(1-\gs_1)z+
\ldots)
 C(y_2+\gs_1\tau_1 y +\gs_1 (1-\tau_1)z + \ldots)|_{y_1=y_2=0}\label{w1w1}\,,
\eee
where $\go$ as well as all contractions  between all $\go$ and $C$
are discarded. The  important feature of this formula is that the
$y$--dependence of both factors of $C$ as well as the term
$\p_{1\ga}\p_2^{\ga}$  in the exponential are accompanied with the
factor of $\tau_1\gs_1 $. As a result,  the dependence on $y$ as
well as all contracting terms between the two fields $C$ disappear
from $h_{-\infty}(\hmt_{0} ( \go *s_1 ) *\hmt_{0}( \go *s_1 ))$.

Indeed, $\hmt_{0} ( \go *s_1 ) *\hmt_{0} ( \go *s_1 ) $ can be
represented in the form
\be
\hmt_{0} ( \go *s_1 ) *\hmt_{0} ( \go *s_1 ) = \hmt_{0} ( \go *s_1
) *\hmt_{0} ( \go *s_1 )\Big |_0 + \hmt_{0} ( \go *s_1 ) *\hmt_{0}
( \go *s_1 )\Big |_1\,,
\ee
where
\bee
\hmt_{0} ( \go *s_1 ) *\hmt_{0} ( \go *s_1 )\Big |_0
\sim\eta^2\int_0^1 \dr\tau_1 \int_0^1 \dr\gs_1 (1-\tau_1)(1-\gs_1)
\exp i[\tau_1\circ\gs_1 z_\ga y^\ga ]\times\nn\\   \times C( -
\tau_1 z+ \ldots)
 C( \gs_1 z + \ldots)\label{|0}\,
\eee
is a part of (\ref{w1w1}) with all terms containing a factor of $\tau_1\gs_1$
set to zero while
\bee
\hmt_{0} ( \go *s_1 ) *\hmt_{0} ( \go *s_1 )\Big |_1
\sim\eta^2\int_0^1 \dr\tau_1 \int_0^1 \dr\gs_1
(1-\tau_1)(1-\gs_1)\gs_1\tau_1
\exp i[\tau_1\circ\gs_1 z_\ga y^\ga ]   C C+\dots\label{|1}
\eee
contains the rest. The ultra-local term $\hmt_{0} ( \go *s_1 )
*\hmt_{0} ( \go *s_1 )\Big |_0$, which is free from the $y$
dependence in the arguments of $C$ as well as from the
contractions between the two factors of $C$, does contribute to
the final result.  On the other hand the term $\hmt_{0} ( \go *s_1
) *\hmt_{0} ( \go *s_1 )\Big |_1$ does not if the limiting
projector $h_{0,\,-\infty}$ is used.

Indeed, using that
\be
 {\gt_1(1-\gs_1)}\le{\gt_1\circ\gs_1} \q
 {\gs_1(1-\gt_1)}\le{\gt_1\circ\gs_1} \q
 \ee
we observe that
\bee
\hmt_{0} ( \go *s_1 ) *\hmt_{0} ( \go *s_1 )\Big |_1 \leq \int_0^1
\dr\tau_1 \int_0^1 \dr\gs_1 (\gs_1\circ\tau_1)^2
\exp i[\tau_1\circ\gs_1 z_\ga y^\ga ]   C(   \ldots)
 C(  \ldots)\,.
\eee

 Using (\ref{hgb0})
and   observing that any factor of  $\displaystyle{\ff{\tau_1\circ\gs_1}{(1-\tau_1\circ\gs_1\gb)}
\leq \ff{1}{-\gb}
}\to 0$ at $\gb\to-\infty$ one can see that analogously to (\ref{hlim'})
\be
\lim_{\gb\to -\infty}h_{0,\,\gb} (\hmt_{0} ( \go *s_1 ) *\hmt_{0}
( \go *s_1 ))\Big |_1 =0\,.
\ee

 Thus the expression $h_{0,\,-\infty}
(\hmt_{0} ( \go *s_1 ) *\hmt_{0}( \go *s_1 ))$ indeed turns out to
be ultra-local \ie local and with no $y$--dependence  in the
arguments of $C$. Let us stress the crucial role  of the factor of
$(1-\tau_1)(1-\gs_1)$ in the integration measure as well as of the
coefficient $\tau_1\gs_1$ in front of the $y$-dependence in $C$
and the contracting terms $\p_1\p_2$ in the exponential.

Note that one should not be confused that the projector
$h_{0,\,-\infty}$ is applied to a $\gb$--independent expression
$W_1*W_1$. This is because the first-order expression for $W_1$
considered in this example is $\gb$--independent as a consequence
of (\ref{kleinpr}). Other contributions to equations (\ref{ver1})
like those generated by $W_2$ are $\gb$--dependent. Some of them
originating from $S_2$ disappear in the $\gb\to-\infty$ limit upon
local field redefinition. The resolution of identity implies that
a single cohomology projector $h_{0,\,\gb}$ should be applied to
HS vertex generating equation \eqref{HS1} or \eqref{HS3} as a
whole (\ie it makes no sense to apply different projectors to
different terms in these equations since each of them is not
$z$--independent while the full expression is). Remarkably, a
similar phenomenon of the suppression of non-local terms in the
limit $\gb\to-\infty$ occurs in  other contributions to the
vertices in question as we explain now. That is why the limit
$\gb\to-\infty$  plays a distinguished role in the locality
analysis.

\section{$C^2\go^2$ vertices}
\label{results}
\subsection{Generalities}
Our goal is to extract $\Upsilon(\go,\go,C,C)$ vertex in
\eqref{ver1} from HS  equations \eqref{HS1}-\eqref{HS5}.
To do so we need to have all master fields to be solved up to the order
$C^2$. We label them as $W_2, S_2$ and $B_2$. Those that
are linear in $C$ are $W_1$, $S_1$ and $C$ itself. In terms of these fields the
desired vertex appears in parenthesis
\be\label{2nd}
\dr_x\go+\go*\go=-\{\go, W_1\}_*-(\dr_x W_1+\dr_x
W_2+W_1*W_1+\{\go,W_2\}_*)+ O(C^3)\,.
\ee
Note that $\dr_x W_1$ contributes to $C^2$--terms too\footnote{{In
general $\dr_x W_1$ and  $\dr_x W_2$ contain contribution of
arbitrary high order.}} through the second-order corrections
obtained in \cite{Didenko:2018fgx} (see Eqs.~\eq{go1},\,
\eq{C2gen}). Now, in solving for $W_1$ and
 $W_2{}\big|_{\eta\eta}$ we
will use \CH  $\hmt_{q,\,\gbzer}$.

 The local frame
is prescribed by PLT theorem \cite{Gelfond:2018vmi}. It applies
differently for different PLT field parity. Namely,
(anti)holomorphic $W$ belongs to even sector of HS equations and
the locality requirement \eqref{even} is fulfilled for instance
for $q=0$. So, solving  for  holomorphic $W$ and $S$ we can use
$\hmt_{0,\,\gbzer}$ with some  $\gbzer$. In other words,
$W_1=\hmt_{0,\,\gbzer}(\dots)$ and $W_2{}\big|_{\eta\eta}=\hmt_{0,\,\gbzer}(\dots)$,
where the precise expressions will be specified in what follows.

Generic $\go$ vertex $\Upsilon(\go,\go,C)$ calculated in
\cite{Didenko:2018fgx} and presented in Appendix~C  
turned out to be
ultra-local. To proceed further we need explicit expressions for
$W_1$ and $W_2$ to plug them into \eqref{2nd}. Once it is done we
take the limit $\gbzer\to-\infty$ to get the final result for the
local vertex that takes  form \eqref{verin} with

\begin{align}
&\Upsilon^{\eta\eta}(\go,\go,C,C)=\Upsilon^{\eta\eta}_{\go\go CC}+
\Upsilon^{\eta\eta}_{\go C\go C}+ \Upsilon^{\eta\eta}_{C\go\go C}+
\Upsilon^{\eta\eta}_{C\go C\go}+
\Upsilon^{\eta\eta}_{CC\go\go}+ \Upsilon^{\eta\eta}_{\go CC\go}\,,\label{eta}\\
&\Upsilon^{\bar\eta\bar\eta}{}(\go,\go,C,C)=\Upsilon^{\bar\eta\bar\eta}_{\go\go
CC}+ \Upsilon^{\bar\eta\bar\eta}_{\go C\go C}+
\Upsilon^{\bar\eta\bar\eta}_{C\go\go C}+
\Upsilon^{\bar\eta\bar\eta}_{C\go C\go}+
\Upsilon^{\bar\eta\bar\eta}_{CC\go\go}+
\Upsilon^{\bar\eta\bar\eta}_{\go CC\go}\,,\label{etabar}\\
&\Upsilon^{\eta\bar\eta}(\go,\go,C,C)=\Upsilon^{\eta\bar\eta}_{\go\go CC}+
\Upsilon^{\eta\bar\eta}_{\go C\go C}+
\Upsilon^{\eta\bar\eta}_{C\go\go C}+ \Upsilon^{\eta\bar\eta}_{C\go
C\go}+ \Upsilon^{\eta\bar\eta}_{CC\go\go}+
\Upsilon^{\eta\bar\eta}_{\go CC\go}\,,\label{etaeta}
\end{align}
where the lower labels refer to different orderings of the factors $\go$ and $C$
in the vertex.

 Calculation
of HS vertices is simplified by the following trick.
Since for any $z$, $\theta$--independent function $f(y)$,
$h_{q,\,\gbzer} f(y)= f(y)$, one can apply any projector  $h_{q,\,\gbzer}$
 to each
  $Z$, $\theta^A$--independent part on the \rhs of \eq{2nd}
  like   $\Upsilon^\eta_{C\go C}$, $\Upsilon^{\eta\eta}_{C\go C\go}$,
 $\Upsilon^{\eta\bar{\eta}}_{\go C\go C}$,
 {\it etc.}

Applying $h_{0,\,\gbzer}$ to $(\dr_x W_1+\dr_x W_2+W_1*W_1+\{\go,W_2\})
\big|_{\eta\eta}$ one arrives at
\be\label{2nd+}
\dr_x\go+\go*\go=\Upsilon (\go,\go,C)+ \Upsilon^{\eta\eta}(\go,\go,
C,C)+\dots\,,
\ee
where
 \be\label{ver}
\Upsilon^{\eta\eta}(\go,\go,C,C)=-h_{0,\,\gbzer}(W_1*W_1+\{\go,W_2\})
\big|_{\eta\eta}\,.
\ee
The convenience of this particular projector is that it annihilates
contributions from $\dr_x W_1$ and  $\dr_x W_2$. Indeed, by virtue of
\eqref{hhmt} and (\ref{dh})  one has
\be
h_{0,\,\gbzer}(\dr_x W_1)\big|_{\eta\eta}=h_{0,\,\gbzer}(\dr_x W_2)\big|_{\eta\eta}=0\,.
\ee

Now we are in a position to present the final results for all
vertices (\ref{eta})-(\ref{etaeta}) leaving the explanation of the
details of their derivation for
Sections \ref{noninvariant} and \ref{invariant}.

\subsection{$\eta^2$ vertices}

Using notations  \cite{Gelfond:2018vmi,Didenko:2018fgx} with $\bar{*}$ denoting the star product for the barred variables $\bar y$
and \bee
\label{CgoCgoCpt}
 && p_j{}_\ga := - i\ff{\p}{\p y^j{}^\ga} \q   \bar p_j{}_\pa :=  -i\ff{\p}{\p \by^j{}^\pa} \,,
 \\\label{t}
&& t_j{}_\ga := - i\ff{\p}{\p y{}^j{}^\ga}\q  \bar t_j{}_\pa :=
-i\ff{\p}{\p \by{}^j{}^\pa} \,,
 \eee
where $p_j{}_\ga$ and  $t_j{}_\ga$ denote derivatives with respect to the unbarred arguments
of zero-forms $C$ and one-forms $\go$, respectively, counted from the left to the right (indices $j=1,2$)
  we have
  \bee\label{verwwcc}
\Upsilon^{\eta\eta}_{\go\go C C} =
-\ff{\eta^2}{4}\int_{[0,1]^2}\dr\gs\dr\gs'\,\gs\gs'\int\dr_{\Delta}^3\tau\,
(t_{1}{}_\ga t_{2}^\ga)^2\, \rule{60pt}{0pt}
\\ \nn
\tau_1\,\exp\big[i{ (\tau_2\gs+\tau_3\gs'+\tau_1\gs\gs')  t_{1}{}_\ga t_{2}^\ga}\big]
 \go(y\!-\!(1\!-\!\tau_3)\gs y,\bar y; K)\,\bar{*}
\,\go(\tau_3\gs' y,\bar y;K)\bar{*}
\\
  \bar{*}\,C( \tau_2\gs   t_{1} \!+\! (1\!-\!\gs'(1\!-\!\tau_2))  t_{2} ,\bar
y;K)\,\bar{*}\,C(\!-\!\tau_1\gs   t_{1} \!-\!  \gs'\tau_1   t_{2} ,\bar
y;K)\,,\nn
\eee
 \bee\label{verccww}
 \Upsilon^{\eta\eta}_{CC\go\go
}=\ff{\eta^2}{4}\int_{[0,1]^2}\dr\gs\dr\gs'\,\gs\gs'\int\dr_{\Delta}^3\tau\,
(t_{1}{}_\ga t_{2}^\ga)^2\,
\rule{67pt}{0pt}\\ \nn
\tau_1\,\exp\big[-i  (\tau_2\gs+\tau_3\gs'+\tau_1\gs\gs')   t_{1}{}_\ga t_{2}^\ga\big]
  C( \tau_1\gs'   t_{1} +\gs\tau_1  t_{2} ,\bar
y;K)\bar{*}\\\bar{*}C(\!-\!\tau_2\gs   t_{2} \!-\! (1-\gs'(1-\tau_2))   t_{1} ,\bar
y;K)\bar{*}
 \go(\tau_3\gs' y,\bar y; K)\bar{*}\,\go(y-\gs(1-\tau_3) y,\bar y;
K)\,,\nn
\eee
\bee\label{vercwwc}
\Upsilon^{\eta\eta}_{C\go\go
C}=\ff{\eta^2}{4}\int_{[0,1]^2}\dr\gs\dr\gs'\,\gs\gs'\int\dr_{\Delta}^3\tau\,
(t_{1}{}_\ga t_{2}^\ga) ^2\,
\rule{108pt}{0pt}\\ \nn
 \tau_3\,
\exp\big[-i{(\tau_2\gs+\tau_1\gs'+\tau_3\gs\gs')  t_{1}{}_\ga t_{2}^\ga}\big]
 C(\!-\!\tau_1\gs'   t_{2} \!-\!(1-\gs(1-\tau_1))  t_{1} ,\bar
y;K)\bar{*}\go(-\tau_3\gs y,\bar y; K)\bar{*}\\
\bar{*}\,\go(-\tau_3\gs' y,\bar y;
K)\bar{*}C( \tau_2\gs  t_{1} \!+\!(1-\gs'(1-\tau_2))   t_{2} ,\bar
y;K)\,,\nn
\eee
\bee\label{verwcwc}
\Upsilon^{\eta\eta}_{\go C\go
C}\! \!=\! \ff{\eta^2}{4}\int_{[0,1]^2}\dr\gs\dr\gs'\,\gs\gs'\int\dr_{\Delta}^3\tau\,
(   t_{1}{}_\ga t_{2}^\ga  )^2\,
\rule{118pt}{0pt}\\ \nn
  \Big\{\tau_1
  \exp\big[i{ (\!-\!\tau_2\gs\!+\!\tau_3\gs'\!+\!\tau_1\gs\gs')  t_{1}{}_\ga t_{2}^\ga}\big]
 \go(y\!-\!(1\!-\!\tau_3)\gs y,\bar y;
K)\bar{*}C( \tau_2\gs   t_{1} \!-\!(1\!-\!\gs'(1\!-\!\tau_2))   t_{2} ,\bar
y;K)\,\bar{*}\,\nn\\\nn
  \quad\bar{*}
\go(\!-\!\tau_3\gs' y,\bar y;K) \bar{*}C(\!-\!\tau_1\gs  t_{1}
+\gs'\tau_1   t_{2} ,\bar y;K)
\nn\\
  \!-\! \tau_1
\exp\big[ i{ ( \tau_2\gs\!+\!\tau_3\gs'\!-\!\tau_1\gs\gs')  t_{1}{}_\ga t_{2}^\ga}\big]
  \go(y\!-\!(1\!-\!\tau_3)\gs y,\bar y;K)
 \bar{*}
C(\tau_1\gs  t_{1} \!-\!\gs'\tau_1   t_{2} ,\bar
y;K)\bar{*}\nn\\\nn
  \quad\bar{*}\go(\!-\!\tau_3\gs' y,\bar y;K)
\bar{*}C(\!-\!\tau_2\gs   t_{1} +(1\!-\!\gs'(1\!-\!\tau_2))  t_{2} ,\bar
y;K)\\
  -\tau_3
\exp\big[ i{ (\tau_2\gs\!-\!\tau_1\gs'+\tau_3\gs\gs')  t_{1}{}_\ga t_{2}^\ga}\big]
 \go(\tau_3\gs y,\bar y;
K)\bar{*}
\nn C(\!-\!\gs'\tau_1   t_{2} +(1\!-\!\gs(1\!-\!\tau_1))  t_{1} ,\bar
y;K)\bar{*}\\
\nn
  \quad\bar{*}
\go(\!-\!\tau_3\gs' y,\bar y;K)
  \bar{*}C(\!-\!\tau_2\gs  t_{1} +(1\!-\!\gs'(1\!-\!\tau_2)) t_{2} ,\bar
y;K)\Big\}\,,\nn
\eee
\bee\label{vercwcw}
\Upsilon^{\eta\eta}_{C\go C\go
}\! \!=\! \!\ff{\eta^2}{4}\int_{[0,1]^2}\dr\gs\dr\gs'\,\gs\gs'\int\dr_{\Delta}^3\tau\,
(t_{1}{}_\ga t_{2}^\ga) ^2\,
\rule{110pt}{0pt}\\ \nn
\Big\{\tau_1\,\exp\big[i{ (\tau_2\gs+\tau_3\gs'-\tau_1\gs\gs')  t_{1}{}_\ga t_{2}^\ga}\big]
C(\tau_2\gs  t_{2} -(1-\gs'(1-\tau_2))  t_{1} ,\bar
y;K)\bar{*}\go(-\tau_3\gs' y,\bar
y;K)\bar{*}
\\\nn
  \quad\bar{*}C(-\tau_1\gs   t_{2} +\gs'\tau_1  t_{1} ,\bar
y;K)\bar{*}\go(y-(1-\tau_3)\gs y,\bar y; K) \\ \nn -\tau_1\,
\exp\big[-i{ (\tau_2\gs-\tau_3\gs'-\tau_1\gs\gs')  t_{1}{}_\ga
t_{2}^\ga}\big]
 C( \tau_1\gs   t_{2}-\gs'\tau_1  t_{1} ,\bar
y;K)\bar{*}\go(-\tau_3\gs' y,\bar y;K) \bar{*}
\nn\\\nn
  \quad\bar{*}C(\!-\!\tau_2\gs  t_{2} \!+\!(1-\gs'(1-\tau_2))  t_{1} ,\bar
y;K)\bar{*}\go(y-(1-\tau_3)\gs y,\bar y;
K)\nn\\
-\tau_3\,
 \exp\big[ i{ (-\tau_2\gs+\tau_1\gs'+\tau_3\gs\gs')  t_{1}{}_\ga t_{2}^\ga}\big]
C( \gs'\tau_1  t_{2} -(1-\gs(1-\tau_1))  t_{1} ,\bar
y;K)\bar{*}\go(-\tau_3\gs y,\bar y;
K)\bar{*}\nn\\
  \quad\bar{*}C( \tau_2\gs  t_{1} -(1-\gs'(1-\tau_2))  t_{2} ,\bar
y;K)\bar{*}\go(\tau_3\gs' y,\bar y;K) \nn
\Big\}\,, \eee
\bee\label{verwccw}
\Upsilon^{\eta\eta}_{\go CC\go
}=\ff{\eta^2}{4}\int_{[0,1]^2}\dr\gs\dr\gs'\,\gs\gs'\int\dr_{\Delta}^3\tau\,
(t_{1}{}_\ga t_{2}^\ga) ^2\, \rule{90pt}{0pt}\\ \nn
\Big\{\tau_1\,
\exp\big[-i{ (-\tau_2\gs+\tau_3\gs'-\tau_1\gs\gs')  t_{1}{}_\ga t_{2}^\ga}\big]
  \,\go(y-(1-\tau_3)\gs y,\bar
y;K)\bar{*}C( \tau_1\gs  t_{1} +\gs'\tau_1   t_{2} ,\bar
y;K)\bar{*}\\
\bar{*}C(-\tau_2\gs   t_{1} -(1-\gs'(1-\tau_2))   t_{2} ,\bar
y;K)\bar{*}\go(\tau_3\gs' y,\bar y; K)
\nn\\
-
\tau_1\,
  \exp \big[i{ (\tau_2\gs-\tau_3\gs'+\tau_1\gs\gs')  t_{1}{}_\ga t_{2}^\ga}\big]
\, \go(\tau_3\gs' y,\bar y;K)
\bar{*}C( \tau_2\gs  t_{2} +(1-\gs'(1-\tau_2))  t_{1} ,\bar
y;K)\bar{*}\nn\\
\bar{*}C(-\tau_1\gs   t_{2}-\gs'\tau_1   t_{1} ,\bar
y;K)\bar{*}\go(y-(1-\tau_3)\gs y,\bar y;
K)\nn\\
+\tau_3\,
 \exp \big[i{ (\tau_2\gs+\tau_1\gs'-\tau_3\gs\gs')  t_{1}{}_\ga t_{2}^\ga}\big]
 \, \go(-\tau_3\gs y,\bar y;
K)\bar{*}C( \gs'\tau_1   t_{2} +(1-\gs(1-\tau_1))  t_{1} ,\bar
y;K)\bar{*}\nn\\
\bar{*}C(-\tau_2\gs   t_{1} -(1-\gs'(1-\tau_2))  t_{2} ,\bar
y;K)\bar{*}\go(\tau_3\gs' y,\bar y;K)\Big\}\,.\nn
\eee

All  vertices \eqref{verwwcc}-\eqref{verwccw} are local because neither exponentials nor
arguments of  $C$-fields contain $C$-derivatives $p_{j\ga}$. As discussed in
Section \ref{pert}, in this case the $\bar *$ product over the antiholomorphic
variables $\bar y$ does not spoil the spin-locality of the vertices.  Moreover
all obtained vertices are  ultra-local in the sense of \cite{Didenko:2018fgx}. Indeed fields $C$ in \eqref{verwwcc}-\eqref{verwccw} carry no dependence on $y$ just as
 for central on-shell theorem that describes propagation
of free fields or its generalization via vertex \eqref{goUps1}-\eqref{goUps3}. This implies
that at the next order they will not produce non-localities via
star product.

Antiholomorphic vertex \eqref{etabar} results from \eqref{eta} by
swapping barred and unbarred  sectors including $\eta$ and
$\bar\eta$.

Another property of the resulting $\eta^2$ and $\bar\eta^2$ vertices
\eqref{eta} and \eqref{etabar} is
\be
\Upsilon^{\eta\eta}=\Upsilon^{\bar{\eta}\bar\eta}{}=0\quad\textnormal{for}\quad
\go=\go_{s\leq2}\,,
\ee
where $\go_{s\leq2}$ denotes the part of $\go$ that is at most bilinear
in the $Y$--variables and commutative with respect to possible color indices
to correspond to the gravitational spin-two sector.
Indeed, since all vertices \eqref{verwwcc}-\eqref{verwccw} contain the
prefactor of
four $\go$-derivatives $(t_1\cdot t_2)^2$ it makes the whole vertex
vanish for  commutative $\go$'s  at most bilinear in oscillators.
Particularly, it vanishes for $\go$ describing $AdS_4$ background.
This fact is in agreement with the result shown in
\cite{Gelfond:2017wrh}  that local HS current
interactions are independent of the phase of $\eta$
that guarantees in particular positivity of the gravitational
constant identified with the coefficient in front of the
stress tensor.  It also agrees with the $AdS_4/CFT_3$
prediction on the parity dependence  of three-point correlation
functions \cite{Maldacena:2011jn, Vasiliev:2016xui}.

\subsection{$\eta\bar\eta$ vertices}
\label{pinvv} Using for brevity the notation \eqref{oldhom} used in
\cite{Didenko:2018fgx},
the $\eta\bar\eta$ vertices are
\begin{align}\label{mixwwCC}
&\Upsilon^{\eta\bar{\eta}}_{\omega \omega C
C}=-\frac{\eta\bar\eta}{16}\omega* \omega* C * C *
h_{p_2}\hmt_{p_1 +2p_2}\hmt_{p_1+2p_2+t_2}\gamma*
\bar{h}_{\bar{p}_2}
\bar{\hmt}_{\bar{p}_1+\bar{p}_2+\bar{t}_1+\bar{t}_2}
\bar{\hmt}_{\bar{p}_1+\bar{p}_2+\bar{t}_2}\bar{\gamma}\nn\\
&+\delta \Upsilon_{\omega \omega C C}^{\eta\bar{\eta}}+h.c.\,,
\end{align}
\begin{multline}
\Upsilon^{\eta \bar{\eta}}_{C\omega\omega C}=
\frac{\eta\bar\eta}{16}C* \omega*\omega* C*\\
\Big[\frac{1}{2}h_{p_1+2p_2+t_1+t_2}\hmt_{p_1+2p_2+t_1+2t_2}\hmt_{p_1+2p_2+2t_1+2t_2}\gamma*
\bar{h}_{\bar{p}_1+2\bar{p}_2+\bar{t}_1+\bar{t}_2}
\bar{\hmt}_{\bar{p}_1+2\bar{p}_2+\bar{t}_1+2\bar{t}_2}
\bar{\hmt}_{\bar{p}_1+2\bar{p}_2+2\bar{t}_1+2\bar{t}_2}\bar{\gamma}\\
-h_{p_1+2p_2+t_1+t_2}\hmt_{p_1+2p_2+t_1+2t_2}\hmt_{p_1+2p_2+2t_1+2t_2}\gamma*
\bar{h}_{\bar{p}_1+\bar{p}_2+\bar{t}_1+\bar{t}_2}
\bar{\hmt}_{\bar{p}_1+2\bar{p}_2+2\bar{t}_1+2\bar{t}_2}\bar{\hmt}_{\bar{p}_2}\bar{\gamma}\\
+\big(h_{p_2}-h_{p_1+2p_2+t_1+2t_2}\big)\hmt_{p_2+t_2}\hmt_{p_1+2p_2+t_1+t_2}\gamma*
\bar{h}_{\bar{p}_1+\bar{p}_2+\bar{t}_1+\bar{t}_2}
\bar{\hmt}_{\bar{p}_1+2\bar{p}_2+\bar{t}_1+2\bar{t}_2}
\bar{\hmt}_{\bar{p}_1+2\bar{p}_2+2\bar{t}_1+2\bar{t}_2}\bar{\gamma}
\Big]+h.c.\,,
\end{multline}
\begin{multline}
\Upsilon^{\eta\bar \eta}_{\omega C\omega C}=\frac{\eta\bar\eta}{16}\omega* C*
\omega* C*\\
\Big[\big(h_{p_2}-h_{p_1+2p_2+t_1+2t_2}\big)\hmt_{p_2+t_2}\hmt_{p_1+2p_2+t_1+t_2}\gamma*
 \bar{h}_{\bar{p}_1+\bar{p}_2+\bar{t}_1+\bar{t}_2}\bar{\hmt}_{\bar{p}_2}
 \bar{\hmt}_{\bar{p}_1+2\bar{p}_2+\bar{t}_1+2\bar{t}_2}\bar{\gamma}\\
-h_{p_1+2p_2+t_1+2t_2}\hmt_{p_2+t_2}\hmt_{p_1+2p_2+t_2}\gamma*
\bar{h}_{\bar{p}_1+\bar{p}_2+\bar{t}_1+\bar{t}_2}\bar{\hmt}_{\bar{p}_2+\bar{t}_2}
\bar{\hmt}_{\bar{p}_2}\bar{\gamma}\\
+\big(h_{p_2}-h_{p_1+2p_2+2t_2}\big)\hmt_{p_2+t_2}\hmt_{p_1+2p_2+t_2}\gamma*
\big(\bar{h}_{\bar{p}_1+\bar{p}_2+\bar{t}_1+\bar{t}_2}-\bar{h}_{\bar{p}_1+2\bar{p}_2+2\bar{t}_2}\big)
\bar{\hmt}_{\bar{p}_1+\bar{p}_2+\bar{t}_2}\bar{\hmt}_{\bar{p}_2}\bar{\gamma}\\
-\frac{1}{2} h_{p_1+2p_2+t_1+t_2}\hmt_{p_1+2p_2+t_1+2t_2}\hmt_{p_2}\gamma*
\bar{h}_{\bar{p}_1+2\bar{p}_2+\bar{t}_1+\bar{t}_2}\bar{\hmt}_{\bar{p}_1+2\bar{p}_2+t_1+2\bar{t}_2}
\bar{\hmt}_{\bar{p}_2}\bar{\gamma}\\
+\frac{1}{2}h_{p_1+2p_2+t_2}\hmt_{p_1+2p_2+2t_2}\hmt_{p_2}\gamma*
\bar{h}_{\bar{p}_1+2\bar{p}_2+\bar{t}_2}\bar{\hmt}_{\bar{p}_1+2\bar{p}_2+2\bar{t}_2}
\bar{\hmt}_{\bar{p}_2}\bar{\gamma}\\
+\frac{1}{2}h_{p_1+2p_2+t_1+2t_2}\hmt_{p_2+t_2}\hmt_{p_2}\gamma*
\bar{h}_{\bar{p}_1+2\bar{p}_2+\bar{t}_1+2\bar{t}_2}
\bar{\hmt}_{\bar{p}_2+\bar{t}_2}\bar{\hmt}_{\bar{p}_2}\bar{\gamma}\\
-\frac{1}{2}h_{p_1+2p_2+2t_2}\hmt_{p_2+t_2}\hmt_{p_2}\gamma*
\bar{h}_{\bar{p}_1+2\bar{p}_2+2\bar{t}_2}\bar{\hmt}_{\bar{p}_2+\bar{t}_2}\bar{\hmt}_{\bar{p}_2}
\bar{\gamma}\Big]+h.c.\,,
\end{multline}
\begin{multline}
\Upsilon^{\eta\bar \eta}_{\omega C C \omega}=\frac{\eta\bar\eta}{16}\omega*
\ C* C* \omega*\\
\Big[\frac{1}{2} h_{p_2}\hmt_{p_2+2t_2}\hmt_{p_1+2p_2+2t_2}\gamma*
\bar{h}_{\bar{p}_2+\bar{t}_2}\bar{\hmt}_{\bar{p}_2+2\bar{t}_2}
\bar{\hmt}_{\bar{p}_1+2\bar{p}_2+\bar{t}_1+2\bar{t}_2}\bar{\gamma}\\
-\frac{1}{2}h_{p_1+2p_2+2t_2}\hmt_{p_2+t_2}\hmt_{p_2+2t_2}\gamma*
\bar{h}_{\bar{p}_1+2\bar{p}_2+2\bar{t}_2}\bar{\hmt}_{\bar{p}_2+\bar{t}_2}
\bar{\hmt}_{\bar{p}_2+2\bar{t}_2}\bar{\gamma}\\
+h_{p_2+2t_2}\hmt_{p_2+t_2}\hmt_{p_1+2p_2+t_1+2t_2}\gamma*
\bar{h}_{\bar{p}_1+2\bar{p}_2+\bar{t}_1+2\bar{t}_2}\bar{\hmt}_{\bar{p}_2+\bar{t}_2}
\bar{\hmt}_{\bar{p}_1+\bar{p}_2+\bar{t}_1+\bar{t}_2}\bar{\gamma}\\
+h_{p_2+2t_2}\hmt_{p_2+t_2}\hmt_{p_1+2p_2+2t_2}\gamma*
\big(\bar{h}_{\bar{p}_1+\bar{p}_2+\bar{t}_2}-\bar{h}_{\bar{p}_2+\bar{t}_2}\big)
\bar{\hmt}_{\bar{p}_1+\bar{p}_2+\bar{t}_1+\bar{t}_2}\bar{\hmt}_{\bar{p}_1+2\bar{p}_2+2\bar{t}_2}
\bar{\gamma}\\
+h_{p_2+2t_2}\hmt_{p_1+2p_2+2t_2}\hmt_{p_1+2p_2+t_1+2t_2}\gamma*
\big(\bar{h}_{\bar{p}_1+\bar{p}_2+\bar{t}_1+2\bar{t}_2}-\bar{h}_{\bar{p}_2+\bar{t}_2}\big)
\bar{\hmt}_{\bar{p}_1+\bar{p}_2+\bar{t}_1+\bar{t}_2}\bar{\hmt}_{\bar{p}_2+2\bar{t}_2}
\bar{\gamma}\Big]+\delta \Upsilon_{\omega  C
C\omega}^{\eta\bar{\eta}}+h.c.\,,
\end{multline}
\begin{multline}\label{mixCwCw}
\Upsilon^{\eta\bar \eta}_{C \omega C  \omega }=\frac{\eta\bar\eta}{16}C* \omega *  C* \omega*\\
\Big[\big(h_{p_1+2p_2+2t_1+2t_2}-h_{p_2+2t_2}\big)\hmt_{p_2+t_1+2t_2}\hmt_{p_1+2p_2+t_1+2t_2}\gamma*\bar{h}_{\bar{p}_1+\bar{p}_2+\bar{t}_1+
2\bar{t}_2}\bar{\hmt}_{\bar{p}_1+\bar{p}_2+\bar{t}_1+\bar{t}_2}\bar{\hmt}_{\bar{p}_1+2\bar{p}_2+2\bar{t}_1+2\bar{t}_2}\bar{\gamma}\\
-h_{p_2+2t_2}\hmt_{p_2+t_2}\hmt_{p_1+2p_2+t_1+2t_2}
\gamma*\bar{h}_{\bar{p}_1+\bar{p}_2+\bar{t}_1+\bar{t}_2}
\bar{\hmt}_{\bar{p}_1+2\bar{p}_2+\bar{t}_1+2\bar{t}_2}
\bar{\hmt}_{\bar{p}_1+2\bar{p}_2+2\bar{t}_1+2\bar{t}_2}\bar{\gamma}\Big]+h.c.\,,
\end{multline}
\begin{multline}
\label{mixCCww}
\Upsilon^{\eta\bar{\eta}}_{C C \omega \omega }=\frac{\eta\bar\eta}{16}C* C* \omega*\omega*\\
 h_{p_2+2t_1+2t_2}\hmt_{p_2+t_1+2t_2}\hmt_{p_1+2p_2+2t_1+2t_2}\gamma* \bar{h}_{\bar{p}_1+\bar{p}_2+\bar{t}_1+2\bar{t}_2}
 \bar{\hmt}_{\bar{p}_1+2\bar{p}_2+2\bar{t}_1+2\bar{t}_2}\bar{\hmt}_{\bar{p}_1+
 \bar{p}_2+\bar{t}_1+\bar{t}_2}\bar{\gamma}+ \delta \Upsilon_{C C\omega \omega
 }^{\eta\bar{\eta}}+h.c.\,,
\end{multline}
where
\begin{multline}\label{dver1}
\delta \Upsilon_{\omega \omega C C}^{\eta\bar{\eta}}= \dfrac{i\eta
\bar{\eta}}{8}\int_0^1\mathrm{d} \tau\,
e^{i(t_1+t_2)^\alpha (\tau p_1+(\tau-1)p_2-t_1)_{\al}}\omega(y,\bar{y})\bar{*}\omega(y,\bar{y})\bar{*}C(\tau y,\bar{y})\bar{*}C((\tau-1)y,\bar{y})k\ast\\
\ast
\bar{h}_{\bar{p}_1+\bar{p}_2+\bar{t}_1+\bar{t}_2}\bar{\hmt}_{\bar{p}_1+\bar{p}_2+\bar{t}_2}\bar{\hmt}_{\bar{p}_1+\bar{p}_2}\bar{\gamma}
\end{multline}
\begin{multline}\label{dver2}
\delta \Upsilon^{\eta\bar{\eta}}_{\omega C C \omega}= \dfrac{i\eta
\bar{\eta}}{8}\int_0^1 \mathrm{d} \tau\,
e^{i(t_1+t_2)^\alpha (\tau p_1+(\tau-1)p_2+t_1)_\alpha}\omega(y,\bar{y})\bar{*}C(\tau y,\bar{y})\bar{*}C((\tau-1)y,\bar{y})\bar{*}\omega(-y,\bar{y})k\ast\\
\ast (\bar{h}_{\bar{p}_1+\bar{p}_2+\bar{t}_1+\bar{t}_2}-
\bar{h}_{\bar{p}_1+\bar{p}_2+2\bar{t}_2})\bar{\hmt}_{\bar{p}_1+\bar{p}_2+\bar{t}_1+2\bar{t}_2}\bar{\hmt}_{\bar{p}_1+\bar{p}_2+\bar{t}_2}\bar{\gamma}
\end{multline}
\begin{multline}\label{dver3}
\delta\Upsilon^{\eta \bar{\eta}}_{CC\omega \omega}=\dfrac{i\eta
\bar{\eta}}{8}\int_0^1 \mathrm{d} \tau\,
e^{i(t_1+t_2)^\alpha (\tau p_1+(\tau-1)p_2-t_1)_{\al}} C(\tau y,\bar{y})\bar{*}C((\tau-1)y,\bar{y})\bar{*}\omega(-y,\bar{y})\bar{*}\omega(-y,\bar{y})k\ast\\
\ast
\bar{h}_{\bar{p}_1+\bar{p}_2+\bar{t}_1+\bar{t}_2}\bar{\hmt}_{\bar{p}_1+\bar{p}_2+2\bar{t}_1+2\bar{t}_2}\bar{\hmt}_{\bar{p}_1+\bar{p}_2+
\bar{t}_1+2\bar{t}_2}\bar{\gamma}\,,
\end{multline}
where $\delta\Upsilon^{\eta \bar{\eta}}$'s show up from specific
local and $z$--independent field redefinition of field $B$
designed to make lower order vertices to contain minimal number of
derivative contractions. The redefinition of this type is
introduced in section \ref{B2}.

Using also that $f(y)* \bar f(\bar y) = f(y) \bar f(\bar y)$ for
any $f(y)$ and $\bar f(\bar y)$ and, hence,
\be
h_a\hmt_b\hmt_c \gga*\bar{h}_{\bar a} \bar{\hmt}_{\bar b}\bar{\hmt}_{\bar{c}}\bar{\gamma}=
h_a\hmt_b\hmt_c\gga \,\bar{h}_{\bar a} \bar{\hmt}_{\bar b}\bar{\hmt}_{\bar{c}}\bar{\gamma}
\ee
we observe that the resulting $\eta\bar\eta$ vertices are free from contractions between spinor indices
 of the zero-forms $C$ in either
holomorphic or antiholomorphic sectors (or both). This can be
easily seen from the fact shown in \cite{Didenko:2018fgx} that
the  expression
\be\label{homac}
C*C*h_{a_1p_1+a_2p_2}\hmt_{b_1p_1+b_2p_2}\hmt_{c_1p_1+c_2p_2}\gga\,
\ee
 is local provided that
\be\label{locgen}
a_2-a_1=b_2-b_1=c_2-c_1=1\,.
\ee
One can see that all  vertices (\ref{mixwwCC})-(\ref{mixCCww})
meet this condition.

\section{Derivation details}
\label{derivation}

It is natural to expect that calculation of vertices
\eqref{eta}-\eqref{etaeta} should be quite involved. To compute
$W_2$ one needs $B_2$ and $S_2$ which potentially leads to
complicated analysis. Vertex $\Upsilon^{\eta\bar\eta}$ however
comes from the cross-product of holomorphic and anti-holomorphic
first-order fields and therefore is simpler. In particular, in
this sector the analysis is $\gb$--independent because of the
properties considered in Section \ref{gammact}. The challenge is
to calculate \eqref{eta}. Somewhat surprisingly the analysis in
the holomorphic sector simplifies due to a remarkable cancellation
that takes place for the $S_2$ contribution to the vertices. The
cancellation is related to a structure relation found in
\cite{GV}. We will show that in the limit $\gb\to-\infty$ one can
ignore (anti)holomorphic part of the $S$ field to the second order
as it turns out to give no contribution to (anti)holomorphic
vertices provided one fixes $B_2$ corresponding to the minimal
local couplings in $\Upsilon(\go,C,C)$ found in
\cite{Vasiliev:2016xui}.

\subsection{First order }
\label{First-order fields}

From \eqref{HS4} we have at first order
\be\label{S1eq}
-2i\dr_Z S_1=i\eta\, C*\gga+i\bar\eta\, C*\bar\gga\,,
\ee
which is solved as
\be\label{S1om}
S_1=-\ff{\eta}{2}\hmt_{0,\,\gbzer}(C*\gga)+h.c.=-\ff{\eta}{2}C*\hmt_{p,\,0}\gga+h.c.\,,
\ee
where in the last line we made use of \eqref{leftd} and
\eqref{kleinpr}. We see that though  \CH \eqref{res} depends
on $\gbzer$ the result for $S_1$ is $\gbzer$-independent thanks to
special properties (\ref{kleinpr}) of central elements \eqref{klein}. Thus obtained
$S_1$ is therefore identical to the one resulting from the
conventional  \CH  $\hmt_{0,\,0}$.

 The situation with $W_1$ is analogous. From \eqref{HS2} at first
order we have
\be\label{W1eq}
2i\dr_ZW_{1}=\dr_x S_{1}+\go *S_1 +S_1 *\go+h.c.\,.
\ee
Solving it using $\hmt_{0,\,\gbzer}$ {(recall that due to
\eqref{0gb} and \eqref{acom}, \eqref{S1om}
$\hmt_{0,\,\gbzer}(\dr_x S_{1})=0$) }
\be\label{W1conv}
W_{1}=\ff{1}{2i}\hmt_{0,\,\gbzer}(\go *S_1 +S_1 *\go)+h.c.
\ee
one finds upon using \eqref{leftd}, \eqref{rightd} and
\eqref{kleinpr} that
\be\label{W1om}
W_1=-\ff{\eta}{4i}\left( C*\go*\hmt_{p+t,\,0}\hmt_{p+2t,\,0}\gga-
\go*C*\hmt_{p+t,\,0}\hmt_{p,\,0}\gga\right)+h.c.\,.
\ee
Again, $W_1$ is $\gbzer$-independent and equal to the one obtained
via conventional contracting homotopy. This makes in particular
vertex $\Upsilon(\go,\go,C)$ $\gb$--independent. Note that for
$\gb=1$ the invariance of on-shell theorem was also demonstrated
in \cite{DeFilippi:2019jqq}.

\subsection{Second order. $\eta^2$ sector}
 \label{noninvariant}
\subsubsection{Solving for $B_2$}
\label{B2}
  To find $B_2$ following \cite{Didenko:2018fgx} we use \eqref{HS5} up to the second order which
amounts to
\be
-2i\dr_z B_2+[S_1,C]_*=0\,.
\ee
Substituting $S_1$ from \eqref{S1om} and using \eqref{leftd} one
gets (similarly in the $\bar\eta$ -- sector)
\be
-2i\dr_z
B_2=-\ff{\eta}{2}C*C*\hmt_{p_2,\,0}\gga+\ff{\eta}{2}C*\hmt_{p_1,\,0}\gga*C=
-\ff{\eta}{2}C*C*(\hmt_{p_2,\,0}-\hmt_{p_1+2p_2,\,0})\gga\,,
\ee
where $p_1$ and $p_2$ are derivatives (\ref{CgoCgoCpt}) of the
first and second factors of $C$, respectively.

The field $B$ belongs to the PLT-odd sector of HS equation in
nomenclature of \cite{Gelfond:2018vmi}. Therefore, \CH corresponding to the local frame should satisfy \eqref{odd}. At
this order this leads us to use
$\hmt_{(1-\gb)(v_1p_1+v_2p_2),\,\gbzer}$ for solving $B_2$, where
$v_2-v_1=1$. The overall  prefactor of $(1-\gb)$ appears
due to $\gb$-induced rescaling \eqref{redef} (see \cite{GV} for
the corresponding PLT modification). One can choose any value for
the remaining $v_1+v_2$ parameter. In \cite{Didenko:2018fgx} it
was shown that the resulting $B_2$ is independent of that free
parameter. So, we can take $v_2=1$ and $v_1=0$ for convenience
\be
B_2=\ff{\eta}{4i}\hmt_{(1-\gb)p_2,\,\gbzer}(C*C*(\hmt_{p_2,\,0}-\hmt_{p_1+2p_2,\,0})\gga)\,.
\ee
Again, using \eqref{leftd} and \eqref{kleinpr} we end up with the
$\gbzer$-independent result from \cite{Didenko:2018fgx}
\be\label{b2}
B_2=\ff{\eta}{4i}C*C*\hmt_{p_1+2p_2,\,0}\hmt_{p_2,\,0}\gga+h.c.\,.
\ee
This entails that vertex $\Upsilon(\go,C,C)$ is also $\gb$
independent (see \eqref{CUps1loc}-\eqref{ccCUps3loc}).

An important comment is as follows. While $B_2$ does reproduce
local vertices in accordance with the PLT they do not carry
minimal amount of derivatives. It was shown in
\cite{Vasiliev:2016xui} that the number of derivatives can be
reduced for
\begin{align}\label{B2min}
&B_{2}^{min}=B_2+\gd B_2(y)\,,\\
&\gd B_2(y)=\frac{\eta}{2}\int_0^1 \dr \tau \,C\left(\tau
y,\bar{y};
K\right)\,\overline{*}\,C\left(\left(\tau-1\right)y,\bar{y};
K\right)k+h.c.\label{dB2}
\end{align}
with $B_2$ \eqref{b2}. {It is interesting to know if the field
redefinition $\gd B_2$ admits any shifted homotopy representation.
In section \ref{S2sec} it will be shown that some limiting
representation does exist.}

\subsubsection{Solving for $S_2$ and $W_2$}
Equation for $S_2$ resulting from \eqref{HS4} in the
$\eta^2$ sector is
\begin{equation}
\label{dS2hol} \dr_z S_2=\frac{i}{2}(i\eta B_2* \gamma-S_1*
S_1)\,.
\end{equation}
The $\eta^2$ part of $S_2$ belongs to PLT-even locality class.
Therefore, it can be solved using $\hmt_{0,\,\gb}$
\be\label{S2hol}
S_2=\frac{i}{2}\hmt_{0,\,\beta}(i\eta B_2* \gamma-S_1* S_1)\,.
\ee
In \cite{GV} it is shown, that so obtained $S_2$ \eqref{S2hol}
belongs to a specific class of functions  $\Sp^{+0}$ which
contribution to dynamical equations of motion \eqref{HS1} is
ultra-local in the limit $\beta \rightarrow -\infty$ in the given
order of perturbation theory. Moreover, we will show in Section \ref{S2sec}  that such
contribution is just zero for $B_2$ chosen from \eqref{B2min}.
This implies in particular that in solving for holomorphic part of
$W_2$ one can ignore the contribution due to $S_2$.

Equation for $W_2$ has the form
\begin{equation}\label{dzW2}
\dr_z W_2=\frac{1}{2i}(\dr_x S_1+\dr_x S_2+W_1* S_1+S_1*
W_1+\omega * S_2+S_2 * \omega)\,.
\end{equation}
Again, the $\eta^2$--part of $W_2$ belongs to PLT-even class and
can be solved using $\hmt_{0,\,\gb}$.
This results in that
$\dr_x S_1$ as well as $\dr_x S_2$ vanish after applying
$\hmt_{0,\gbzer}$  because, as discussed in Section \ref{prop}, $\lbrace
\dr_x,\hmt_{0,\gbzer}\rbrace=0$ and $\hmt_{0,\gbzer}\hmt_{0,\gbzer}=0$.
As argued above, terms that result
from $S_2$ can be omitted. Hence the part of $W_2$ to be
taken into account analyzing the dynamical equations is
\be
W_2\simeq\ff{1}{2i}\hmt_{0,\,\beta} \left(W_1*
S_1+S_1*W_1\right)\,,
\ee
where $\simeq$ means equality up to terms coming from $S_2$, which
we collectively denote by $W_2'$
\be\label{W2'}
W_2'=\ff{1}{2i}\hmt_{0,\,\beta} \left(\go*
S_2+S_2*\go\right)\,.
\ee
This piece will be shown in Section \ref{S2sec} not to contribute to the final vertex
provided $S_2$ itself is solved for using \eqref{B2min}.
Substituting $W_1$ from \eqref{W1om} and $S_1$ from \eqref{S1om}
we obtain the final result which we leave for Appendix~B4.

Vertex $\Upsilon(\go,\go,C,C)$ comes about in two pieces
\eqref{ver}. Each contains
$\Upsilon^{\eta\eta}$, $\Upsilon^{\bar\eta\bar\eta}{}$ and
 $\Upsilon^{\eta\bar\eta}$.  We start with
$\Upsilon^{\eta\eta}$.
From that $\Upsilon^{\bar\eta\bar\eta}{}$ can be obtained by
swapping $y\leftrightarrow\bar y$, $\eta\leftrightarrow\bar\eta$.

\subsubsection{$\eta^2$ vertex $\Upsilon^{\eta\eta}$}
\label{noninvariant verts}

Vertex $\Upsilon^{\eta\eta}$ contains six structures  \eqref{eta}. Computation of
each of these is pretty similar. Some of the structures result from
one of the two pieces \eqref{ver}. Namely,
$\Upsilon^{\eta\eta}_{CC\go\go}$ and $\Upsilon^{\eta\eta}_{\go\go
CC}$ belong to $h_{0,\,\gbzer}\{\go,W_2\}_*$, while
$\Upsilon^{\eta\eta}_{C\go\go C}$ is in $h_{0,\,\gbzer}(W_1*W_1)$.
The rest acquire contributions from the two simultaneously. To simplify
presentation we calculate here $\Upsilon^{\eta\eta}_{C\go\go C}$
and $\Upsilon^{\eta\eta}_{CC\go\go}$ only each coming from a
single piece of different origin.

We begin with $\Upsilon^{\eta\eta}_{CC\go\go}$. From \eqref{ver}
this one is given by
\be\label{verdef1}
\Upsilon^{\eta\eta}_{CC\go\go}(\gbzer)=-h_{0,\,\gbzer}(W_2*\go)\Big|_{CC\go\go}\simeq
-\ff{1}{2i}h_{0,\,\gbzer}(\hmt_{0,\,\gbzer}(S_1*W_1)*\go)\Big|_{CC\go\go}\,,
\ee
where $\simeq$ means that  terms that do not
contribute to the vertex at $\gbzer\to-\infty$ are discarded. For the exact
equality one should add contribution from $S_2$ to $W_2$, but that
just vanishes in the vertex $\Upsilon(-\infty)$ as shown in  Section \ref{S2sec}.

Using \eqref{S1exp}, \eqref{W1exp} and \eqref{hwd} we find the
following result\footnote{Expressions containing $\bar{*}$ should
be understood as follows. Star product $\bar *$ involves
antiholomorphic oscillators $\bar y$ only, while fields $\go$ and
$C$ upon being differentiated are taken at $y=0$. {For example
below in \eqref{ccwwbeta} $C$'s and $\go$'s on the left should be
understood as $C(0,\bar y)$ and $\go(0, \bar y)$ while their
$y$-dependence is placed into integrals by means of the identity
$f(y)=f(0)e^{-ip^{\al}y_{\al}}$.}}
\bee\label{ccwwbeta}
&&\Upsilon^{\eta\eta}_{CC\go\go}(\gbzer)
=-\ff{\eta^2}{4}C\bar{*}C\bar{*}\go\bar{*}\go \int_{0}^{1}\dr\gs
\int_{0}^{1}\dr\tau_1 \int\dr^3_{\Delta}\tau'
\,\,\,\ff{1-\gbzer}{\tau_1'\xi^2}t_2^{\al}t_1^{\gb}\ff{\p}{\p
p_1^{\al}}\ff{\p}{\p p_2^{\gb}}\times\\
&&\exp i({ y^{\al}a_{\al}+a_1t_1^{\al}t_{2\al}+
a_2t_2^{\al}p_{1\al}+a_3t_1^{\al}p_{2\al}+a_4t_2^{\al}p_{2\al}
+ a_5t_1^{\al}p_{1\al}+a_6p_1^{\al}p_{2\al}}),\nn
\eee
\begin{align}
&a^{\al}=\ff{1}{\xi}(\tau_1\tau_1'(p_1+p_2+t_1)+(1-\tau_1)\tau_3't_1+
(1-\gs\tau_{\circ})t_2-\gbzer\tau_{\circ}(1-\gs)t_2)^{\al}\,,\\
&a_1=\ff{1}{\xi}(\tau_1\tau_1'+\tau_3'(1-\tau_1))+
\ff{1-\gbzer}{\xi}\gs(\tau_1'(1-\tau_1)+ \tau_1\tau_3')\,,\\
&a_2=\ff{1-\gbzer}{\xi}\gs\tau_1(1-\tau_1')-\ff{1}{\xi}\tau_1\tau_1'\,,\\
&a_3=\ff{\gbzer}{\xi}\tau_1'\tau_3'(1-2\tau_1)+\tau_3'-1\,,\\
&a_4=-\ff{\tau_1\tau_1'}{\xi}-\ff{1-\gbzer}{\xi}\gs\tau_1'(1-\tau_1)\,,\\
&a_5=-\ff{\tau_1\tau_1'}{\xi}+\ff{1-\gbzer}\xi\tau_1\tau_3'\,,\\
&a_6=\ff{\tau_1\tau_1'}{\xi}\,,\qquad
\end{align}
\be\label{kruzhochek}\tau_{\circ}:=\tau_1\circ\tau_1'=\tau_1+\tau_1'-2\tau_1\tau_1',\ee   \be
\label{betaznamen}\xi=1-\gbzer\tau_{\circ},\ee where $p_i$, $t_j$
are defined in (\ref{CgoCgoCpt}), (\ref{t}). Homotopy integration
variable $\gs$ appears from the application of $\hmt_{0,\,\gbzer}$
\eqref{res}, $\tau_1$  from $S_1$ and $\tau_{1-3}'$ from $W_1$.
Since we consider the vertex in the limit $\gbzer\to-\infty$ it
was not necessary to carry out exact calculation \eqref{ccwwbeta}
as lots of terms vanish in this limit. However we would like to
demonstrate that the expression provided is perfectly well defined
for any $\gbzer<1$. For finite $\gbzer<1$ the integrand has no
poles and therefore the result is finite. For $\gbzer\to-\infty$
it is easy to see that each $a_i$ in the exponential is bounded by
a $\gbzer$--independent constant due to inequalities
\be\label{ineq}
\tau_3'\leq 1-\tau_1'\,,\quad \tau_1(1-\tau_1')\leq\tau_{\circ}\,,\quad
\tau_1'(1-\tau_1)\leq\tau_{\circ}\,.
\ee

Hence, the integration behavior depends on the pre-exponential
that forms in \eqref{ccwwbeta} after differentiation over
$p_1$ and $p_2$. It acquires the following leading form at
large $\gbzer$
\be
\ff{\gbzer^3\tau\tau_1\tau_3'(1-\tau_1')(1-\tau_1)}{(1+\gbzer\tau_{\circ})^4}\leq
\ff{\gbzer^2}{(1-\gbzer\tau_{\circ})^3}\,,
\ee
which is integrable and has finite limit at $\gbzer\to-\infty$.

To take the limit we first integrate \eqref{ccwwbeta} over
$\tau_2'$ which is only  present in the measure
$\dr^3_{\Delta}\tau'$. It gives the integration domain for
$\tau_3'\in[0,1-\tau_1']$ which makes it convenient introducing
$\tau_3'=(1-\tau_1')\gs'$, where $\gs'\in[0,1]$ is the new
integration variable. The vertex then reduces to
\begin{align}\label{ccwwbigbeta}
\Upsilon^{\eta\eta}_{CC\go\go}(\gbzer)
=-\ff{1}{4}C\bar{*}C\bar{*}\go\bar{*}\go
\int_{[0,1]^2}\dr\gs\,\dr\gs'\,\gs\gs'
\int_{[0,1]^2}\dr\tau_1\,\dr\tau_1'
\ff{\gbzer^3\tau_1(1-\tau_1')^2(1-\tau_1)}{(1-\gbzer\tau_{\circ})^4}\times\\
\times (t_1t_2)^2\, \exp i\left({-\ff{ \gbzer}{\xi}\tau_1(1-\tau_1')
A-\ff{ \gbzer}\xi\tau_1'(1-\tau_1)B+\ff{1}{\xi}C+D}\right)+O(1/\gbzer)\,,\nn
\end{align}
where
\begin{align}
&A=\gs\gs' t_1^{\al}t_{2\al}+\gs t_2^{\al}p_{1\al}
+\gs't_1^{\al}p_{1\al}-\gs y^{\al}t_{2\al}\,,\label{A}\\
&B=\gs t_1^{\al}t_{2\al}-\gs' t_1^{\al}p_{2\al}-\gs
t_{2}^{\al}p_{2\al}-\gs y^{\al}t_{2\al}\,,\label{B}\\
&C=\gs' t_1^{\al}t_{2\al}+\gs'y^{\al}t_{1\al}\,,\quad
D=y^{\al}t_{2\al}+(\gs'-1)t_1^{\al}p_{2\al}\label{CD}\,.
\end{align}
The last step is to take limit $\gbzer\to-\infty$ which we do
using \eqref{I1inf}. This gives
\begin{align}
\Upsilon^{\eta\eta}_{CC\go\go}
=\ff{1}{4}C\bar{*}C\bar{*}\go\bar{*}\go\, (t_1t_2)^2
\int_{[0,1]^2}\dr\gs\,\dr\gs'\,\gs\gs'\int\dr^3_{\Delta}\tau\,\tau_1
\exp i({ \tau_1A+ \tau_2B+ \tau_3C+ D})\,.
\end{align}
Recalling the definition of   $p_i$ \eqref{CgoCgoCpt} one rewrites
the result in the form \eq{verccww}.

Vertex $\Upsilon^{\eta\eta}_{C\go\go C}$ can be worked out
analogously. One calculates
\be\label{cwwcbeta}
\Upsilon^{\eta\eta}_{C\go\go
C}(\gbzer)=-h_{0,\,\gbzer}(W_1*W_1)\Big|_{C\go\go C}
\ee
at finite $\gbzer$ and then sends $\gbzer\to-\infty$. Again,
direct computation using \eqref{W1exp} and \eqref{h} gives well
defined expression for any $\gbzer<1$. Up to terms that vanish for
large $-\gbzer$ the final result reads
\begin{align}\label{cwwcbigbeta}
\Upsilon^{\eta\eta}_{C\go\go C}(\gbzer)=\ff14
C\bar{*}\go\bar{*}\go\bar{*} C\,(t_1t_2)^2
\int_{[0,1]^2}\dr\gs\dr\gs'\,\gs\gs'
\int_{[0,1]^2}\dr\tau_1\dr\tau_1'\,
\ff{\gbzer^2(1-\tau_1)(1-\tau_1')}{(1-\gbzer\tau_{\circ})^4}\times\\
\times \exp i\left({-\ff{ \gbzer}{\xi}\tau_1(1-\tau_1')
\tilde{A}-\ff{ \gbzer}\xi\tau_1'(1-\tau_1)\tilde B+\ff 1\xi \tilde
C+ \tilde D}\right)+O(1/\gbzer)\,\nn
\end{align}
with
\begin{align}
&\tilde A=-(\gs t_1+\gs' t_2)^{\al}(p_1+t_1+\gs'
t_2)_{\al}\,,\quad \tilde B=(\gs
t_1+\gs' t_2)^{\al}(p_2+t_2+\gs t_1)_{\al}\,,\label{AB}\\
&\tilde C=(\gs t_1+\gs' t_2)^{\al}y_{\al}\,,\quad \tilde
D=p_1^{\al}t_{1\al}-p_2^{\al}t_{2\al}-t_2^{\al}t_{1\al}\gs\gs'\label{CD'}\,.
\end{align}
Taking the limit $\gbzer\to-\infty$ using \eqref{I2inf} we find
\be
\ls\Upsilon^{\eta\eta}_{C\go\go C}
=\ff{1}{4}C\bar{*}\go\bar{*}\go\bar{*}C\, (t_1{}_\ga t_2^\ga)^2
\int_{[0,1]^2}\dr\gs\,\dr\gs'\,\gs\gs'\int\dr^3_{\Delta}\tau\,\tau_3
\exp i({\tau_1\tilde A+\tau_2\tilde B+\tau_3\tilde C+\tilde D})\,,
\ee
which can be rewritten as \eqref{vercwwc}.

 The rest of the
structures in \eqref{eta} and \eqref{etabar} are calculated
analogously. In all these cases one encounters the following
integrals that should be calculated at $\gb\to-\infty$
\begin{align}
&I_1(\gbzer)=-\int_{[0,1]^2}\dr\tau\,\dr\tau'
\ff{\gbzer^3\tau(1-\tau')^2(1-\tau)}{(1-\gbzer\tau\circ \tau')^4}
\exp i\left({-\ff{\gbzer}{\xi}\tau (1-\tau')A-\ff{\gbzer}\xi\tau '(1-\tau)B+\ff {1}{ \xi} C+ D}\right)\,,\\
 &I_2(\gbzer)=\int_{[0,1]^2}\dr\tau\,\dr\tau'
\ff{\gbzer^2(1-\tau')(1-\tau)}{(1-\gbzer\tau\circ \tau')^4}\exp
i\left({-\ff{\gbzer}{\xi}\tau (1-\tau')
A-\ff{\gbzer}\xi\tau'(1-\tau)B+\ff 1 \xi C+ D}\right)\,,
\end{align}
where  $\xi=1-\gbzer\tau\circ \tau'$   and $A$, $B$, $C$
and $D$ do not depend on $\tau$ and $\tau'$. As proven in Appendix~B3, 
 the limiting values
of these are given in terms of integrals over a simplex
\begin{align}
&I_1(-\infty)=\int\dr^3_{\Delta}\tau\,\tau_1
\exp i\left({\tau_1A+\tau_2B+\tau_3C+D}\right)\,,\\
&I_2(-\infty)=\int\dr^3_{\Delta}\tau\,\tau_3
\exp i\left({\tau_1A+\tau_2B+\tau_3C+D}\right)\,.
\end{align}

Let us stress that the resulting expressions turn out to be
ultra-local because the coefficients $A$, $B$, $C$ and $D$
(\ref{A})-(\ref{CD}) and $\tilde A$, $\tilde B$, $\tilde C$ and
$\tilde D$ (\ref{AB}), (\ref{CD'}) do not contain $p_{1\ga}
p_2{}^\ga$, that implies locality, and  $p_{i\ga} y^\ga$, that
implies that arguments of the zero-forms $C$ are $y$-independent.

\subsection{Second order. ${\eta\bar\eta}$ sector}
\label{invariant}

Our strategy for extracting $\Upsilon^{\eta\bar\eta}$ will be as
follows. We first solve equations for fields $S_2$ and $W_2$ that
depend on $B_2$ using \eqref{b2}. Having expressions for $W_1$ and
$W_2$ we then calculate the mixed vertices
$\Upsilon^{\eta\bar\eta}$ via \eqref{2nd}. This way we get the
result that does not account for the local field redefinition
\eqref{B2min}. Its effect \eqref{dB2} leads to the change of
$W_2\to W_2+\gd W_2$ in both holomorphic and mixed sectors and
therefore changes vertices $\Upsilon^{\eta\bar\eta}\to
\Upsilon^{\eta\bar\eta}+\gd\Upsilon^{\eta\bar\eta}$. That change
can be easily taken into account provided $\gd W_2$ is solved by
using conventional contracting homotopy $\hmt_0$. The choice of
$\hmt_0$ is just a matter of convenience since for any local field
redefinitions any homotopy results in a local contribution to the
final vertex at given order. This allows one using \eqref{0gb} to
extract $\gd\Upsilon^{\eta\bar\eta}$ from \eqref{2nd} as
\be\label{dver}
\gd\Upsilon^{\eta\bar\eta}=-h_0\{\go, \gd W_2\}_*\,.
\ee
Recall also that since second order $\eta\bar\eta$ -- sector
originate from products of first order holomorphic$\times$
antiholomorphic contributions parameter $\gb$ drops out and one
can set $\gb=0$ in that calculation.

 \subsubsection{Mixed   contracting homotopy}
\label{mixed}
To evaluate
$\Upsilon^{\eta\bar{\eta}}$ one needs to solve equation of the type
\eqref{steq} with  \rhs  containing one-forms
proportional to  $\theta^\alpha$ and  $\bar{\theta}^{\pa}$ or two-forms
proportional to  $\theta^\alpha \bar{\theta}^{\pa}$.
There are several approaches that can be used to solve the
problem. One is to apply a total  \CH  $\hmt^{tot}$ with
respect to $Z^A=(z^\alpha,\bar{z}^{\dot{\alpha}})$ and
$\Theta^A=(\theta^\alpha,\bar{\theta}^{\dot{\alpha}})$. Another
one is to use the spectral sequence approach with respect to the
 shifted \CH  $\hmt_q$ \eq{oldres}, $\bar\hmt_{\bar q}$ and
cohomology projectors $h_{q}\equiv h_{q\,,\gbzer}$  and $\bar
h_{\bar q}$, introduced in \cite{Didenko:2018fgx}
\be\label{oldproj}
h_{q}J(z,y;\theta)= J(-q, y; 0)\,.
 \ee
This means that starting from, say, holomorphic sector with no
$\bar \theta_\ga$ dependence one reconstructs
$\theta^p\bar\theta^{\bar{p}}$-forms  by solving
equation
\be
\dr_z f^{p,\bar{p}}  =g^{p+1,\bar{p}} -\dr_{\bar z}f^{p+1, \bar{p}-1}
\ee
with the help of some $\hmt_q$ step by step  increasing $\bar p$ up to the
last step in the anti-holomorphic sector with the equation
\be
\label{spec} \dr_{\bar z}f^{0, \bar{p}-1}= g^{0,\bar{p}}
\ee
to be solved with the help of $\bar{\hmt}_{\bar{q}}$ for some $\bar{q}$.

Both of these approaches are inconvenient for one reason or
another. The total \CH  ${\hmt}_{tot}$ does not preserve the
PLT structure  with respect to holomorphic and anti-holomorphic
variables separately. The spectral sequence approach (\ref{spec})
treats the holomorphic and anti-holomorphic sectors
asymmetrically.

It is more convenient to use a  mixed holomorpic-antiholomorphic
symmetric operator
\be
\tilde\hmt_{q\bar q}  := \half(\hmt_q+\bar{\hmt}_{\bar{q}})\,.
\ee
Using formulas from \cite{Didenko:2018fgx} we have
\begin{equation}
\label{qres} \lbrace \mathrm{d}_Z, \tilde\hmt_{q\bar
q}\rbrace=1-\tilde{h}_{q\bar q}\,,
\ee
where
\be
\tilde{h}_{q\bar q}:=\frac{1}{2}(h_q+\bar{h}_{\bar{q}})\,.
\end{equation}
Here \CH $\bar{\hmt}_{\bar{q}}$ and respective projector $\bar{h}_{\bar{q}}$
can be obtained from $\hmt _{q}$ \eqref{oldres} $h_{q}$ and
\eqref{oldproj} by swapping $z\to \bar z$, $q\to \bar q$.

 As  follows from its definition, $\tilde{h}_{q\bar q}$ is {\it
not} a projector to $\mathrm{d}_Z$-cohomology. Nevertheless,
equations of the type \eqref{steq} can be solved  by \eqref{qres}
if the \rhs is annihilated by $\tilde{h}_{q\bar q}$. As we shall
see, for this to be true, the \CH indices $q$ and $\bar q$
should be chosen appropriately. Note that the analysis of the
mixed sector is $\gbzer$-independent.

For the future convenience we present here
the following useful formulae derived in \cite{Didenko:2018fgx}:
\be\label{hmtd}
 \hmt_b  -  \hmt_a =[\dr_z,\hmt_a  \hmt_b]  +  h_a \hmt_b\q
  \ee
\be\label{hmtdgga}
 (\hmt_b  -  \hmt_a)\gga = \dr_z \hmt_a  \hmt_b \gga\q
   \ee
\be\label{3iden}
(h_a\hmt_b\hmt_c-h_a\hmt_b\hmt_d-h_a\hmt_d\hmt_c+h_b\hmt_d\hmt_c)\gamma=0
\q\ee
 \be \label{4iden}
(\hmt_a-\hmt_b)(\hmt_c-\hmt_d)\gamma=(h_a-h_b)\hmt_c\hmt_d\gamma\,.\qquad
\end{equation}

 \subsubsection{Solving for $S_2$ and $W_2$}
\label{s2w2mi}

Equation for $S_2$ resulting from \eqref{HS4} in the mixed sector
is
\begin{equation}\label{S2mixeq}
\dr_Z S_2|_{\eta\bar\eta}=\frac{i}{2}(i\bar\eta B_2^{\eta}*
\bar\gamma+i\eta  B_2^{\bar\eta}* \gamma - S_1^{\bar \eta}*
S_1^{\eta}-S_1^{\eta}* S_1^{\bar \eta})\,\,
\end{equation}
with  $S_1$ \eq{S1om} and  $B_2$ \eq{b2}.
Using star-exchange formulae \eqref{leftd}-\eqref{righth} with $\gbzer=0$,
equation for $S_2$ in the $\eta\bar\eta$ sector can be brought
to the form
\begin{multline}
\label{S2mix}
\mathrm{d}_Z S_2|_{\eta\bar\eta}=-\frac{i\eta \bar{\eta}}{8}C* C*\Big[\hmt_{p_2}\hmt_{p_1+2p_2}
\gamma* \bar{\gamma}+\bar{\hmt}_{\bar{p}_2}\bar{\hmt}_{\bar{p}_1+2\bar{p}_2}\bar{\gamma}* \gamma\\
+\hmt_{p_1 +2p_2}\gamma*
\bar{\hmt}_{\bar{p}_2}\bar{\gamma}-\hmt_{p_2}\gamma *
\bar{\hmt}_{\bar{p}_1+2\bar{p}_2}\bar{\gamma}\Big]\,.
\end{multline}
According to  \cite{Gelfond:2018vmi}
the \rhs of \eq{S2mix} is {\it totally PLT-odd}, \ie PLT-odd  with respect to both holomorphic and  anti-holomorphic variables.
Hence, the proper choice  demands both $q$ and $\bar q$ be
PLT-odd obeying \eqref{odd} and the conjugated condition, respectively.
Remarkably,  the \rhs of Eq.~\eq{S2mix} is annihilated by $\tilde{h}_{q'\bar q'}$ if
\be\label{oddqbarq}
 {q'}=-\mu p_2 -(1+\mu) p_1 \q \bar {q}'= -\bar\mu \bar{p}_2 -(1+\bar\mu) \bar{p}_1
\q \forall \mu, \bar\mu\in \mathbb{C}
\ee
as one can see using the following formula that holds for any $\mu$   \cite{Didenko:2018fgx}
\begin{equation}
h_{(1+\mu)q_2-\mu q_1}\hmt_{q_1} \hmt_{q_2}\gga=0
\end{equation}
along with the following corollary of \eq{lefth}
\be \tilde{h}_{q'\bar q'}C* C*(...)= C* C*
\tilde{h}_{(q'+p_1+p_2)(\bar q'+\bar p_1+\bar p_2)}(...)\,. \ee
Evidently $q'$ and $\bar {q}'$ (\ref{oddqbarq}) are  PLT-odd.

 Hence one can solve
\eqref{S2mix} using $ \tilde{\hmt}_{q'\bar q'}$, namely
\begin{multline}
\label{S2mixmunu=}
  S_2|^{\mu\bar\mu}_{\eta\bar\eta}=-\frac{i\eta \bar{\eta}}{16}C* C*
  \big[{\hmt}_{(1-\mu) p_2 - \mu  p_1}+\bar{\hmt}_{(1-\bar\mu) \bar{p}_2 - \bar\mu  \bar{p}_1}\big]
  \Big[\hmt_{p_2}\hmt_{p_1+2p_2}\gamma* \bar{\gamma}+\bar{\hmt}_{\bar{p}_2}\bar{\hmt}_{\bar{p}_1+2\bar{p}_2}\bar{\gamma}* \gamma\\
+\hmt_{p_1 +2p_2}\gamma*
\bar{\hmt}_{\bar{p}_2}\bar{\gamma}-\hmt_{p_2}\gamma *
\bar{\hmt}_{\bar{p}_1+2\bar{p}_2}\bar{\gamma}\Big]\,.
\end{multline}

We set $\mu=\bar\mu=-1 $ since some of the formulas simplify for this choice. This yields
\begin{equation}
S_2|_{\eta\bar\eta}=-\frac{i\eta \bar{\eta}}{8}C* C*\Big[
\hmt_{p_2}\hmt_{p_1+2p_2}\gamma*
\bar{\hmt}_{\bar{p}_1+2\bar{p}_2}\bar{\gamma}+\hmt_{p_1+2p_2}\gamma*
\bar{\hmt}_{\bar{p}_2}\bar{\hmt}_{\bar{p}_1+2\bar{p}_2}\bar{\gamma}\Big].
\end{equation}
Equation for $W_2$ resulting from \eqref{HS4} in the mixed sector is
\begin{equation}
\label{W_2mix} \mathrm{d}_Z W_2^{\eta\bar\eta}=-\frac{i}{2}(\mathrm{d}_x S_1+\mathrm{d}_x
S_2+W_1* S_1+S_1*W_1+\omega* S_2+S_2*\omega) \big
|_{\eta\bar\eta}\,.
\end{equation}
$W_2^{\eta\bar\eta}{}$ contains three types of terms
\be\label{W2sum}
W_2^{\eta\bar\eta}{} = W_{2\,\go CC}^{\eta\bar\eta} +W_{2\, C\go C}^{\eta\bar\eta} + W_{2\, CC\go}^{\eta\bar\eta}\q
\ee
where   the lower label refers to the ordering of fields $\omega$ and $C$.
 From \eq{W_2mix} it follows\begin{multline}
\label{preW2wCCmix}
\mathrm{d}_Z W_{2\,\omega C C}^{\eta\bar\eta}=\frac{ \eta \bar{\eta}}{16}\omega* C* C*
\Big[\underbrace{h_{p_2}\hmt_{p_1+2p_2}\hmt_{p_1+2p_2+t}\gamma*
\bar{\hmt}_{\bar{p}_1+\bar{p}_2+\bar{t}}\bar{\gamma}}_{\mathrm{d}_x S_1}\\
\underbrace{+\hmt_{p_2}\hmt_{p_1+2p_2+t}\gamma*
\bar{\hmt}_{\bar{p}_1+2\bar{p}_2+\bar{t}}\bar{\gamma}}_{\mathrm{d}_xS_2}
+\underbrace{\hmt_{p_1+2p_2+t}\hmt_{p_1+2p_2}\gamma*
\bar{\hmt}_{\bar{p}_2}\bar{\gamma}}_{W_1{}_{\,\omega C}*
S_1}-\underbrace{\hmt_{p_2}\hmt_{p_1+2p_2}\gamma*
\bar{\hmt}_{\bar{p}_1+2\bar{p}_2}\bar{\gamma}}_{\omega*
S_2}\Big]+h.c.,
\end{multline}
\begin{multline}
\label{preW2CwCmix}
  \mathrm{d}_Z W_{2\, C\omega C} ^{\eta\bar\eta}=\frac{ \eta \bar{\eta}}{16}C*
\omega * C*\Big[\underbrace{(h_{p_1+2p_2+2t}-h_{p_2})\hmt_{p_2+t}\hmt_{p_1+2p_2+t}\gamma*
\bar{\hmt}_{\bar{p}_1+\bar{p}_2+\bar{t}}\bar{\gamma}}_{\mathrm{d}_x S_1}\\
\underbrace{-\hmt_{p_2}\hmt_{p_1+t+2p_2}\gamma*
\bar{\hmt}_{\bar{p}_1+2\bar{p}_2+\bar{t}}
\bar{\gamma}+\hmt_{p_2+t}\hmt_{p_1+2p_2+2t}\gamma*
\bar{\hmt}_{\bar{p}_1+2\bar{p}_2+2\bar{t}}\bar{\gamma}}_{\mathrm{d}_x S_2}\\
\underbrace{-\hmt_{p_1+2p_2+t}\hmt_{p_1+2p_2+2t}\gamma*
\bar{\hmt}_{\bar{p}_2}\bar{\gamma}}_{W_{1\,C\omega}*
S_1}-\underbrace{\hmt_{p_2+t}\hmt_{p_2}\gamma*
\bar{\hmt}_{\bar{p}_1+2\bar{p}_2+2\bar{t}}\bar{\gamma}}_{S_1*
W_1{}_{\,\omega C}}\Big]+h.c.,
\end{multline}
\begin{multline}
\label{preW2CCwmix}
\mathrm{d}_Z W_{2\, CC\omega}^{\eta\bar\eta}=\frac{ \eta \bar{\eta}}{16} C* C* \omega*
\Big[\underbrace{h_{p_2+2t}\hmt_{p_2+t}\hmt_{p_1+2p_2+2t}\gamma *
\bar{\hmt}_{\bar{p}_1+\bar{p}_2+\bar{t}}\bar{\gamma}}_{\mathrm{d}_x S_1}\\
\underbrace{-\hmt_{p_2+t}\hmt_{p_1+2p_2+2t}\gamma*
\bar{\hmt}_{\bar{p}_1+2\bar{p}_2+2\bar{t}}\bar{\gamma}}_{\mathrm{d}_x S_2}
+\underbrace{\hmt_{p_2+t}\hmt_{p_2+2t}\gamma*
\bar{\hmt}_{\bar{p}_1+2\bar{p}_2+2\bar{t}}\bar{\gamma}}_{S_1* W_1{}_{\,C\omega}}\\
\underbrace{+\hmt_{p_2+2t}\hmt_{p_1+2p_2+2t}\gamma*
\bar{\hmt}_{\bar{p}_1+2\bar{p}_2+2\bar{t}}\bar{\gamma}}_{S_2*\omega}\Big]+h.c.\,.
\end{multline}
Notation $\underbrace{P}_{Q} $  specifies  the part of  $P$ coming
from $Q$.

Note that all  terms    of the form
$\hmt_a\hmt_b(\gga)*\bar\hmt_c\bar\hmt_d (\bar\gga)$ in
\eq{preW2wCCmix}-\eq{preW2CCwmix} are   totally PLT-odd while
those  of the form $h_a\hmt_b\hmt_c(\gga)*\bar\hmt_c\bar\hmt_d
(\bar\gga)$ are PLT-even with respect to the barred variables and
PLT-odd with respect to the unbarred ones. From here it follows
that for any $q^U\,,\bar{q }^U$
\be \tilde{h}_{q^U\bar{q }^U}\mathrm{d}_Z W^{\eta\bar\eta}_{2\,U}\ne0\,,\qquad U=\{CC\go,\,\, \go CC ,\,\, C\go C\}.
\ee

However it turns out  that each equation
\eqref{preW2wCCmix}-\eqref{preW2CCwmix} can be rewritten as
 \be \label{DzU}
  \mathrm{d}_Z W^{\eta\bar\eta}_{2\,U}=  \mathrm{d}_Z F_{U}+ G_{U}\q
\ee
  where all $G_{U}$ are  totally PLT-odd and there exist
such
   shifts $q_{U} $ and $\bar{q}_{U} $ that
   \be \label{Uexact}\tilde{h}_{q_{U}\bar{q}_{U}}G_{U}=0.\ee
Hence  \eq{DzU} is solved by
\be\label{solU}
W^{\eta\bar\eta}_{2\,U}=F^{U}+\tilde{\hmt}_{q_{U}\bar{q}_{U}}G^{U}.
\ee

Consider firstly  $W^{\eta\bar\eta}_{{2\,\go CC}}$ \eqref{preW2wCCmix}.
 Eq.~\eq{hmtdgga} yields for any $\bar{q}$
\bee    && h_{p_2}\hmt_{p_1+2p_2}\hmt_{p_1+2p_2+t}\gamma*
\bar{\hmt}_{  \bar{p}_1+  \bar{p}_2+ \bar{t}}\bar{\gamma}  =\\ \nn
 &&\dr_Z \big( h_{p_2}\hmt_{p_1+2p_2}\hmt_{p_1+2p_2+t}\gamma*
\bar{\hmt}_{\bar q}\bar{\hmt}_{  \bar{p}_1+  \bar{p}_2+ \bar{t}}
\bar{\gamma}\big)
+  h_{p_2}\hmt_{p_1+2p_2}\hmt_{p_1+2p_2+t}\gamma*
\bar{\hmt}_{\bar q}\bar{\gamma}.\eee

Denoting
\bee
 {F}_{q,\bar q\,\,}{}_{\omega C C}&=&
\frac{ \eta \bar{\eta}}{16} \omega* C* C*
h_{p_2}\hmt_{p_1+2p_2}\hmt_{p_1+2p_2+t}\gamma* \bar{\hmt}_{\bar
q}\bar{\hmt}_{  \bar{p}_1+  \bar{p}_2+ \bar{t}} \bar{\gamma} +
h.c. \q\\ \nn G_{q,\bar q\,\,}{}_{\omega C C}&=&\mathrm{d}_Z
W^{\eta\bar\eta}_{2\,\omega C C}- \mathrm{d}_Z F_{q,\bar
q\,\,}{}_{\omega C C}
 \eee
we have to find such $Q,\bar Q$ and $q,\bar q$  that
\be\label{wccclosB}
  \tilde{h}_{Q-p_1-p_2-t,\bar Q-\bar{p}_1-\bar{p}_2-\bar{t}} G_{q,\bar q\,\,}{}_{\omega C C}=0\,.
\ee
By virtue of \eqref{lefth}  this equation 
demands in particular
\begin{multline}
\nn
  \Big[
 h_{p_2}\hmt_{p_1+2p_2}\hmt_{p_1+2p_2+t}\gamma*
\bar{\hmt}_{\bar q}\bar{\gamma}\\
+{h_Q\hmt_{p_2}\hmt_{p_1+2p_2+t}\gamma*
\bar{\hmt}_{\bar{p}_1+2\bar{p}_2+\bar{t}}\bar{\gamma}}
+{h_Q\hmt_{p_1+2p_2+t}\hmt_{p_1+2p_2}\gamma*
\bar{\hmt}_{\bar{p}_2}\bar{\gamma}}
 -{h_Q\hmt_{p_2}\hmt_{p_1+2p_2}\gamma*
\bar{\hmt}_{\bar{p}_1+2\bar{p}_2}\bar{\gamma}}   \Big]=0\,.
\end{multline}
 There are three evident solutions to this equation
 \be\label{3solQ}
 \{ Q=p_1+2p_2+t\,,\,\,   \bar q= \bar p_1+2\bar p_2\}\,,\quad 
\{ Q =p_1+2p_2\,,   \bar q=   \bar p_1+2\bar p_2+\bar t\}\,,\quad 
 \{ Q =  p_2\,,\,\,  \bar q=  \bar p_2\}
 \,.\qquad\ee
Choosing for simplicity $Q=q=  p_2\,,\,\,  \bar q=\bar Q =  \bar p_2$   by virtue of Eq.~\eq{hmtdgga}
   one obtains
\be\nn {F}_{p_2,\bar p_2}{}_{\omega C C}=
\frac{ \eta \bar{\eta}}{16} \omega* C* C*
h_{p_2}\hmt_{p_1+2p_2}\hmt_{p_1+2p_2+t}\gamma* \bar{\hmt}_{\bar
p_2}\bar{\hmt}_{  \bar{p}_1+  \bar{p}_2+ \bar{t}}
\bar{\gamma}+h.c.\q\ee
\be\label{wccclosB33}
  \tilde{h}_{ -p_1 -t,   -\bar{p}_1 -\bar{t}} G_{p_2,\bar p_2}{}_{\omega C C}=0\,.
\ee
     Hence, by virtue of (\ref{solU}),
 \begin{multline}\label{W2wcc=}
W^{\eta\bar\eta}_{2\,\omega C C}=-\frac{\eta\bar{\eta}}{16}\omega* C* C*
\Big[h_{p_2}\hmt_{p_1+2p_2}\hmt_{p_1+2p_2+t}\gamma*\bar{\hmt}_{\bar{p}_2}
\bar{\hmt}_{\bar{p}_1+\bar{p}_2+\bar{t}}\bar{\gamma}\\
+\frac{1}{2}\hmt_{p_2}\hmt_{p_1+2p_2+t}\gamma*
\bar{\hmt}_{\bar{p}_2}\bar{\hmt}_{\bar{p}_1+2\bar{p}_2+\bar{t}}\bar{\gamma}
+\frac{1}{2}\hmt_{p_2}\hmt_{p_1+2p_2}\gamma*
\bar{\hmt}_{\bar{p}_1+2\bar{p}_2}\bar{\hmt}_{\bar{p}_2}\bar{\gamma}
\Big]+h.c.
\end{multline}
solves \eq{preW2wCCmix}.

Consideration of (\ref{preW2CwCmix}) is analogous. Setting
\bee
 {F}_{q,\bar q\,\,}{}_{C\omega  C}&=&\frac{\eta\bar{\eta}}{16}C*
 \omega* C*
 \big[(h_{p_1+2p_2+2t}-h_{p_2})\hmt_{p_2+t}\hmt_{p_1+2p_2+t}\gamma*
 \bar{\hmt}_{\bar q}\bar{\hmt}_{\bar{p}_1+\bar{p}_2+\bar{t}}
 \bar{\gamma}\big]+h.c.   \q\\ \nn
G_{q,\bar q\,\,}{}_{C\omega  C}&=&\mathrm{d}_Z
W^{\eta\bar\eta}_{2\,C\omega  C}- \mathrm{d}_Z F_{q,\bar
q\,\,}{}_{C\omega  C}
 \eee
one can see that the equation \be\label{cwcclosB}
  \tilde{h}_{Q-p_1-p_2-t,\bar Q-\bar{p}_1-\bar{p}_2-\bar{t}} G_{q,\bar q\,\,}{}_{C\omega  C}=0\,
\ee admits a solution
 $  q={p}_1+2{p}_2+2{t}$, $\bar q=\bar{p}_1+2\bar{p}_2+2\bar{t}$,\,
 $Q={p}_1+2{p}_2+ {t}$,\, $\bar Q=\bar{p}_1+2\bar{p}_2+ \bar{t}$.
The respective solution to \eq{preW2CwCmix}  of the form \eq{solU}
is\begin{multline}\label{W2cwc=}
W^{\eta\bar\eta}_{2\,C\omega C}=-\frac{\eta\bar{\eta}}{16}C*\omega*
C*\Big[(h_{p_1+2p_2+2t}-h_{p_2})\hmt_{p_2+t}\hmt_{p_1+2p_2+t}\gamma *\bar{\hmt}_{p_1+2p_2+2t}\bar{\hmt}_{p_1+p_2+t}\bar{\gamma}\\
-\frac{1}{2}\hmt_{p_2}\hmt_{p_1+2p_2+t}\gamma* \bar{\hmt}_{\bar{p}_1+2\bar{p}_2+2\bar{t}}\bar{\hmt}_{\bar{p}_1+2\bar{p}_2+
\bar{t}}\bar{\gamma}-\frac{1}{2}\hmt_{p_1+2p_2+t}\hmt_{p_1+2p_2+2t}\gamma*\bar{\hmt}_{\bar{p}_1+2\bar{p}_2+2\bar{t}}\bar{\hmt}_{\bar{p}_2}\bar{\gamma}\\
+\frac{1}{2}h_{p_1+2p_2+t}\hmt_{p_1+2p_2+2t}\hmt_{p_2}\gamma*
\bar{\hmt}_{\bar{p}_1+2\bar{p}_2+t}\bar{\hmt}_{\bar{p}_1+2\bar{p}_2+2\bar{t}}\bar{\gamma}\Big]+h.c.\,.
\end{multline}
Finally, from \eq{preW2CCwmix} it follows that
\begin{equation}
\label{W2CCwmix} \mathrm{d}_Z
W^{\eta\bar\eta}_{2\,CC\omega}=-\frac{\eta\bar{\eta}}{16}C* C* \omega*
\mathrm{d}_Z\big(h_{p_2+t}\hmt_{p_2+2t}\hmt_{p_1+2p_2+2t}\gamma*
 \bar{\hmt}_{\bar{p}_1+2\bar{p}_2+2\bar{t}}\bar{\hmt}_{\bar{p}_1+\bar{p}_2
 +\bar{t}}\bar{\gamma}\big)+h.c.\,.
\end{equation}
Hence
\begin{equation}\label{W2ccw=}
W^{\eta\bar\eta}_{2\,CC\omega}=  \frac{\eta\bar{\eta}}{16}C*
C* \omega* h_{p_2+
t}\hmt_{p_2+2t}\hmt_{p_1+2p_2+2t}\gamma*\bar{\hmt}_{p_1+2p_2+2t}\bar{\hmt}_{p_1+p_2+t}\bar{\gamma}+h.c.
\end{equation}
solves \eq{preW2CCwmix}.

The following comment is now in order. The   choice of  $\mu=\bar\mu=-1$
of the \CH parameters
$\displaystyle
 \half\big[{\hmt}_{(1-\mu) p_2 - \mu  p_1}+\bar{\hmt}_{(1-\bar\mu) \bar{p}_2 - \bar\mu  \bar{p}_1}\big]$
 in \eq{S2mixmunu=}   leads to  asymmetric result
for $W_2^{\eta\bar{\eta}}$ as can be seen from the fact that $W_{2\,CC\omega}^{\eta\bar{\eta}}$
\eqref{W2ccw=} is  simpler than
  $W_{2\,\go CC}^{\eta\bar{\eta}}$ \eq{W2wcc=}.
  As  shown below, this leads to essentially different formulae for components of the respective vertices
  $\Upsilon^{\eta\bar{\eta}}({ \go,  \omega, C,C})$.

The   choice  $\mu=\bar\mu=0$ in \eq{S2mixmunu=} yields
\be S_2=
-\frac{i\eta \bar{\eta}}{8}C* C*
   \Big[ {\hmt}_{ p_2  }\gamma*\bar{\hmt}_{\bar{p}_2}\bar{\hmt}_{\bar{p}_1+2\bar{p}_2}\bar{\gamma}
   +{\hmt}_{ p_2}\hmt_{p_1 +2p_2}\gamma*\bar{\hmt}_{\bar{p}_2}\bar{\gamma}
 \Big]\,
\,
\ee
  leading to   simplification of   $W_{2\,\go CC}^{\eta\bar{\eta}}$ and
  complication of $W_{2\,CC\omega}^{\eta\bar{\eta}}$.
To obtain a form of $W_2^{\eta\bar{\eta}}$ and,
hence, $\Upsilon^{\eta\bar{\eta}}({ \go,  \omega, C,C})$ symmetric
 with respect to reordering of $C$ and $\go$ one can  take an averaged
 sum of   $\mu=0\,,\,\,
\bar\mu=-1$
\begin{multline}
S^{\eta\bar \eta}_2=
-\frac{i\eta \bar{\eta}}{16}C* C*
    \Big[\hmt_{p_2}\hmt_{p_1+2p_2}\gamma* \bar{\hmt}_{ \bar{p}_1+2 \bar{p}_2}\bar{\gamma}
  + {\hmt}_{ p_2  }\gamma*\bar{\hmt}_{\bar{p}_2}\bar{\hmt}_{\bar{p}_1+2\bar{p}_2}\bar{\gamma}\\
+\hmt_{p_1 +2p_2}\gamma*\bar{\hmt}_{\bar{p}_2}\bar{\hmt}_{ \bar{p}_1+2 \bar{p}_2}\bar{\gamma}
+{\hmt}_{ p_2  }\hmt_{p_1 +2p_2}\gamma*\bar{\hmt}_{\bar{p}_2}\bar{\gamma}
 \Big]\,
\,.
\end{multline}

Note that all  resulting vertices $\Upsilon^{\eta\bar\eta}$ will be local.

\subsubsection{$\eta\bar\eta$ vertex $\Upsilon^{\eta\bar\eta}$}
From \eqref{HS3} we have 
\begin{equation}
\label{a}
\Upsilon^{\eta\bar\eta}(\go,\go,C,C)=
-\big(\mathrm{d}_x
W_1+W_1* W_1+\mathrm{d}_x W_2+\omega* W_2+W_2*
\omega\big)\big|_{{\eta\bar\eta}}.
\end{equation}
Plugging the obtained expressions for $W_1$ and  $W_2$ into
\rhs of Eq.~\eq{a}
one can calculate $\eta\bar\eta$ vertices in this order of
perturbation theory. For the reader's convenience, the
 expressions for $\dr_x C$ and $\dr_x \go$  in terms of lower order corrections
obtained in \cite{Didenko:2018fgx}
are collected in Appendix~C.

The procedure goes as follows. Since \rhs
of \eqref{a} is by construction $z,\bar{z}$--independent
one can apply any projector   $h_q \bar h_{\bar q}$ ($q$ and $\bar q $ are not necessarily complex conjugated).

As an example,
consider vertex $\Upsilon^{\eta\bar\eta}_{C\omega C \omega}$.
Eq.~\eqref{a} yields
\begin{multline}
\label{preVert}  \Upsilon_{C\omega C\omega}= -\big(\mathrm{d}_x
W_2 {}_{C\omega\underline{C}{}}+\mathrm{d}_x W_2
{}_{C\underline{C}\omega} +\mathrm{d}_x W_2 {}_{\underline{C} C
\omega}+\mathrm{d}_x W_1{}_{\,C\underline{\omega}} +\mathrm{d}_x
W_1{}_{\,\underline{C}\omega} \big)\big|_{C\omega C \omega}\\  -
W_1 {}_{\,C\omega}* W_1 {}_{\,C\omega}-W_2 {}_{C\omega C}* \omega
  \q\end{multline}
where underlined labels refer to  the fields $\mathrm{d}_x$ is
acting on.
For instance,   using formula \eq{W2ccw=} for $W^{\eta\bar\eta}_{2\,C {C}\omega}$   one has
(discarding the conjugated  terms for brevity)
\bee\label{fr}
\mathrm{d}_x W^{\eta\bar\eta}_{2\,C\underline{C}\omega}=:
-\frac{\eta\bar{\eta}}{16}C*(\mathrm{d}_x C)* \omega*
h_{p_2+2t}\hmt_{p_2+t}\hmt_{p_1+2p_2+2t}\gamma*
\bar{\hmt}_{p_1+2p_2+2t}\bar{\hmt}_{p_1+p_2+t}\bar{\gamma}  \q
\eee
whence it follows by virtue of  \eq{C2gen}
\be\label{preVert1}
\mathrm{d}_x W_2^{\eta\bar\eta}{}_{C\underline{C}\omega}
=\frac{\eta \bar{\eta}}{16} C* {\omega * C} * \omega *
h_{p_2+t_1+2t_2}\hmt_{p_2+t_1+t_2}\hmt_{p_1+2p_2+2t_1+2t_2}\gamma*
\bar{\hmt}_{\bar{p}_1+2\bar{p}_2+2\bar{t}_1+2\bar{t}_2}
\bar{\hmt}_{\bar{p}_1+\bar{p}_2+\bar{t}_1+\bar{t}_2}\bar{\gamma}
\,.\qquad\qquad
\ee
{In obtaining this and similar expressions one should be careful
in keeping track of the homotopy parameters. For example in
\eqref{fr} $p_2$ acts on $\dr_x C$ which after substitution of
\eq{C2gen} contributes $\go*C$ implying that one should replace
$p_2\to p_2+t_1$ and rename $t\to t_2$ since there are now two
$\go$'s.}

Analogously one obtains from \eq{W2cwc=}, \eq{W2ccw=} and
\eq{W1om} by virtue of  \eq{go1}-\eq{ccCUps2loc}
 \bee \label{preVert2}  \!\!
\mathrm{d}_x W^{\eta\bar\eta}_{2\,C\omega\underline{C}{}}\
&\!\!=\!\!& -\frac{\eta \bar{\eta}}{16} C* \omega * {C
* \omega} *\\ &&   \nn
\Big[(h_{p_1+2p_2+2t_1+2t_2}-h_{p_2+t_2})\hmt_{p_2+t_1+t_2}\hmt_{p_1+2p_2+t_1+2t_2}\gamma*
\bar{\hmt}_{\bar{p}_1+2\bar{p}_2+2\bar{t}_1+2\bar{t}_2}
\bar{\hmt}_{\bar{p}_1+\bar{p}_2+\bar{t}_1+\bar{t}_2}\bar{\gamma}\\
&&   \nn -\frac{1}{2}\hmt_{p_2+t_2}\hmt_{p_1+2p_2+t_1+2t_2}\gamma*
\bar{\hmt}_{\bar{p}_1+2\bar{p}_2+2\bar{t}_1+2\bar{t}_2}
\bar{\hmt}_{\bar{p}_1+2\bar{p}_2+\bar{t}_1+2\bar{t}_2}\bar{\gamma}\\
&&   \nn
-\frac{1}{2}\hmt_{p_1+2p_2+t_1+2t_2}\hmt_{p_1+2p_2+2t_1+2t_2}\gamma
*
\bar{\hmt}_{\bar{p}_1+2\bar{p}_2+2\bar{t}_1+2\bar{t}_2}\bar{\hmt}_{\bar{p}_2+\bar{t}_2}\bar{\gamma}\\
&&   \nn
+\frac{1}{2}h_{p_1+2p_2+t_1+2t_2}\hmt_{p_1+2p_2+2t_1+2t_2}\hmt_{p_2+t_2}\gamma*
\bar{\hmt}_{\bar{p}_1+2\bar{p}_2+\bar{t}_1+2\bar{t}_2}
\bar{\hmt}_{\bar{p}_1+2\bar{p}_2+2\bar{t}_1+2\bar{t}_2}\bar{\gamma}\Big]\,, \\
\label{preVert21}\!\!
 \ls \mathrm{d}_x W^{\eta\bar\eta}_{2\,\underline{C} C \omega}
\!\!\! &\!\!\! \!\!=\!&\!-\!\frac{\eta \bar{\eta}}{16} {C* \omega}
\!* C * \omega \!*\!
\Big[h_{p_2+2t_2}\hmt_{p_2+t_2}\hmt_{p_1+2p_2+t_1+2t_2}\gamma\!*\!
\bar{\hmt}_{\bar{p}_1+2\bar{p}_2+\bar{t}_1+2\bar{t}_2}
\bar{\hmt}_{\bar{p}_1+\bar{p}_2+\bar{t}_1+\bar{t}_2}\bar{\gamma}\Big]
\!,\rule{31pt}{0pt}\eee \bee
 \label{preVert22}   \!\!
  \mathrm{d}_x W^\eta_{1\,C\underline{\omega}} &=&\frac{\eta \bar{\eta}}{16} C* {\omega * C *
\omega} *
\Big[\big(h_{p_2+t_1+t_2}\hmt_{p_2+t_1+2t_2}\hmt_{p_2+t_2}\gamma
+h_{p_2+t_1+2t_2}\hmt_{p_2+2t_2}\hmt_{p_2+t_2}\gamma\big)* \,\,\,\qquad \\ && \nn
*\bar{\hmt}_{\bar{p}_1+\bar{p}_2+\bar{t}_1+\bar{t}_2}
\bar{\hmt}_{\bar{p}_1+2\bar{p}_2+2\bar{t}_1+2\bar{t}_2}\bar{\gamma}\Big]
\,,\qquad \\  \label{preVert23}
 \mathrm{d}_x W^\eta_{1\,\underline{C}\omega} &=&\frac{\eta \bar{\eta}}{16} {C* \omega * C} * \omega *
\Big[(h_{p_1+2p_2+2t_1+2t_2}-h_{p_2+2t_2})\hmt_{p_2+t_1+2t_2}\hmt_{p_1+2p_2+t_1+2t_2}\gamma*
\qquad\\ && \nn
*
\bar{\hmt}_{\bar{p}_1+\bar{p}_2+\bar{t}_1+\bar{t}_2}
\bar{\hmt}_{\bar{p}_1+\bar{p}_2+\bar{t}_1+2\bar{t}_2}\bar{\gamma}\Big]
 \,.\eee
  Using star-exchange formulae  \eqref{leftd}-\eqref{righth}
one obtains from \eq{W1om}  and \eq{W2cwc=}
\bee
\label{preVert3}
W^\eta_{1\,C\omega}* W^{\bar{\eta}}_{1\,C\omega}&\!\!=\!\!&
\frac{\eta \bar{\eta}}{16} C* \omega * C * \omega *
\Big[\hmt_{p_1+2p_2+t_1+2t_2}\hmt_{p_1+2p_2+2t_1+2t_2}\gamma*
\bar{\hmt}_{\bar{p}_2+\bar{t}_2}\bar{\hmt}_{\bar{p}_2+2\bar{t}_2}\bar{\gamma}\Big]\,,
 \\
\label{preVert31}
 W^{\eta\bar\eta}_{2\,C\omega C}* \omega\quad &\!\!=\!\!&
\frac{\eta \bar{\eta}}{16} C* \omega * C * \omega *
\Big[(h_{p_1+2p_2+2t_1+2t_2}-h_{p_2+2t_2})\times\\ &\!\!\!\!& \nn
 \hmt_{p_2+t_1+2t_2}\hmt_{p_1+2p_2+t_1+2t_2}\gamma*
 \bar{\hmt}_{\bar{p}_1+2\bar{p}_2+2\bar{t}_1+2\bar{t}_2}
 \bar{\hmt}_{\bar{p}_1+\bar{p}_2+\bar{t}_1+2\bar{t}_2}\bar{\gamma}\\ &\!\!\!\!& \nn
-\frac{1}{2}\hmt_{p_2+2t_2}\hmt_{p_1+2p_2+t_1+2t_2}\gamma*
\bar{\hmt}_{\bar{p}_1+2\bar{p}_2+2\bar{t}_1+2\bar{t}_2}
\bar{\hmt}_{\bar{p}_1+2\bar{p}_2+\bar{t}_1+2\bar{t}_2}\bar{\gamma}\\
&\!\!\!\!& \nn
-\frac{1}{2}\hmt_{p_1+2p_2+t_1+2t_2}\hmt_{p_1+2p_2+2t_1+2t_2}\gamma*
\bar{\hmt}_{\bar{p}_1+2\bar{p}_2+2\bar{t}_1+2\bar{t}_2}\bar{\hmt}_{\bar{p}_2+2\bar{t}_2}\bar{\gamma}
\\ &\!\!\!\!& \nn
+\frac{1}{2}h_{p_1+2p_2+t_1+2t_2}\hmt_{p_1+2p_2+2t_1+2t_2}\hmt_{p_2+2t_2}\gamma*
\bar{\hmt}_{\bar{p}_1+2\bar{p}_2+\bar{t}_1+2\bar{t}_2}
\bar{\hmt}_{\bar{p}_1+2\bar{p}_2+2\bar{t}_1+2\bar{t}_2}\bar{\gamma}\Big]\,.\nn
\eee
Substitution of \eq{preVert1}-\eq{preVert31} into \eq{preVert}
and application of the cohomology projector
\be h_{p_1+2p_2+2t_1+2t_2}
\bar{h}_{\bar{p}_1+2\bar{p}_2+2\bar{t}_1+2\bar{t}_2}\ee 
yields $\Upsilon^{\eta\bar \eta}_{C \omega C  \omega }$  \eqref{mixCwCw}.

Other vertices are extracted from Eq.~\eq{a} analogously  by virtue of
Eqs.~\eq{W1om}, \eq{W2wcc=}-\eq{W2ccw=}
 taking into account Eqs.~\eq{go1}-\eq{ccCUps2loc}. To bring them
to the form \eqref{mixwwCC}-\eqref{mixCCww}   identity  \eq{3iden}
can be useful.

So far the vertices were calculated using \eqref{b2}. Now in order
to account for local field redefinition \eqref{dB2} one has to add
\eqref{dver} to them. To do so we calculate $\gd S_2$ using
\eqref{S2mixeq} and eventually $\gd W_2$ using \eqref{W_2mix} via
contracting homotopy $\hmt_0$
\begin{align}
&\gd S_2=-\frac{1}{2}\hmt_0(\bar\eta\, \gd B_2^{\eta}*
\bar\gamma+\eta\, \gd B_2^{\bar\eta}* \gamma)\,,\\
&\gd W_2=-\frac{i}{2}\hmt_0(\omega* \gd S_2+\gd S_2*\omega) \big
|_{\eta\bar\eta}\,,\label{dW2}
\end{align}
where $\gd B_2$ is explicitly given in \eqref{dB2}. Substituting
\eqref{dW2} into \eqref{dver} we eventually find
$\gd\Upsilon^{\eta\bar\eta}$ in the form
\eqref{dver1}-\eqref{dver3}.

\section{$\gb$-shift as star-product re-ordering}

The local framework for analysis of HS equations
\eqref{HS1}-\eqref{HS5} elaborated in \cite{Gelfond:2018vmi,
Didenko:2018fgx, GV} and in this paper rests on the specific
choice of \CH operators in solving for $z$-dependence of
master fields. This choice is driven by PLT theorem that places
constraint on homotopy shifts as in \eqref{even} and \eqref{odd}
in the PLT  -even and -odd sectors, respectively.

As shown in \cite{GV},  PLT  extends to extra derivative shift
\eqref{yshift} with any  $\gb<1$, preserving shift conditions \eq{even}
and \eq{odd} upon overall $(1-\gb)$--rescaling in \eqref{redef}.
  Still, the PLT requirement alone is insufficient for
locality:  to obtain local vertices in the one-form sector one has
to take the limit $\gb\to-\infty$. On the computational side,
$\gb\to-\infty$ limit cuts off non-local contributions keeping the
local ones.

\subsection{Star-product re-ordering}
\label{order}

To understand the $\gb\to -\infty$ limit better we note that $\gb$--extended
\CH \eqref{res} is related to that with $\gb=0$ via $\gb$-dependent
star-product re-ordering. (That such a map exists for
$\gb=1$ was in fact  shown in \cite{DeFilippi:2019jqq}.) Indeed, consider
the following re-ordering operator that maps symbol $f(z,y)$ of
original ordering \eqref{star} to the $\gb$--deformed one
\begin{align}
&O_{\gb}f(z,y)=\int\ff{\dr u\dr v}{(2\pi)^2} f(z+v,y+\gb u)\exp (iu_{\al}v^{\al})\,,\label{ord}\\
&O^{-1}_{\gb}f(z,y)=O_{-\gb}f(z,y)=\int\ff{\dr u\dr
v}{(2\pi)^2}f(z+ v,y-\gb u)\exp (iu_{\al}v^{\al})\,.
\end{align}
It can be shown then by direct computation that
\be
O_{\gb}\hmt_{q,\,\gb}=\hmt_{q,\,0}O_{\gb}
\ee
and
\be\label{hb0}
h_{q,\,\gb}=h_{q,\,0}O_{\gb}\,.
\ee
In other words, $\hmt_{q,\,\gb}$ and cohomology projector
$h_{q,\,\gb}$ appear from the re-ordering similarity transform of
$\hmt_{q,\,0}$ and $h_{q,\,0}$ correspondingly, \be\label{simhom}
\hmt_{q,\,\gb}=O_{\gb}^{-1}\hmt_{q,\,0}O_{\gb}\q
h_{q,\,\gb}=h_{q,\,0}O_{\gb}\,. \ee It is easy to find star
product $\star_\gb$ that corresponds to new ordering \eqref{ord}
from \be\label{startr} f\star_{\gb}
g=O_{\gb}(O^{-1}_{\gb}f*O^{-1}_{\gb}g)\,, \ee where $*$ is the
original star product \eqref{star}. An elementary calculation
gives \be\label{gepstar} f\star_{\gb} g= \int\ff{\dr u\dr u'\dr
v\dr v'}{(2\pi)^4} f(z+u', y+u)g(z-(1-\gb)v-v',y+v+(1-\gb)v')
\exp({iu_{\al}v^{\al}+iu'_{\al}v'^{\al}})\,. \ee

The new star product contains extra integration over $u'$ and
$v'$. There are two points $\gb=0$ and $\gb=2$ that reduce it down
to the normal \eqref{star} and anti-normal orderings of $y\pm z$
operators correspondingly and a point in between, $\gb=1$,
corresponding to their Weyl ordering. For an arbitrary $\gb$ one
can obtain from \eqref{gepstar} the following product rules
\begin{align}
&y\star_\gb =y+i\ff{\p}{\p y}-i(1-\gb)\ff{\p}{\p z}\,,\qquad
\star_\gb y=y-i\ff{\p}{\p y}-i(1-\gb)\ff{\p}{\p
z}\,,\\
&z\star_\gb =z-i\ff{\p}{\p z}+i(1-\gb)\ff{\p}{\p y}\,,\qquad
\star_\gb z=z+i(1-\gb)\ff{\p}{\p y}+i\ff{\p}{\p z}\,,
\end{align}
which show that $y$ and $z$ still commute and their commutators
with star-product elements $f(z,y)$ remain undeformed for any $\gb$
\be\label{com}
[y_{\al},f]_{\star_\gb}=2i\ff{\p}{\p y^{\al}}f\,,\qquad
[z_{\al},f]_{\star_\gb}=-2i\ff{\p}{\p z^{\al}}f\,.
\ee

Since star product defined in \eqref{gepstar} is associative and
space-time independent one can consider HS equations
\eqref{HS1}-\eqref{HS5} in the $\gb$--ordering. This amounts to
simply replacing original star product \eqref{star} by
\eqref{gepstar} and modifying the central element \eqref{klein} by
(cf, Eqs.(3.44) of \cite{DeFilippi:2019jqq})
\be\label{gbga}
\gga_{\gb}=O_{\gb}(\exp({iz_{\al}y^{\al}}))k\theta^{\al}\theta_{\al}=
\ff{1}{(1-\gb)^2}\exp\left({\ff{i}{1-\gbzer}z_{\al}y^{\al}}\right)
k\theta^{\al}\theta_{\al}\,.
\ee

We observe that, within the $\gb$--reordered HS equations, local
HS interactions in the one-form sector are recovered in terms of
conventional \CH  in the limit $\gb\to-\infty$. Indeed, from
(\ref{ord}) it follows that the reordering procedure does not
affect $Y$- or $Z$- independent functions and, hence,
$Z$-independent HS fields $\go(Y,K|x)$ and $C(Y,K|x)$. In
particular, HS vacuum \eqref{S0} is not going to change in the new
ordering. Therefore the perturbative expansion remains the same.
For example, vertex $\Upsilon^{\eta\eta}_{C\go\go C}(\gbzer)$ is
given by
\be\label{verb}
\Upsilon^{\eta\eta}_{C\go\go
C}(\gbzer)=-h_{0,\,0}(W_1{}_{\,\gb}\star_{\gb}W_1{}_{\,\gb})\Big|_{C\go\go
C}\,,
\ee
where $W^{\gb}_{1}$ is the one-form first-order correction
calculated in the $\gb$--ordering using conventional \CH
\eqref{oldres} with $q=0$ and, as opposed to \eqref{cwwcbeta}, we
use conventional projector $h_{0,\,0}$. One can see that
\eqref{verb} is exactly equal to the one in \eqref{cwwcbeta}.
Indeed, using \eqref{ord}  and \eqref{hb0} we have
\be
h_{0,\,0}(W_1{}_{\,\gb}\star_{\gb}W_1{}_{\,\gb})=
h_{0,\,0}O_{\gb}(W_1*W_1)=h_{0,\,\gb}(W_1*W_1)
\ee
reproducing this way \eqref{cwwcbeta}.

Therefore, one may argue that $\gb\to-\infty$ \CH \eqref{res} manifests a way to implement specific re-orderings of
operators that renders HS vertices local. The details of such
localization is yet to be understood. Put it differently, instead
of using $\gb$--shifted \CHs \eqref{res} one could have used the
$\gb$--independent ones  at the price of using modified star
product $\star_\gb$ \eqref{gepstar}.

Note, however, that the limiting points $\gb=1$ and $\gb=-\infty$
lead to singularities in (\ref{gbga}). In fact,  $\gb=1$
corresponds to the Weyl ordering prescription analyzed in
\cite{Vasiliev:2015wma,Bonezzi:2017vha,Iazeolla:2017dxc,DeFilippi:2019jqq} (and
references therein). The singularity in (\ref{gbga}) expresses the
fact that inner Klein operators become $\delta$-functions in the
Weyl ordering \cite{BS,Didenko:2009td}. The analysis of
\cite{Vasiliev:2015wma} indicated that it is hard to handle the
resulting divergencies. On the other hand, recent results of
\cite{DeFilippi:2019jqq} show that one can proceed at least in the
lowest order. From the perspective of this paper this can be
understood as a consequence of the fact that, in the lowest order,
$\gb$--shifted contracting homotopies are equivalent to the
conventional ones with $\gb=0$. It would be interesting to see to
which extent the results of \cite{DeFilippi:2019jqq} can be
extended to higher orders. In our analysis we were not able to
make use of the $\gb\to 1$   limit.

The local limit singularity at $\gb\to-\infty$ in the star product
is less clear demanding further investigation. In this case,
formula (\ref{ord}) as well as $\star_{-\infty}$ to be well
defined require precise specification of the functional class
used. Leaving detailed analysis of this issue for the future, at
this stage we therefore prefer to analyze the problem in terms of
the standard HS star product (\ref{star}) but using
$\hmt_{q,-\infty}$.

Practically, the use of the reordering operator \eqref{ord}
sometimes simplifies calculations that include analysis of several
\CH  operators. Particularly, it is useful for the analysis of
$S_2$ contribution to (anti)holomorphic vertices as we   show now.

\subsection{$S_2$ contribution to
$\Upsilon^{\eta\eta}$}\label{S2sec} It was argued already that
contribution to (anti)holomorphic part of the vertices from $S_2$
can be made to vanish at $\gb\to-\infty$. Let us demonstrate this
in some detail. This contribution comes from the second term on
the \rhs of \eqref{ver} with the part of $W_2$ taken from
\eqref{W2'}. It potentially contributes to the three vertex
structures $\Upsilon^{\eta\eta}_{\go\go CC}$,
$\Upsilon^{\eta\eta}_{CC\go\go }$ and $\Upsilon^{\eta\eta}_{\go
CC\go}$. Let us consider $\Upsilon^{\eta\eta}_{\go\go CC}$ for
example,
\be -\hhmt_{0,\,\gb}(\go*W_2')\big|_{\go\go
CC}=\ff{1}{4}h_{0,\,\gb}(\go*\hmt_{0,\,\gb}(\go*\hmt_{0,\,\gb}(S_1*S_1-i\eta
B_2*\gga))\,.
\ee
Using   star-exchange relation \eqref{lexch} one brings it to
\be\label{Over}
-\hhmt_{0,\,\gb}(\go*W_2')\big|_{\go\go CC}=
\ff14\go*\go*h_{(1-\gb)(t_1+t_2),\,\gb}\hmt_{(1-\gb)t_2,\,\gb}\hmt_{0,\,\gb}(S_1*S_1-i\eta
B_2*\gga)\,.
\ee
A convenient way to tackle three consecutive $\gb$-homotopies is
to use the similarity reordering identities \eqref{simhom} which
allows one to reduce them to the much simpler $\gb$-independent
ones. This gives \be
\ff14\go*\go*h_{(1-\gb)a}\hmt_{(1-\gb)b}\hmt_{0}O_{\gb}(S_1*S_1-i\eta
B_2*\gga)\,,\qquad a=t_1+t_2\,,\quad b=t_2\,, \ee where $O_{\gb}$
is given by \eqref{ord}. Using the scaling property of
$h_{a}\hmt_b\hmt_c$ \eqref{scale} to calculate the vertex at
$\gb\to-\infty$ it suffices to analyze the following limit
\be\label{lim} \lim_{\gb\to-\infty}(1-\gb)^2O_{\gb}(S_1*S_1-i\eta
B_2*\gga)\Big|_{z\to(1-\gb) z}\,. \ee The vertex is then
reproduced by applying
$\ff{1}{4}\go*\go*h_{a}\Delta_{b}\Delta_{0}$ to the resulting
limiting expression. This way we first of all find using the
explicit \eqref{B2exp} and changing the integration variables that
\begin{align}
&(1-\gb)^2O_{\gb}(B_2*\gga)\Big|_{z\to(1-\gb)
z}=-\ff{i\eta}{2}\theta^2 C\bar{*}C\nn\int \dr^3_{\triangle}\tau
(p_1+p_2)^{\al}(y+p_1)_{\al}\times\\
& \exp i \left({(\tau_1+\tau_2)z_{\al}y^{\al}-
 (\tau_1p_2-\tau_2p_1)^{\al}z_{\al}-
 (1-\tau_1-\tau_2)(p_1+p_2)^{\al}(y+p_1)_{\al}}\right)\label{OB2}
\end{align}
is $\gb$ -- independent.

Contribution from $S_1*S_1$ is found from \eqref{S1exp} by
performing star-product integration
\begin{align}
&S_1*S_1=-\ff{\eta^2}{2}
\theta^2C\bar{*}C\gep^{\al\gb}\p_{1\al}\p_{2\gb}
\int_{[0,1]^2}\dr\tau_1\dr\tau_2
\times\\
&\exp i \left({ \tau_\circ z_\al y^{\al}
- \tau_1\tau_2(p_1+p_2)^{\al}y_{\al}- (\tau_2(1-\tau_1)p_2-\tau_1(1-\tau_2)p_1)^{\al}z_{\al}
+ \tau_1\tau_2 p_{1}^{\al}p_{2\al}}\right)\,,\nn\\
\nn&\tau_{\circ}=\tau_1(1-\tau_2)+\tau_2(1-\tau_1)\,,
\end{align}
where $\p_{1,2}{}_\ga$ denote differentiation with respect to
$p^\ga_{1,2}$. Applying to it \eqref{ord} one finds
\begin{align}
&O_{\gb}(S_1*S_1)=-\ff{\eta^2}{2} \theta^2C\bar{*}C\gep^{\al\gb}
 \int_{[0,1]^2}\dr\tau_1\dr\tau_2\p_{1\al}\p_{2\gb}\ff{1}{(1-\gb\tau_\circ)^2}\times\\&\exp i\left({\ff{1}{1-\gb\tau_\circ}(\tau_\circ
z_\al y^{\al}
-\tau_1\tau_2(p_1+p_2)^{\al}y_{\al}- (\tau_2(1-\tau_1)p_2-\tau_1(1-\tau_2)p_1)^{\al}z_{\al}
+\tau_1\tau_2 p_{1}^{\al}p_{2\al})}\right)\,.\nn
\end{align}
We now take the limit
\begin{align}\label{S1S1}
&\lim_{\gb\to-\infty}(1-\gb)^2O_{\gb}(S_1*S_1)\Big|_{z\to(1-\gb)
z}=-\ff{\eta^2}{2}
\theta^2C\bar{*}C\times\nn\\
&\int
d^3_{\triangle}\tau[(\tau_1+\tau_2)z_{\al}y^{\al}-z_{\al}(\tau_1p_2-\tau_2
p_1)^{\al}-(1-\tau_1-\tau_2)(p_1+p_2)^{\al}(y+p_1)_{\al}-2i]\times\\
&\exp i\left({ (\tau_1+\tau_2)z_{\al}y^{\al}-
 (\tau_1p_2-\tau_2p_1)^{\al}z_{\al}-
 (1-\tau_1-\tau_2)(p_1+p_2)^{\al}(y+p_1)_{\al}}\right)\,\nn
\end{align}
and note that the exponential parts of \eqref{S1S1} and
\eqref{OB2} identically coincide. Pre-exponentials are different
but when combined together according to \eqref{lim} $B_2*\gga$
cancels out with 1 in
$(1-\tau_1-\tau_2)(p_1+p_2)^{\al}(y+p_1)_{\al}$ part of the
pre-exponential of \eqref{S1S1} bringing us to
\begin{align}
&\lim_{\gb\to-\infty}(1-\gb)^2O_{\gb}(S_1*S_1-i\eta
B_2*\gga)\Big|_{z\to(1-\gb) z}=-\ff{\eta^2}{2}
\theta^2C\bar{*}C\times\\
\nn&\int
d^3_{\triangle}\tau[(\tau_1+\tau_2)z_{\al}y^{\al}-z_{\al}(\tau_1p_2-\tau_2
p_1)^{\al}+(\tau_1+\tau_2)(p_1+p_2)^{\al}(y+p_1)_{\al}-2i]\times\\
\nn&\exp i\left({ (\tau_1+\tau_2)z_{\al}y^{\al}-
 (\tau_1p_2-\tau_2p_1)^{\al}z_{\al}-
 (1-\tau_1-\tau_2)(p_1+p_2)^{\al}(y+p_1)_{\al}}\right)\,.
\end{align}
The latter can be rewritten as
\begin{align}
&\ff{i\eta^2}{2} \theta^2C\bar{*}C\int
d^3_{\triangle}\tau[\tau_1\ff{\p}{\p\tau_1}+\tau_2\ff{\p}{\p\tau_2}+2]\times\\
\nn&\exp i\left({ (\tau_1+\tau_2)z_{\al}y^{\al}-
 (\tau_1p_2-\tau_2p_1)^{\al}z_{\al}-
 (1-\tau_1-\tau_2)(p_1+p_2)^{\al}(y+p_1)_{\al}}\right)\,.
\end{align}
Partial integration amounts to
\begin{align}\label{part}
&-\ff{i\eta^2}{2} \theta^2C\bar{*}C\int
\dr^3\tau(\tau_1\ff{\p}{\p\tau_1}+\tau_2\ff{\p}{\p\tau_2})X\times\\
&\exp i\left({ (\tau_1+\tau_2)z_{\al}y^{\al}-
 (\tau_1p_2-\tau_2p_1)^{\al}z_{\al}-
 (1-\tau_1-\tau_2)(p_1+p_2)^{\al}(y+p_1)_{\al}}\right)\,,
\nn
\end{align}
where (recall the definition of measure $\dr^3_{\triangle}$ in
\eqref{mera3simplex})
\be
X=\gd(1-\sum\tau_i)\theta(\tau_1)\theta(\tau_2)\theta(\tau_3)
 \,.
\ee
Noting that
\be
\ff{\p}{\p\tau_1}\gd(\dots)=\ff{\p}{\p\tau_2}\gd(\dots)=\ff{\p}{\p\tau_3}\gd(\dots)
\ee
and $\tau_1+\tau_2=1-\tau_3$ we further rewrite \eqref{part}
in the following form
\begin{align}
&-\ff{i\eta^2}{2} \theta^2\int
\dr^3\tau\ff{\p}{\p\tau_3}(\gd(1-\sum\tau_i)(1-\tau_3))
\theta(\tau_1)\theta(\tau_2)\theta(\tau_3)
 \times\\
\nn
&\exp i\left({(\tau_1+\tau_2)z_{\al}y^{\al}-
 (\tau_1p_2-\tau_2p_1)^{\al}z_{\al}-
 (1-\tau_1-\tau_2)(p_1+p_2)^{\al}(y+p_1)_{\al}}\right)\,.
\end{align}

Yet another partial integration along $\tau_3$ yields the final
result
\begin{align}\label{limS}
&\lim_{\gb\to-\infty}(1-\gb)^2O_{\gb}(S_1*S_1-i\eta
B_2*\gga)\Big|_{z\to(1-\gb) z}=\nn\\
&=\ff{i\eta^2}{2}
\theta^2C\bar{*}C\int_{[0,1]^2}\dr\tau_1\dr\tau_2
\gd(1-\tau_1-\tau_2)\exp i \left({ z_{\al}y^{\al}- (\tau_1
p_2-\tau_2 p_1)^{\al}z_{\al}}\right)\,.
\end{align}
{Using \eqref{lexch} and denoting by $a=p_1+2p_2$ and $b=p_2$ one
can rewrite \eqref{limS} in the form
\be
\lim_{\gb\to-\infty}\gb^2O_{\gb}(\hmt_a\gga*\hmt_b\gga-\hmt_a\hmt_b\gga*\gga)\Big|_{z\to(1-\gb)
z}=\phi(y;k)*\gga\,,
\ee
where $\phi(y;k)$ is the $z$--independent function given by
\be
\phi(y;k)=2i\int_{[0,1]^2}\dr\tau_1\dr\tau_2
\gd(1-\tau_1-\tau_2)e^{i(\tau_1 a+\tau_2 b)^{\al}y_{\al}} k\,.
\ee
}
 Let us note that this cancellation mechanism can be
analyzed without precise calculation of the limit $\gb\to-\infty$
rather using a language of the functional classes of \cite{GV},
where it is shown that $S_1*S_1-i\eta B_2*\gga$ enjoys a
remarkable structure relation. The corresponding contribution to
the vertex appears from application of
$\go*\go*h_{t_1+t_2}\hmt_{t_2}\hmt_0$ to \eqref{limS} and thanks
to \eqref{def1.0dok0} is obviously ultra-local.

We now note that the non-zero result \eqref{limS} can be
completely eliminated by the following local $z$--independent
field redefinition
\begin{align}
&B_2\to B_2+\gd B_2(y) =B_2^{min}\,,\\
\gd
&B_2=\ff{\eta}{2}C\bar{*}C\int_{[0,1]^2}\dr\tau_1\dr\tau_2\gd(1-\tau_1-\tau_2)\exp
\left({-i(\tau_2p_1-\tau_1p_2)^{\al}y_{\al}}\right)k \,,
\end{align}
which is nothing but \eqref{B2min} corresponding to the minimal
derivative vertex $\Upsilon(\go,C,C)$ found in
\cite{Vasiliev:2016xui}. Indeed, the effect of such
$z$--independent field redefinition for $B_2$ amounts to the
corresponding change in fields $S_2$ and $W_2$ according to
\eqref{S2hol} and \eqref{W2'}. This results in the change of the
vertex \eqref{Over}
\be
\gd \Upsilon^{\eta\eta}_{\go\go CC}=
\ff14\go*\go*h_{(1-\gb)(t_1+t_2),\,\gb}\hmt_{(1-\gb)t_2,\,\gb}\hmt_{0,\,\gb}(-i\eta
\gd B_2*\gga)\,.
\ee
To analyze its limiting value at $\gb\to-\infty$ one again uses
the scaling property \eqref{scale} which leads to the following
expression
\be\label{lim1}
(1-\gb)^2O_{\gb}(-i\eta\, \gd B_2*\gga)\Big|_{z\to(1-\gb)
z}=-i\eta\gd B_2*\gga\,,
\ee
where the \rhs is due to $z$--independence of $\gd B_2$. Up to
an overall sign it equals \eqref{limS} thus cancelling it out.

Similar consideration applies to the rest of contributions to
$\Upsilon^{\eta\eta}_{CC\go\go }$ and $\Upsilon^{\eta\eta}_{\go
CC\go}$ which are expressed as well via \eqref{lim} followed by
some $h_a\hmt_b\hmt_c$ action. Therefore, we conclude that
$\gb\to-\infty$ contribution from $S_2$ to (anti)holomorphic
vertices is ultra-local in accordance with the general
consideration of \cite{GV}. Moreover it vanishes at all for the
minimally coupled $B_{2}^{min}$ \eqref{B2min}.

\section{Conclusion}
\label{conc}

In this paper we identified the homotopy procedure that leads to local HS vertices
not only in the zero-form sector of the HS equations as in \cite{Didenko:2018fgx}
but also in the one-form sector.
Specifically we have evaluated all vertices that are bilinear both in the zero-form
 fields $C$ and in the one-form fields $\go$ in the  sector of equations (\ref{ver2}) for
 $\go$. Since $\go$ describes dynamical HS fields (not, say,
 just $AdS_4$ background) the corresponding vertices describe HS interactions
 of massless fields of all spins up to
 the quintic order at the Lagrangian level. All these vertices are shown to be spin-local.
 This conclusion is in agreement with the conjecture of \cite{Gelfond:2018vmi} that
 HS gauge theories are spin-local not only in the lowest order but in the higher orders as well.

The same time, the obtained vertices agree with the lower-order
results obtained in \cite{Vasiliev:2016xui}-\cite{Gelfond:2017wrh}
by less sophisticated means.
 In particular, in \cite{Gelfond:2017wrh} it was shown that $\eta^2$ and $\bar\eta^2$
vertices in the one-form sector vanish in the $AdS_4$ background.
Though higher-order vertices obtained in this paper are non-zero
in the $\eta^2$ and $\bar \eta^2$ sectors, their restrictions to
$AdS_4$ background one-form fields indeed vanish that provides a
highly nontrivial test of the obtained results. This implies in
turn a very important property of HS equations that the resulting
currents on the \rhs of the HS equations are local and the
coupling constants in front of them only depend on $\eta\bar \eta$
having definite signs independent of the phase of $\eta$. In
particular, the gravitational  constant as a coefficient in front
of the stress tensor is positive. Moreover $\eta^2$ and $\bar
\eta^2$ vertices vanish for one-forms that are no more than
bilinear in $Y's$, particularly, for any gravitational or
electromagnetic background.

To obtain corrections to the \rhss  of the Fronsdal equations one
has to project all obtained vertices to the sector of frame-like
fields with the equal numbers of $y$ and $\bar y$ in the bosonic
case and differing by 1 in the fermionic. (For more detail see
e.g. \cite{Vasiliev:1999ba} and
\cite{Gelfond:2010pm,Misuna:2017bjb} for analogous analysis of HS
interactions.) After expanding in powers of $y$ and $\bar y$, one
has to evaluate integrals over the homotopy parameters like $\gs$
and $\tau$ in the $\eta^2$ sector or $\bar\gs$, $\bar\tau$ in the
$\bar\eta^2$-sector and $\tau$, $\bar \tau$ in the $\eta\bar\eta$
sector, that enter through the \CHs $\hmt$, $\bar \hmt$ and
cohomology projectors $h$, $\bar h$. This will result in
nontrivial coefficients in front of different types of vertices in
terms of fields. Being straightforward,
 derivation of their explicit form is
beyond the scope of this paper.

The homotopy  procedure used in our analysis is based on the limit
$\gbzer \to -\infty$ with respect to the parameter $\gbzer$ that
enters the shifted \CH $\hmt_{0,\gbzer}$. Remarkably, the limit is
well defined and the terms with infinite towers of higher
derivatives vanish in this limit so that the resulting vertices
become spin-local. This phenomenon, which is general in nature, is
anticipated to work in higher orders as well. The technique
appropriate for this analysis is developed further in \cite{GV}
based on the previous analysis of locality in
\cite{Vasiliev:2015wma}. {In particular, in \cite{GV} it is shown
that the shifted homotopy technique is well defined for any
$-\infty<\gb <1$ and a sufficient condition for the limit $\gb\to
-\infty$ be well defined is found and shown to be fulfilled for
the vertices analysed in this paper.} At any rate, the limiting
$\gb$--shifted \CH identified in this paper is unique for higher
orders as well since it makes no sense to choose any other value
of $\gbzer$ except for the limiting one  $\gbzer \to -\infty$. The
reason why the lowest-order analysis of locality in
\cite{Gelfond:2018vmi,Didenko:2018fgx}, which was based on the
shifted homotopy with $\gbzer=0$, was successful is that, as shown
in this paper, the relevant lower-order terms turn out to be
$\gbzer$-independent. This phenomenon does not take place at
higher-orders where the limit $\gbzer\to -\infty$ becomes fully
significant.

The obtained vertices naturally split into two classes of $\eta^2$, $\bar\eta^2$  and
$\eta\bar\eta$ vertices that originate from different
 \CHs prescribed by PLT. This led to their different locality
properties. While the $\eta^2$, $\bar\eta^2$ vertices turn out to be spin
ultra-local, in accordance with their current
interaction structure, the $\eta\bar\eta$ ones
 are spin-local\footnote{This situation is similar
to what happens at lower orders. Vertex $\Upsilon(\go,\go,C)$
stems from PLT-even sector and is spin-ultra-local, however
$\Upsilon(\go,C,C)$ originates from the PLT-odd one and is
spin-local.} but $z$-dominated in the terminology of
\cite{Gelfond:2018vmi}.

 It should be stressed that there is of course a great freedom in
performing spin-local field redefinitions that preserve the class of spin-local
vertices. The class of these field redefinitions is defined the same way as
spin-local vertices as being free from contractions between different factors
of zero-forms $C$ in either holomorphic or antiholomorphic sector (or both).
Spin-local vertices form an equivalence class with respect to spin-local field redefinitions.
So in this paper we found a particular representative of the class of spin-local
vertices that may or may not be most useful for applications. For instance the
$\eta\bar \eta$ vertices presented in Section \ref{pinvv} do not respect the
invariance of the whole setup under the fundamental anti-automorphism $\rho$ of the
HS theory that relates opposite  orderings of the field product factors. As such, it is not most convenient for
the analysis of the minimal HS model resulting from the truncation induced by
$\rho$ (for more detail see \cite{Vasiliev:1999ba}).
As explained in Section \ref{s2w2mi}, this can however be easily cured by a minor modification of the \CH scheme.

A related comment is that while all possible local field
redefinitions lead to HS vertices with different number of
derivatives it might be crucial for higher-order locality to
single out those that give minimal derivative couplings. We have
shown that the local field redefinition \eqref{dB2} that makes
lower order vertex $\Upsilon(\go, C,C)$ to contain minimal number
of derivatives results in an extra ultra-local cancellations
within $\Upsilon(\go,\go, C,C)$ structures. Particularly, in this
case field $S$ second order contribution completely vanishes
leading among other things to the vanishing of (anti)holomorphic
vertices on any gravitational backgrounds.

The obtained results illustrate  high efficiency
of the nonlinear HS equations of \cite{more} as compared to
 other approaches available in the literature. (It is not clear whether at all
it is possible to compute quintic vertices with arbitrary parity breaking
parameter $\phi$ in $\eta =|\eta| \exp i\phi$ for infinite towers of spins
by other means like, e.g.,
holographic reconstruction). Though in this paper it has been developed
specifically for the $4d$ HS theory, our approach is applicable to the analysis
of $3d$ theory
of \cite{prok} (to large extent this analysis is contained in the $\eta^2$ sector of the $4d$ theory), HS theory in any dimension of
\cite{Vasiliev:2003ev} and, most important, to the multi-particle theory of
\cite{Vasiliev:2018zer}, conjectured to be related to String Theory.

Of course, it remains to be checked how our prescription works for
other higher-order vertices, namely those containing more than two
zero-forms $C$. The simplest vertex of this type is
$\Upsilon(\go,C,C,C)$  in  (\ref{ver2}). In particular, it is this
vertex that contains the scalar field self-interaction
corresponding to the quartic Lagrangian spin-zero vertex. The fact
that $\eta^2$ and $\bar \eta^2$ vertices appear in the ultra-local
form severely constrains potential non-localities of the
$\Upsilon(\go,C,C,C)$ vertex. Indeed, had the $\eta\eta$ and
$\bar\eta\bar\eta$ vertices been just spin-local rather than
ultra-local the integrability condition for that quartic vertex
would resulted in star products of spin-local pieces which are
generally non-local. This however never happens to ultra-local
terms.

It would be interesting to compare the vertex of this type
resulting from HS equations with those  elaborated in
\cite{Bekaert:2015tva,Sleight:2017pcz,Ponomarev:2017qab}. This is
the most urgent problem on the agenda that we leave for a future
publication. The technical tools for the analysis of this problem
are elaborated in this paper and in \cite{GV}.

While the mechanism that brings vertices to their local form in
the limit $\gb\to-\infty$ is essentially the suppression of higher
derivative terms it is not entirely clear why it works in a so
delicately fine tuned way.  We have shown here that the effect of
parameter $-\infty <\gb< 1$ is equivalent to a star-product
re-ordering. An interesting problem for the future is to
understand to which extent the limiting $\gb\to -\infty$ star
product is defined on its own right.

An interesting feature of the developed formalism is that it treats differently
one-forms $\go$ and zero-forms $C$. In the sector of higher spins this is just
what is needed given that these are zero-forms $C$ that contain infinite tails of higher derivatives
of Fronsdal fields.
However, the most general version of the $4d$ HS theory \cite{more} (see also \cite{Vasiliev:1999ba}) contains the sector
of topological (Killing-like) fields, each carrying at most a finite number of
degrees of freedom. In this sector the roles of one-forms and zero-forms
are swapped: zero-forms $C^{top}$ carry a finite number  of derivatives
of the topological fields while one-forms $\go^{top}$ contain infinite towers
of derivatives. This can affect the analysis of locality if the HS and topological sectors
 get interacting that can happen if some of the topological fields
acquire nontrivial VEVs. In particular this happens in the $3d$ HS theory of
\cite{prok} where the topological sector is related to the dynamical one.

An important related issue is the study of the HS black hole
solutions found in
\cite{Didenko:2009td,Iazeolla:2011cb,Iazeolla:2012nf}.
 Analysis of field fluctuations around the BH solutions demands accurate
choice of the field variables and specification of the relevant classes of functions.
(Otherwise it is hard to distinguish between dynamical and pure gauge degrees of
freedom \cite{DVU}.) A related point is that topological fields play a
role of chemical potentials in the analysis of invariant functionals associated with
boundary charges \cite{Didenko:2015pjo}. This suggests that the effects of (non)locality in presence of topological fields can in particular
 play a role in the analysis  of HS black holes.

\section*{Acknowledgments}
We are grateful to Ruslan Metsaev, Ergin Sezgin, Alexey Sharapov,
Oleg Shaynkman, and especially to Carlo Iazeolla, David De
Philippi and Per Sundell for helpful discussions. {Finally, we
would like to thank the Referee for making numerous useful remarks
on the manuscript.} The work was supported by the Russian Science
Foundation grant 18-12-00507 in association with Lebedev Physical
Institute. We thank Erwin Schr\"{o}dinger Institute in Vienna for
hospitality during the program ``Higher Spins and Holography''
where this work was completed.

\newcounter{appendix}
\setcounter{appendix}{1}
\renewcommand{\theequation}{\Alph{appendix}.\arabic{equation}}
\addtocounter{section}{1} \setcounter{equation}{0}
 \renewcommand{\thesection}{\Alph{appendix}.}

 \addcontentsline{toc}{section}{\,\,\,\,\,\,\,Appendix A.  Derivation of
 contracting homotopy formula}
 \section*{Appendix A. Derivation of contracting homotopy formula}
\label{AppA}

Applying
(\ref{res}) to (\ref{f}) we obtain
\bee
 &&\hmt_{0,\gbzer} f(z,y, \theta) =\ff{1}{(2\pi)^2}\int \dr^2 u \dr^2 v  \int_0^1 \dr\tau\int_0^1 \dr t t^{p-1}
 \exp i[v_{\al} u^\al+\tau (tz+(1-t) u)_{\al} (\gb v+y)^\al ]\nn\\&&
 \times (z-u )^\ga\frac{\p}{\p \theta^\ga}
  \phi(\tau( tz+(1-t) u),(1-\tau)(\gb v+y), \gt\theta,\tau )\,.
\eee
Introducing new integration variables,
\be
\tau_1 = t\tau\q \tau = \tau_1+\tau_2\q 1-\tau = \tau_3\,
\ee
with the Jacobian
\be
\det \Big | \frac{\p (\tau, t)}{\p \tau_i}\Big |=(\tau_1
+\tau_2)^{-1}
\ee
we obtain
\bee
 &&\hmt_{0,\,\gbzer} f(z,y, \theta) =\ff{1}{(2\pi)^2}\int \dr^2 u \dr^2 v  \int \dr^3_{\triangle} \tau
 \frac{\tau_1^{p-1}}{(\tau_1 +\tau_2)^{p}}\delta(1-\sum_{i=1}^3\tau_i)\\&& \nn
 \exp i[v_{\gb} u^\gb+ (\tau_1 z+\tau_2 u)_{ \gb } (\gb v+y)^\gb ]
  (z-u )^\ga\frac{\p}{\p \theta^\ga}
  \phi(\tau_1 z+\tau_2 u,\tau_3(\gb v+y),(\tau_1 +\tau_2) \theta,\tau_1 +\tau_2 )\,.
\eee

Then, the shift of the integration variables,
\be
u_\ga \rightarrow u_\ga+ \frac{\tau_1\gbzer}{1-\tau_2\gbzer} z_\ga\q v_\ga\rightarrow (1-\tau_2\gbzer)^{-1}
( v_\ga +\tau_2 y_\ga)
\ee
yields using definition \eq{mera3simplex}
\bee
 &&\hmt_{0,\,\gbzer} f(z,y, \theta) =
 \ff{1}{(2\pi)^2}\int \dr^2 u \dr^2 v  \int\dr_{\Delta}^3\tau
  (\tau_1)^{p-1}  (1-\gbzer \tau_2)^{-3}
 \exp i\left[v_{ \gb } u^\gb + \frac{\tau_1}{(1-\gbzer\tau_2)}z_{\gb } y^\gb \right]\qquad \nn\\&&\ls
  ((1-(\tau_1 +\tau_2)\gbzer) z- (1-\gbzer \tau_2) u )^\ga\frac{\p}{\p \theta^\ga}
  \phi\left(\frac{\tau_1}{(1-\gbzer\tau_2)} z+\tau_2 u,\frac{\tau_3}{1-\gbzer\tau_2}(y+\gb v),
   \theta,\tau_1 +\tau_2\right)\!.\,
\qquad\eee

To reduce this expression to the desired form (\ref{f}) we finally change variables to
\be\label{1'}
\tau_1' = \frac{\tau_1}{1-\gbzer\tau_2}\q \tau_3' = \frac{\tau_3}{1-\gbzer\tau_2}\q\tau_2' =
\frac{(1-\gbzer)\tau_2}{1-\gbzer\tau_2}\,.
\ee
This change of variables, which we call {\it simplicial map}, is remarkable in several respects.
Firstly, all $\tau_i'\in [0,1]$, taking into account that
\be\label{sum}
\sum_{i=1}^3 \tau_i=1.
\ee
 Secondly, it preserves the class of simplices of unit perimeter since
 \be
 \sum_{i=1}^3 \tau'_i=1
 \ee
 as a consequence of (\ref{sum}). The Jacobian is
 \be
 \det \Big |\frac{\p \tau'_i}{\p \tau_j}\Big |= \frac{1-\gbzer}{(1-\gbzer \tau_2 )^3}\,.
 \ee
Using also that
\be
1-\gbzer\tau_2 = \frac{1-\gbzer}{1-\gbzer (1-\tau_2')}
\ee
we finally obtain (\ref{hmtgb0}) upon discarding primes.

 \renewcommand{\theequation}{\Alph{appendix}.\arabic{equation}}
\addtocounter{appendix}{1} \setcounter{equation}{0}
  \addtocounter{section}{1}
\addcontentsline{toc}{section}{\,\,\,\,\,\,\,Appendix B.    Useful formulae}
  \section*{Appendix B.   Useful formulae}
\label{AppB}

\subsection*{Lower order fields}
\label{App1}Here we collect explicit formulae for perturbative master fields.
For $S_1$ from \eqref{S1om} and $W_1$ from \eqref{W1om} it is easy
to obtain
\be\label{S1exp}
S_1^\eta=-i\eta\, \int_{0}^{1}\dr\tau\, C(0,\bar
y)\,\theta^{\al}\ff{\p}{\p p^{\al}} \exp({i\tau
z_{\al}y^{\al}+i\tau p^{\al}z_{\al}})k+h.c.\,
\ee
and
\begin{align}\label{W1exp}
W_1^\eta=\ff{\eta}{2}\int\dr_{\Delta}^3\tau\,
C(0,\bar y)\bar{*}\go(0,\bar y)\ff{1}{\tau_1}t^{\al}\ff{\p}{\p p^{\al}}
\exp({i\tau_{1} z_{\al}y^{\al}+i\tau_3
t^{\al}y_{\al}+i\tau_{1}(p+t)^{\al}z_{\al}+i(1-\tau_3)p^{\al}t_{\al}})k\nn\\
+ \ff{\eta}{2}\int\dr_{\Delta}^3\tau\,\go(0,\bar y)\bar{*}C(0,\bar
y) \ff{1}{\tau_1}t^{\al}\ff{\p}{\p p^{\al}} \exp({i\tau_{1}
z_{\al}y^{\al}-i\tau_3
t^{\al}y_{\al}+i\tau_{1}(p+t)^{\al}z_{\al}-i(1-\tau_3)p^{\al}t_{\al}})k+h.c.\,
\qquad\end{align} For $B_2^\eta$ one finds from \eqref{b2}
\begin{multline}\label{B2exp}
B_2^\eta=\ff{\eta}{2}\int\dr_{\Delta}^{3}\tau\,
C(0,\bar y)\bar{*}C(0,\bar y)\times\\
\ff{\p}{\p\tau_1}
\exp({i(1-\tau_2-\tau_3)z_{\al}y^{\al}+i\tau_1(p_1+p_2)^{\al}(z-p_1)_{\al}+
i(\tau_2p_2-\tau_3p_1)^{\al}y_{\al}})k\,.
\end{multline}
In these expressions field derivatives $p$ and $t$ \eqref{CgoCgoCpt} act
on the left. Equivalently, by explicit action with differential
operators $p$ and $t$ expressions \eqref{S1exp}-\eqref{B2exp} can
be rewritten as follows
\be
\label{S1} S_{1}^{\eta}=\eta\,
\int_{0}^{1}\dr\tau\,\tau\theta^{\al}z_{\al}C(-\tau z, \bar y)
\exp({i\tau z_{\al}y^{\al}})k+h.c.\,,
\ee
\bee
\label{W1} W_{1}^{\eta}=-i\ff{\eta}{2}\int \ff{\dr^2 u\dr^2
v}{(2\pi)^2}\int\dr_{\Delta}^3\tau\,\,\exp({i\tau_{1}
z_{\al}y^{\al}+iu_{\al}v^{\al}})
  z^{\al}\ff{  {\p}}{ {\p} v^\al}
 \rule{180pt}{0pt}\\\nn \Big( C((1\!-\!\tau_3)u\!-\!\tau_1 z,\bar
y)\bar{*}\go(v\!-\!\tau_1 z\!-\!\tau_3 y,\bar
y) +  \go(v\!-\!\tau_1
z+\tau_3 y,\bar y)\bar{*}C((1\!-\!\tau_3)u\!-\!\tau_1 z,\bar y)
\Big)k+h.c.\,,
\rule{20pt}{0pt}\eee
 \be
B_{2}^{\eta}= \ff{\eta}{2}\int \ff{\dr^2 u\dr^2
v}{(2\pi)^2}\int\dr_{\Delta}^{3}\tau\exp
i({\tau_1z_{\al}y^{\al}+u_{\al}v^{\al}}) \ff{\p}{\p\tau_1}\Big(
C_1(\tau_1 u-\tau_1 z+\tau_3 y,\bar y)\bar{*}C_2(v-\tau_1 z-\tau_2
y, \bar y)\Big) k.
\ee

\subsection*{Projector formulas}
\label{App2}In calculation of \eqref{verdef1} and similar it is
convenient to use the following formulae
  \be
h_{0,\,\gbzer}(\go*\hmt_{0,\,\gbzer}(\theta^{\al}f_{\al}(z,y)))=-
\go(0,\bar y)\ff{(1-\gbzer)}{(2\pi)^2}\int \dr^2 u \dr^2  v
\,\exp({iu_{\al}v^{\al}+iy^{\al}t_{\al}}) F^+(f )\q \ee
\be
h_{0,\,\gbzer}(\hmt_{0,\,\gbzer}(\theta^{\al}f_{\al}(z,y))*\go)= -
\ff{(1-\gbzer)}{(2\pi)^2}\int \dr^2 u \dr^2  v\,
\exp({iu_{\al}v^{\al}+iy^{\al}t_{\al}}) F^-(f)\go(0,\bar y)\,\q
\ee
where\be F^\pm(f )=\int_{0}^{1}\dr\gs\,
t^{\al}f_{\al}(v-(1-\gbzer)\gs t, y\pm t+\gbzer u)\,.
\ee

Particularly, for
\be
\tilde{f}_{\al}(z,y)=\phi_{\al} \exp({i\tau_{\circ}
z_{\al}y^{\al}+iA^{\al}y_{\al}+iB^{\al}z_{\al}})\,,
\ee
where $\phi_{\al}$, $A_{\al}$ and $B_{\al}$ are some constant
spinors,
performing Gaussian integration
these amount to
\be
h_{0,\gbzer}(\go*\Delta_{0,\gbzer}(\theta^{\al}\tilde{f}_{\al}(z,y)))\,=\go(0,\bar y) H^{+}\q
h_{0,\gbzer}(\Delta_{0,\gbzer}(\theta^{\al}\tilde{f}_{\al}(z,y))*\go)\,=H^{-} \go(0,\bar y)\q
 \ee  where
\bee \label{hwd}
H^{\pm} &=&-\int_{0}^{1}\dr\tau\,
\ff{1-\gbzer}{(1-\gbzer\tau_{\circ})^2}t^{\al}\phi_{\al}
\\\nn &&
\exp\left({\ff{i}{1-\gbzer\tau_{\circ}}\left[ (A+(1-\gbzer)\tau_{\circ}\tau t)^{\al}(y\pm t)_{\al}-B^{\al}(\gbzer A+(1-\gbzer)\tau
t)_{\al}\right]+iy^{\al}t_{\al}}\right)\,.
\eee

\subsection*{Some homotopy integrals}
\label{App3} In analysis of $\gbzer$--dependent vertices one
encounters the following integrals (cf. \eqref{ccwwbigbeta} and
\eqref{cwwcbigbeta}) that we need to evaluate at
$\gbzer\to-\infty$
\begin{align}
&I_1(\gbzer)=-\int_{[0,1]^2}\dr\tau_1\,\dr\tau_1'
\ff{\gbzer^3\tau_1(1-\tau_1')^2(1-\tau_1)}{(1-\gbzer\tau_{\circ})^4}\times
\exp i\Big({-\ff{\gbzer}{\xi}\tau_1(1-\tau_1')
A-\ff{\gbzer}\xi\tau_1'(1-\tau_1)B+\ff 1\xi C+D}\Big)\,,\label{I1}\\
&I_2(\gbzer)=\int_{[0,1]^2}\dr\tau_1\,\dr\tau_1'
\ff{\gbzer^2(1-\tau_1')(1-\tau_1)}{(1-\gbzer\tau_{\circ})^4}\times
\exp i\Big({-\ff{\gbzer}{\xi}\tau_1(1-\tau_1')
A-\ff{\gbzer}\xi\tau_1'(1-\tau_1)B+\ff 1\xi C+D}\Big)\,,\label{I2}
\end{align}
where
\be
\tau_{\circ}=\tau_1\circ\tau_1':=\tau_1+\tau_1'-2\tau_1\tau_1'\,,\quad
\xi=1-\gbzer\tau_{\circ}\,.
\ee
Using inequalities \eqref{ineq} it is easy to see that both
integrals vanish for $\tau_{\circ}\in[\gep, 1]$ at $\gbzer\to-\infty$,
where $0<\gep<1$ is some fixed $\gbzer$--independent number.
Therefore it  suffices to analyze the integrands in the
vicinity $\tau_{\circ}\to 0$. To do so it is convenient to introduce
\be\label{J}
J^{m_1,m_2}_{n_1,n_2}(\tau)=\int_{[0,1]^2} \dr\tau_1 \dr\tau_2\,
\tau_1^{m_1}\tau_2^{m_2}(1-\tau_1)^{n_1}(1-\tau_2)^{n_2}\gd(\tau-\tau_1\circ\tau_2)\,,
\ee
where $m_{1,2}\geq 0$ and $n_{1,2}\geq 0$. For small $\tau$,
$J(\tau)$ can be easily calculated with the coefficients expressed
via beta-function
\be\label{Jres}
J^{m_1,m_2}_{n_1,n_2}(\tau)=\left\{ \begin{array}{ll}
\ff{m_1!\,m_2!}{(m_1+m_2+1)!} \tau^{m_1+m_2+1}
+o( \tau^{m_1+m_2+1})\,, & m_1+m_2<n_1+n_2\,,\\
 \ff{n_1!\,n_2!}{(n_1+n_2+1)!}
\tau^{n_1+n_2+1} +o( \tau^{n_1+n_2+1})\,, & m_1+m_2>n_1+n_2\,,\\
\ff{2\,m_1!\,m_2!}{(m_1+m_2+1)!} \tau^{m_1+m_2+1} +o(
\tau^{m_1+m_2+1})\,, & m_1+m_2=n_1+n_2\,.
\end{array}\right.
\ee
Plugging  $\gd(\tau-\tau_1\circ\tau_1')$ into \eqref{I1},
\eqref{I2} and using \eqref{Jres} it is easy to take
$\gbzer\to-\infty$ to obtain (after introducing simplicial
integration variables) the following result
\begin{align}
&I_1(-\infty)=\int\dr^3_{\Delta}\tau\,\tau_1
\exp i\big({\tau_1A+\tau_2B+\tau_3C+ D}\big)\,,\label{I1inf}\\
&I_2(-\infty)=\int\dr^3_{\Delta}\tau\,\tau_3
\exp i\big({\tau_1A+\tau_2B+\tau_3C+ D}\big)\,.\label{I2inf}
\end{align}
Let us prove \eqref{I1inf}. Expanding \eqref{I1} in power series
we find the following integrals to be evaluated
\be
-\int \dr\tau_1\dr\tau_1'\ff{\gb^{3+m+n+k}\tau_1^{m+1}
\tau_1'^{n}(1-\tau_1)^{n+1}(1-\tau_1')^{m+2}}{(1-\gb\tau_{\circ})^{4+m+n+k}}\overset{\gb\to-\infty}{\to}
\int_{0}^{\infty}\dr
u\ff{u^{m+n+k+2}}{(1+u)^{m+n+k+4}}\int_{0}^{1}\dr\gl
\gl^{m+1}(1-\gl)^{n}\,,\nn
\ee
where in taking the limit we have introduced $u=-\gb\tau_{\circ}$ and
applied \eqref{Jres} in the form of integration over $\gl$.
Summing up series back into exponential and introducing simplicial
variables
\be
\tau_1=\ff{u\gl}{1+u}\,,\quad \tau_2=\ff{u(1-\gl)}{1+u}\,,\quad
\tau_3=\ff{1}{1+u}\,,\quad \sum_{i}\tau_i=1
\ee
we obtain \eqref{I1inf}. Eq.~\eqref{I2inf} can be worked out
analogously.

\subsection*{$W$ to the second order}
\label{App4}

Here we give explicit formulae used in calculation of the
(anti)holomorphic part of $W_2$ for finite $\gb$
\be
W_2^{\eta\eta}{}= \hmt_{0,\,\gbzer} (\{W_1^\eta{}\,,
S_1^\eta{}\}_*) +\ldots\,,
\ee
where ellipses denotes the terms $DS_2$ that do not contribute to
the final result. \begin{multline} \hmt_{0,\,\gbzer}
\left(W_1^\eta{}{}_{\,\go C}* S_1^\eta{}\right)=
i\ff{\eta^2}{4}\int_0^1 \dr\gr \int\dr_{\Delta}^3\tau \int_0^1 \dr\gs \times\\
\times
\ff{\p }{\p  (  p_{2\gamma})}
\ff{ t_{\alpha}\p }{\p  (  p_{1\alpha})}
\ff{[(1-{\gbzer} \tau_1\circ \gs)z
-{\gbzer}
(\tau_1\gs+\tau_3(1-\gs)t+\tau_1\gs(p_1+p_2))]_\gamma}
{\tau_1 \zeta^3}\times\\
\times \exp \bigg(\frac{i}{\zeta}
\Big[\gr\tau_1\circ \gs z_\ga y^\ga+ \gs(\tau_1-\tau_3) p_2{}_\ga t^\ga
+ \tau_1\gs  p_2{}_\ga p_1^\ga+(-\gr\tau_1(1-\gs)z+\tau_1 \gs y)^\alpha  p_{1\alpha}\\
+(-\gr\tau_1(1-\gs)z-\gr\gs\tau_3 z +\tau_1 \gs y+\tau_3 (1-\gs)y)^\alpha
t_{\alpha}+(\gr\gs(1-\tau_1)z+\tau_1 \gs y)^\alpha p_{2\alpha}\Big]\\
+ i(1-\tau_3)p_1{}_\ga t^\ga- i{\gbzer}\frac{(1-\gr)\tau_1\tau_3(1-2\gs)}{\zeta}
p_1{}_\ga t^\ga 
+i{\gbzer} \frac{(1-\gr)\tau_3 \gs}{\zeta} p_2{}_\ga t^\ga
\bigg)\go CC\,,
\end{multline}
\begin{multline}
\hmt_{0,\,\gbzer} \left(W_1^\eta{}{}_{\,C \go}*
S_1^\eta{}\right)=
i\ff{\eta^2}{4}\int_0^1 \dr\gr \int\dr_{\Delta}^3\tau \int_0^1 \dr\gs \times\\
\times \ff{\p }{\p  (  p_{2\gamma})} \ff{\p }{\p  (  p_{1\alpha})}   t_{\alpha}
\ff{[(1-{\gbzer} \tau_1\circ \gs)z-{\gbzer}((\tau_1\gs-\tau_3(1-\gs)t
+ \tau_1\gs( p_1+ p_2))]_\gamma}{\tau_1 \zeta^3}\times\\
\times \exp \bigg(\frac{i}{\zeta}
\Big[\gr\tau_1\circ \gs z_\ga y^\ga+ \gs(\tau_1+\tau_3) p_2{}_\ga t^\ga +\tau_1\gs  p_2{}_\ga p_1^\ga
+(-\gr\tau_1(1-\gs)z+\tau_1 \gs y)^\alpha   p_{1\alpha}\\
+(-\gr\tau_1(1-\gs)z+\gr\gs\tau_3 z +\tau_1 \gs y
-\tau_3 (1-\gs)y)^\alpha   t_{\alpha}+(\gr\gs(1-\tau_1)z+\tau_1 \gs y)^\alpha  p_{2\alpha}\Big]\\
-i(1-\tau_3)  p_1{}_\ga t^\ga+ i{\gbzer}\frac{(1-\gr)\tau_1\tau_3(1-2\gs)}{\zeta}p_1{}_\ga t^\ga
 - i{\gbzer} \frac{(1-\gr)\tau_3 \gs}{\zeta}p_2{}_\ga t^\ga
\bigg)C \go C\,,
\end{multline}
\begin{multline}
\hmt_{0,\,\gbzer} \left(S_1^\eta{}* W_1^\eta{}{}_{\,\go
C}\right)
=i\ff{\eta^2}{4}\int_0^1 \dr\gr \int\dr_{\Delta}^3\tau \int_0^1 \dr\gs \times\\
\times \ff{\p }{\p  (  p_{1\gamma})} \ff{ t_{\alpha}\p }{\p  (  p_{2\alpha})}
\ff{[(1-{\gbzer} \tau_1\circ \gs)z-{\gbzer}((\tau_1\gs-\tau_3(1-\gs)t
-\tau_1\gs( p_1+p_2))]_\gamma}{\tau_1 \zeta^3}\times\\
\times \exp
\bigg(\frac{i}{\zeta}
\Big[\gr\tau_1\circ \gs z_\ga y^\ga+ \gs(\tau_1+\tau_3)t_\ga  p_1^\ga
+ \tau_1\gs  p_2{}_\ga p_1^\ga+ (-\gr\gs(1-\tau_1)z+\tau_1 \gs y)^\alpha   p_{1\alpha}\\
+(\gr\tau_1(1-\gs)z-\gr\gs\tau_3 z +\tau_1 \gs y-\tau_3 (1-\gs)y)^\alpha    t_{\alpha}
+(\gr\tau_1(1-\gs)z+\tau_1 \gs y)^\alpha  p_{2\alpha}\Big]\\
+i (1-\tau_3)  p_2{}_\ga t^\ga
-i {\gbzer}\frac{(1-\gr)\tau_1\tau_3(1-2\gs)}{\zeta} p_2{}_\ga t^\ga
- i {\gbzer} \frac{(1-\gr)\tau_3 \gs}{\zeta}t_\ga p_1^\ga
\bigg)C \go C\,,
\end{multline}
\begin{multline}
\hmt_{0,\,\gbzer} \left(S_1^\eta{}* W_1^\eta{}{}_{\,C\go}\right)=i\ff{\eta^2}{4}\int_0^1 \dr\gr \int\dr_{\Delta}^3\tau \int_0^1 \dr\gs \times\\
\times \ff{\p }{\p  (  p_{1\gamma})} \ff{\p }{\p
(  p_{2\alpha})}   t_{\alpha}\ff{[(1-{\gbzer} \tau_1\circ \gs)z
+i{\gbzer}((\tau_1\gs+\tau_3(1-\gs)\p _\go+\tau_1\gs(\p _1+\p _2))]_\gamma}{\tau_1 \zeta^3}\times\\
\times \exp \bigg(\frac{i}{\zeta}
\Big[\gr\tau_1\circ \gs z_\ga y^\ga
+ \gs(\tau_1-\tau_3)t_\ga  p_1^\ga + \tau_1\gs  p_2{}_\ga p_1^\ga
+(-\gr\gs(1-\tau_1)z+\tau_1 \gs y)^\alpha  p_{1\alpha}\\
+(\gr\tau_1(1-\gs)z+\gr\gs\tau_3 z +\tau_1 \gs y+\tau_3 (1-\gs)y)^\alpha    t_{\alpha}
+(\gr\tau_1(1-\gs)z+\tau_1 \gs y)^\alpha  p_{2\alpha}\Big]\\
+ i(1-\tau_3)t_\ga  p_2^\ga
- i{\gbzer}\frac{(1-\gr)\tau_1\tau_3(1-2\gs)}{\zeta}t_\ga  p_2^\ga
 +i {\gbzer} \frac{(1-\gr)\tau_3 \gs}{\zeta}t_\ga  p_1^\ga
\bigg)CC \go\,,
\end{multline}
where $\circ$  is defined in \eqref{kruzhochek}, $ p_{1}$ and $p_{2}$ \eq{CgoCgoCpt}  
 differentiate $C's$ as seen from left and
\begin{equation}
\zeta=1-{\gbzer}(1-\gr)(\tau_1\circ \gs)\,.
\end{equation}

\addtocounter{appendix}{1} \setcounter{equation}{0}

  \addtocounter{section}{1}
 \addcontentsline{toc}{section}{\,\,\,\,\,\,\,Appendix C. Lower order vertices  }
 \section*{Appendix C. Lower-order vertices}
\label{AppC}

For the reader's convenience  the formulae for lower-order
vertices  are presented in this Appendix.

 \be\label{go1}
\dr_x\go=-\go*\go+\Upsilon^{\eta}_{\go\go
C}+\Upsilon^{\eta}_{C\go\go}+\Upsilon^{\eta}_{\go C\go}+\Upsilon^{\bar\eta}_{\go\go
C}+\Upsilon^{\bar\eta}_{C\go\go}+\Upsilon^{\bar\eta}_{\go C\go}+\ldots\,,
\ee
\be\label{C2gen}
\dr_x C=-[\go,C]_*+  \Upsilon^{\eta}_{\go CC}+\Upsilon^{\eta}_{CC\go}+\Upsilon^{\eta}_{C\go C}+
 \Upsilon^{\bar\eta}_{\go CC}+\Upsilon^{\bar\eta}_{CC\go}+\Upsilon^{\bar\eta}_{C\go C}+\ldots\,,
\ee
where \cite{Didenko:2018fgx}
\bee
&\Upsilon^{\eta}_{\go\go C}=\ff{\eta}{4i} \go*\go*C* h_{p+t_1+t_2}\hmt_{p}\hmt_{p+t_2
}\gga\,,
\label{goUps1}\\
&\Upsilon^{\eta}_{C\go\go}=\ff{\eta}{4i}C*\go*\go* h_{p+t_1+t_2}\hmt_{p+t_1+2t_2}\hmt_{p+2t_1+2t_2
}\gga\,,
\label{goUps2}\\
&\Upsilon^{\eta}_{\go C\go}=-\ff{\eta}{4i}
\go*C*\go*\big(
h_{p+t_1+t_2}\hmt_{p+t_1+2t_2}\hmt_{p+t_2}\gga
-h_{p+t_1+2t_2}\hmt_{p+2t_2}\hmt_{p+t_2}\gga\,
\big)\,,
\label{goUps3}
\eee
 \bee
&  \Upsilon^{\eta}_{\go CC}=\ff{\eta}{4i}
\go\!*\!C\!*\!C\!*h_{p_2}
\hmt_{p_1+2p_2}\hmt_{p_1+2p_2+t}\gga\,,\label{CUps1loc}
\\
&\Upsilon^{\eta}_{CC\go}=\ff{\eta}{4i}C\!*\!C\!*\!\go\!*h_{p_2+2t}
\hmt_{p_2+t}\hmt_{p_1+2p_2+2t}\gga\,,\label{CUps2loc}\\
&\Upsilon^{\eta}_{C\go C}=\ff{\eta}{4i}C\!*\!\go\!*\!C\!*
(h_{p_1+2p_2+2t}-h_{p_2})\hmt_{p_2+t}\hmt_{p_1+2p_2+t}\gga\label{CUps3loc}\,
\eee
and
\bee
&\Upsilon^{\bar\eta}_{\go\go C}=\ff{\bar{\eta}}{4i} \go*\go*C* \bar h_{\bar p+\bar t_1+\bar t_2}
\bar \hmt_{\bar p}\bar\hmt_{\bar p+\bar t_2}\bar \gga\,,
\label{ccgoUps1}\\
&\Upsilon^{\bar\eta}_{C\go\go}=\ff{\bar{\eta} }{4i}C*\go*\go* \bar{h}_{\bar p+\bar{t}_1
+\bar{t}_2}\bar{\hmt}_{p+\bar{t}_1+2\bar{t}_2}\bar{\hmt}_{p+2\bar{t}_1+2\bar{t}_2
}\bar \gga\,,
\label{ccgoUps2}\\
&\Upsilon^{\bar\eta}_{\go C\go}=-\ff{\bar{\eta}}{4i}
\go*C*\go*\big(
\bar{h}_{\bar p+\bar{t}_1+\bar{t}_2}\bar{\hmt}_{\bar p+\bar{t}_1+2\bar{t}_2}\bar{\hmt}_{\bar p+\bar{t}_2}\bar \gga
-\bar{h}_{\bar p+\bar{t}_1+2\bar{t}_2}\bar{\hmt}_{\bar p+2\bar{t}_2}\bar{\hmt}_{\bar p+\bar{t}_2}\bar \gga\,
\big)\,,
\label{ccgoUps3}
\eee
 \bee
&  \Upsilon^{\bar\eta}_{\go CC}=\ff{\bar{\eta}}{4i} \go\!*\!C\!*\!C\!*\bar{h}_{ \bar p_2}
\bar{\hmt}_{ \bar p_1+2 \bar p_2}\bar{\hmt}_{ \bar p_1+2 \bar p_2+\bar t}\bar \gga\,,\label{ccCUps1loc}
\\
&\Upsilon^{\bar\eta}_{CC\go}=\ff{\bar{\eta}}{4i}C\!*\!C\!*\!\go\!*\bar{h}_{ \bar p_2+2\bar t}
\bar{\hmt}_{ \bar p_2+\bar t}\bar{\hmt}_{ \bar p_1+2 \bar p_2+2\bar t}\bar \gga\,,\label{ccCUps2loc}\\
&\Upsilon^{\bar\eta}_{C\go C}=\ff{\bar{\eta}}{4i}C\!*\!\go\!*\!C\!*
(\bar{h}_{ \bar p_1+2 \bar p_2+2\bar t}-\bar{h}_{ \bar p_2})\bar{\hmt}_{ \bar p_2+\bar t}
\bar{\hmt}_{ \bar p_1+2 \bar p_2+\bar t}\bar \gga\label{ccCUps3loc}\,.
\eee
$\Upsilon(\go,C,C)$ vertices given here are derived for $B$ field
given in \eqref{b2}.

To see that $\Upsilon (\go,\go,C)$ are ultra-local (\ie $C$ is
$y$-independent) one can make sure using Eqs.~\eqref{oldres},
\eqref{oldproj} and \eqref{CgoCgoCpt}
 that for $a,b,c$ independent of  $p_1$
 \bee\nn C(y) *  h_{ p_1{}+c}
   \hmt_{ p_1{}+b} \hmt_{ p_1{}+a}\gga
   =  2\!  \int_{0}^{1}\!\dr_{\Delta}^3\tau C(y)*(b-c)_\gga
 (a-c)^\gga
\exp(  {-i ( p_1{}+\tau_1 c+\tau_2 a+\tau_3 b )_\ga y^\ga})k\,\quad
 \\  \label{AdefHhhgam} =  2
 \int_{0}^{1} \dr_{\Delta}^3\tau\, C(0)*(b-c)_\gga
 (a-c)^\gga
\exp(  {-i (\tau_1 c+\tau_2 a+\tau_3 b )_\ga y^\ga})k\,.
\qquad\eee

\end{document}